\documentclass[12pt,a4paper]{article}

\usepackage{graphicx,psfrag}

\begin{document}


\noindent
DESY 04-177, SFB/CPP-04-49\hfill{\tt hep-lat/0409141\_v2}\\
Sept 2004 -- Jan 2005
\vspace{15pt}

\begin{center}
{\LARGE\bf Gauge action improvement and smearing}
\end{center}
\vspace{5pt}

\begin{center}
{\large\bf Stephan D\"urr}\\[2mm]
DESY Zeuthen, 15738 Zeuthen, Germany\\
(present address: Universit\"at Bern, ITP, Sidlerstr.\ 5, 3012 Bern, Switzerland)
\end{center}
\vspace{5pt}

\begin{abstract}
\noindent
The effect of repeatedly smearing $SU(3)$ gauge configurations is investigated.
Six gauge actions (Wilson, Symanzik, Iwasaki, DBW2, Beinlich-Karsch-Laermann,
Langfeld; combined with a direct $SU(3)$-overrelaxation step) and three
smearings (APE, HYP, EXP) are compared. The impact on large Wilson loops is
monitored, confirming the signal-to-noise prediction by Lepage. The fat-link
definition of the ``naive'' topological charge proves most useful on improved
action ensembles.
\end{abstract}


\newcommand{\pa}{\partial}
\newcommand{\pas}{\partial\!\!\!/}
\newcommand{\Dsl}{D\!\!\!\!/\,}
\newcommand{\hqu}{\hbar}
\newcommand{\ovr}{\over}
\newcommand{\til}{\tilde}
\newcommand{\pri}{^\prime}
\renewcommand{\dag}{^\dagger}
\newcommand{\<}{\langle}
\renewcommand{\>}{\rangle}
\newcommand{\gaf}{\gamma_5}
\newcommand{\lap}{\triangle}
\newcommand{\trc}{\mr{tr}}
\newcommand{\al}{\alpha}
\newcommand{\be}{\beta}
\newcommand{\ga}{\gamma}
\newcommand{\de}{\delta}
\newcommand{\ep}{\epsilon}
\newcommand{\ve}{\varepsilon}
\newcommand{\ze}{\zeta}
\newcommand{\et}{\eta}
\renewcommand{\th}{\theta}
\newcommand{\vt}{\vartheta}
\newcommand{\io}{\iota}
\newcommand{\ka}{\kappa}
\newcommand{\la}{\lambda}
\newcommand{\rh}{\rho}
\newcommand{\vr}{\varrho}
\newcommand{\si}{\sigma}
\newcommand{\ta}{\tau}
\newcommand{\ph}{\phi}
\newcommand{\vp}{\varphi}
\newcommand{\ch}{\chi}
\newcommand{\ps}{\psi}
\newcommand{\om}{\omega}
\newcommand{\psb}{\overline{\psi}}
\newcommand{\etb}{\overline{\eta}}
\newcommand{\psd}{\psi^{\dagger}}
\newcommand{\etd}{\eta^{\dagger}}
\newcommand{\beq}{\begin{equation}}
\newcommand{\eeq}{\end{equation}}
\newcommand{\bdm}{\begin{displaymath}}
\newcommand{\edm}{\end{displaymath}}
\newcommand{\bea}{\begin{eqnarray}}
\newcommand{\eea}{\end{eqnarray}}
\newcommand{\mr}{\mathrm}
\newcommand{\mb}{\mathbf}
\newcommand{\Nf}{N_{\!f}}
\newcommand{\Nc}{N_{\!c}}
\newcommand{\ii}{\mr{i}}
\newcommand{\fm}{\,\mr{fm}}
\newcommand{\MeV}{\,\mr{MeV}}
\newcommand{\GeV}{\,\mr{GeV}}
\newcommand{\Mpi}{M_{\pi}}
\newcommand{\Fpi}{F_{\pi}}


\hyphenation{topo-lo-gi-cal simu-la-tion gra-nu-lar theo-re-ti-cal
granu-la-ri-ty mini-mum fluc-tua-tions}


\section{Introduction}


Smearing of gauge-links has been used on many occasions as a powerful tool to
reduce the UV-fluctuations of a gauge configuration
\cite{ape,hyp,Orginos:1999cr,exp,uvfiltering}.
In principle, such a UV-filtering  may be attempted either in the Monte Carlo
process itself (by complementing the action with contributions from
$W^{1\times2+2\times1}$ and even larger Wilson loops) or in the observable
(where replacing the original links by smeared ones is just the simplest way to
define a filtered observable).
Normally, the motivation for the first option is not to tame the noise in
standard observables, but to reduce cut-off effects.
Nonetheless, an obvious question is whether there is an interplay between the
action used to generate a pure gauge ensemble and the effectiveness of a
smearing recipe to decrease the UV-fluctuations in a given observable.

\begin{table}[b]
\begin{center}
{\small
\begin{tabular}{|c|cccc|cccc|cccc|}
\hline
&1&2&3&4&1&2&3&4&1&2&3&4\\
\hline
1&$\!0.5937\!$&$\!      \!$&$\!      \!$&$\!      \!$&
  $\!0.9509\!$&$\!      \!$&$\!      \!$&$\!      \!$&
  $\!0.0399\!$&$\!      \!$&$\!      \!$&            \\
2&$\!0.3836\!$&$\!0.1902\!$&$\!      \!$&$\!      \!$&
  $\!0.9015\!$&$\!0.7551\!$&$\!      \!$&$\!      \!$&
  $\!0.0073\!$&$\!0.0011\!$&$\!      \!$&            \\
3&$\!0.2527\!$&$\!0.1014\!$&$\!0.0471\!$&$\!      \!$&
  $\!0.8399\!$&$\!0.6205\!$&$\!0.4551\!$&$\!      \!$&
  $\!0.0013\!$&$\!0.0002\!$&$\!0.0000\!$&            \\
4&$\!0.1673\!$&$\!0.0553\!$&$\!0.0229\!$&$\!0.0101\!$&
  $\!0.7843\!$&$\!0.5149\!$&$\!0.3416\!$&$\!0.2357\!$&
  $\!0.0002\!$&$\!0.0000\!$&$\!0.0000\!$&$\!0.0000\!$\\
\hline
\end{tabular}
}
\end{center}
\vspace{-6mm}
\caption{\sl Wilson loops $\<W(R,T)\>$ with $R,T\in\{1,...,4\}$ on a $12^4$
lattice with Wilson glue at $\beta\!=\!6.0$ {\rm (left)}, after one HYP-step
with $\alpha\!=\!(0.3,0.6,0.75)$ and $SU(3)$-projection {\rm (middle)} or the
same $\alpha$ without any projection {\rm (right)}. Statistical errors range
from 1 to 10 in the last digit.}
\label{tab:preview}
\end{table}

Some expectation values depend in a rather sensitive way on the details of the
smearing procedure.
As an example, Table~\ref{tab:preview} contains expectation values of a few
Wilson loops without smearing and after two smearing recipes have been applied.
Obviously, the plaquette and larger Wilson loops take rather different values
in these three cases.
However, since from the difference of the rows one may compute the heavy-quark
(HQ) force, and such a physical observable is supposed to be unaltered by the
smearing, the induced change in the $W(R,T)$ must take place in a coherent
manner for all $R,T$.
Obviously, not only the core recipe (e.g.\ APE, HYP, EXP -- see below) but also
the parameters used and the number of iterations will have an impact on the
quality of the result.
Basic questions that one may have at this point include:
\begin{enumerate}
\vspace{-0mm}
\itemsep -1mm
\item For a given recipe, how can one identify useful combinations of parameter
      values and iteration levels, and is there a specific signature of doing
      ``too much'' smearing ?
\item Optimizing parameter and iteration level w.r.t.\ a certain criterion,
      will the result depend on the gauge action used, or will it be universal
      for a given lattice spacing ?
\item Is there a physics reason why some smearing recipes and/or parameters
      prove more efficient in damping UV-fluctuations than other ?
\vspace{-0mm}
\end{enumerate}

The plan of this article is to first list and review the actions studied.
Then the smearing procedures are discussed, shifting technical
details of the $SU(3)$-projection to the appendix.
The key physics point is an explicit test in the heavy-heavy case of the Lepage
argument which establishes a relationship between the decrease of the signal in
large Wilson loops and the self energy of an infinitely heavy quark.
In order to give a fair comparison of different actions approximately matched
ensembles must be generated and I describe how the associated $\be$-values have
been found with little CPU costs via the local force.
With these at hand, I study of the effect previewed in
Table~\ref{tab:preview}, combining all actions and smearing recipes.
The result is that for Wilson loops at fixed lattice spacing the effect of
smearing is almost independent of the gauge action used.
Finally, I point out that on sufficiently fine lattices smearing allows for a
cheap definition of the field-theoretic topological charge, most notably on
improved glue.


\section{Action details}


We consider gauge actions involving rectangular Wilson loops up to maximum side
length 2.

The action involving nothing but the loop $W^{1\times1}$ is the one due to
Wilson~\cite{Wilson:1974sk}
\begin{equation}
S^{1\times1}=\be\,\sum_{x,\mu<\nu}
1-{1\ovr3}\mr{Re}\,\trc\,W^{1\times1}_{\mu\nu}(x)
\;.
\end{equation}
Under a change of the link $U_\mu(x)$ the acceptance is
determined by the ``local action''
\begin{equation}
S_\mr{loc}^{1\times1}=
-{\be\ovr3}\,\mr{Re}\,\trc\Big(S^{1\times1}_\mu(x)\dag\,U_\mu(x)\Big)
\label{varSmu11}
\end{equation}
where $\mu$ is not summed over and the staple is the sum of all 3-link paths
from $x$ to $x\!+\!\hat\mu$
\begin{equation}
S^{1\times1}_\mu(x)=\sum_{\pm\nu\neq\mu}
U_\nu(x)\,U_\mu(x\!+\!\hat\nu)\,U_\nu(x\!+\!\hat\mu)\dag
\;.
\label{defS11}
\end{equation}
For future reference we define the rescaled Wilson staple
$\til S_\mu^{1\times1}\!=\!{1\ovr6}\,S_\mu^{1\times1}$ where the prefactor
simply reflects the 6 contributions to the sum in (\ref{defS11}).

The class involving the loops $W^{1\times1}$ and
$W^{1\times2+2\times1}=W^{1\times2}\!+\!W^{2\times1}$ has the generic form
\begin{equation}
S^{1\times2+2\times1}=
\be\!\sum_{x,\mu<\nu}
(1\!-\!8c_1)\Big(1-{1\ovr3}\mr{Re}\,\trc\,W^{1\times1}_{\mu\nu}(x)\Big)+
\til c_1\Big(2-{1\ovr3}\mr{Re}\,\trc\,W^{1\times2+2\times1}_{\mu\nu}(x)\Big)
\;.
\end{equation}
Under a change of $U_\mu(x)$ the acceptance is determined by the change of (no
sum over $\mu$)
\begin{equation}
S_\mr{loc}^{1\times2+2\times1}=
-{\be\ovr3}\,\mr{Re}\,\trc\Big(
[(1\!-\!8c_1)S^{1\times1}_\mu(x)+\til c_1S^{1\times2+2\times1}_\mu(x)]\dag
\,U_\mu(x)\Big)
\label{varSmu12Smu21}
\end{equation}
where $S^{1\times2+2\times1}_\mu=S^{1\times2}_\mu\!+\!S^{2\times1}_\mu$
generates all planar 5-link paths from $x$ to $x\!+\!\hat\mu$, viz.\
\begin{eqnarray}
S^{1\times2}_\mu(x)&=&\sum_{\pm\nu\neq\mu}
U_\nu(x)\,U_\nu(x\!+\!\hat\nu)\,U_\mu(x\!+\!2\hat\nu)
\,U_\nu(x\!+\!\hat\mu\!+\!\hat\nu)\dag\,U_\nu(x\!+\!\hat\mu)\dag
\label{def_Smu12}
\\
S^{2\times1}_\mu(x)&=&\sum_{\pm\nu\neq\mu}
U_\nu(x)\,U_\mu(x\!+\!\hat\nu)\,U_\mu(x\!+\!\hat\mu\!+\!\hat\nu)\,
U_\nu(x\!+\!2\hat\mu)\dag\,U_\mu(x\!+\!\hat\mu)\dag
\nonumber
\\
&+&\sum_{\pm\nu\neq\mu}
U_\mu(x\!-\!\hat\mu)\dag\,U_\nu(x\!-\!\hat\mu)\,
U_\mu(x\!-\!\hat\mu\!+\!\hat\nu)\,
U_\mu(x\!+\!\hat\nu)\,U_\nu(x\!+\!\hat\mu)\dag
\;.
\label{def_Smu21}
\end{eqnarray}

The class of actions involving the Wilson loops $W^{1\times1}$ and
$W^{2\times2}$ has the generic form
\begin{equation}
S^{2\times2}=
\be\!\sum_{x,\mu<\nu}
(1\!-\!16c_2)
\Big(1-{1\ovr3}\mr{Re}\,\trc\,W^{1\times1}_{\mu\nu}(x)\Big)+
\til c_2
\Big(1-{1\ovr3}\mr{Re}\,\trc\,W^{2\times2}_{\mu\nu}(x)\Big)
\;.
\end{equation}
Under a change of $U_\mu(x)$ the acceptance is determined by the change of
(no sum over $\mu$)
\begin{equation}
S_\mr{loc}^{2\times2}=
-{\be\ovr3}\,\mr{Re}\,\trc\Big(
[(1\!-\!16c_2)S^{1\times1}_\mu(x)+\til c_2S^{2\times2}_\mu(x)]\dag
\,U_\mu(x)\Big)
\label{varSmu22} 
\end{equation}
where the new staple $S^{2\times2}$ is designed to generate those 7-link paths
from $x$ to $x\!+\!\hat\mu$ which, together with $U_\mu(x)\dag$, form a
$2\!\times\!2$ square, viz.
\begin{displaymath}
S^{2\times2}_\mu(x)=\sum_{\pm\nu\neq\mu}
U_\nu(x)\,
U_\nu(x\!+\!\hat\nu)
U_\mu(x\!+\!2\hat\nu)\,U_\mu(x\!+\!\hat\mu\!+\!2\hat\nu)\,
U_\nu(x\!+\!2\hat\mu\!+\!\hat\nu)\dag\,
U_\nu(x\!+\!2\hat\mu)\dag\,U_\mu(x\!+\!\hat\mu)\dag
\end{displaymath}
\begin{equation}
+\sum_{\pm\nu\neq\mu}
U_\mu(x\!-\!\hat\mu)\dag\,U_\nu(x\!-\!\hat\mu)\,
U_\nu(x\!-\!\hat\mu\!+\!\hat\nu)\,
U_\mu(x\!-\!\hat\mu\!+\!2\hat\nu)\,U_\mu(x\!+\!2\hat\nu)\,
U_\nu(x\!+\!\hat\mu\!+\!\hat\nu)\dag\,
U_\nu(x\!+\!\hat\mu)\dag
\;.
\label{def_Smu22}
\end{equation}

Specific choices for $c_1$ in the $W^{1\times1}\!+\!W^{1\times2+2\times1}$
class include ($\til c_1\!=\!c_1$ unless stated otherwise) 
\begin{equation}
c_1=\left\{
\begin{array}{cl}
-1/12\;\;(\til c_1\!=\!c_1, \til c_1\!=\!c_1/u_0^2)
       &\mbox{Symanzik (tree-level, tadpole) \cite{Symanzik}}\\
-0.331 &\mbox{Iwasaki (RG) \cite{IwasakiTsukubaPreprint}}\\
-1.409 &\mbox{DBW2 (RG) \cite{deForcrand:1999bi}}
\end{array}
\right.
\label{choice_c1}
\end{equation}
with $u_0\equiv({1\ovr3}\mr{Re}\,\trc\<W^{1\times1}\>)^{1/4}$,
while the $c_2$ in the $W^{1\times1}\!+\!W^{2\times2}$ ansatz that I am aware
of are
\begin{equation}
c_2=\left\{
\begin{array}{cl}
-1/48\;\;(\til c_2\!=\!c_2, \til c_2\!=\!c_2/u_0^4)
       &\mbox{Beinlich-Karsch-Laermann (tree-level, tadpole) \cite{twobytwo}}\\
-0.538 &\mbox{Langfeld (HQ potential isotropy) \cite{Langfeld:2004tf}}
\;.
\end{array}
\right.
\label{choice_c2}
\end{equation}

\bigskip

From (\ref{def_Smu12}, \ref{def_Smu21}) and (\ref{def_Smu22}) it follows that
$S_\mu^{1\times2+2\times1}$ receives 18 contributions while $S_\mu^{2\times2}$
is built from 12 terms.
Disregarding the one-third matrix product needed to form the trace in
$\trc(S_\mu\dag U_\mu)$ (no sum) and counting only the multiplications
for the staple, one gets $6\cdot2=12$ operations with the
Wilson action, while the Symanzik-Iwasaki-DBW2 actions lead to
$6\cdot2+18\cdot4=84$ matrix multiplications and for the
Beinlich-Karsch-Laermann-Langfeld actions that figure is
$6\cdot2+12\cdot6=84$, too.
From such a counting one expects that the new actions are roughly a factor 7
more expensive (per sweep) than the Wilson gauge action, but of course the hope
is that this investment pays off by e.g.\ a better scaling behavior.
In any case it is clear that for the new actions the costs for a single-link
update is dominated by the staple.
In such a situation a multihit Metropolis is almost competitive with
a heatbath, since 1 or 10 hits bear almost the same costs and after 10 hits the
link is essentially in equilibrium.

A point worth mentioning is the observation that at standard $\be$-values a
``direct'' overrelaxation step in the full $SU(3)$-group proves remarkably
efficient, if properly implemented.
With the Wilson gauge action the current link is proposed to be replaced
according to
\begin{equation}
U_\mu(x)\quad\longrightarrow\quad U'_\mu(x)=
P_{SU(3)}\{\til S_\mu^{1\times1}(x)\}\,
U_\mu(x)\dag\,
P_{SU(3)}\{\til S_\mu^{1\times1}(x)\},
\label{propoverwils}
\end{equation}
with $\til S_\mu^{1\times1}$ defined below (\ref{defS11}) and $P_{SU(3)}$
specified in the appendix.
Unlike in $SU(2)$ this proposal must be subject to an accept/reject step.
It turns out that at $\be\!=\!6.0$ the acceptance rate in this overrelaxation
step is 99\%.
However, if we would keep the proposal (\ref{propoverwils}) with the
improved actions, the acceptance rate would degrade severely.
Fortunately, the generalized proposal
\begin{equation}
U_\mu(x)\quad\longrightarrow\quad U'_\mu(x)=
P_{SU(3)}\{\tilde S_\mu(x)\}\,
U_\mu(x)\dag\,
P_{SU(3)}\{\tilde S_\mu(x)\},
\label{propoverimpr}
\end{equation}
with the rescaled staple (normalization%
\footnote{With $P_{SU(3)}$ in (\ref{propoverimpr}) as given in the appendix
the rescaling is irrelevant, but with an improper one it matters.}
such that $\tilde S_\mu(x)\!\to\!1$ if formally the links tend to unity)
\begin{equation}
\tilde S_\mu(x)=
\left\{\!\!
\begin{array}{ll}
[(1\!-\!8c_1)S_\mu^{1\times1}(x)+\til c_1 S_\mu^{1\times2+2\times1}(x)]/
[6\!-\!48c_1\!+\!18\til c_1]&
\,\mbox{for}\,W^{1\times1}\!+\!W^{1\times2+2\times1}
\\[2mm]
[(1\!-\!16c_2)S_\mu^{1\times1}(x)+\til c_2 S_\mu^{2\times2}(x)]/
[6\!-\!96c_2\!+\!12\til c_2]&
\,\mbox{for}\,W^{1\times1}\!+\!W^{2\times2}
\end{array}
\right.
\label{propoverdetails}
\end{equation}
works much better.
Typical acceptance rates in this ``direct'' overrelaxation step at standard
$\be$-values (those determined in section 5 to produce matched ensembles)
are collected in Table~\ref{tab:overrelaxrates}.
The rates on the diagonal are satisfactory; it is hence advantageous to flip
in a quenched $SU(3)$-overrelaxation step with respect to the ``effective''
staple (\ref{propoverdetails}) of the gauge action used.

The data in this article have been generated with such a mixture of 4
``direct'' overrelaxation steps per 4-hit Metropolis update, separating
configurations by 10 sweeps.
While this is probably not quite competitive with the standard
Cabibbo-Marinari update algorithm~\cite{heatbathoverrelaxation}, it has the
great advantage of being easy to implement and represents a reasonable choice
whenever the CPU-costs are dominated by the measurement.

\begin{table}[t]
\begin{center}
\begin{tabular}{|l|cccccc|}
\hline
staple inside tr(.)
&$\mr{W}_{\be=6.0}$&$\mr{S}_{\be=4.439}$&$\mr{I}_{\be=2.652}$
&$\mr{DBW2}_{\be=1.047}$&$\mr{BKL}_{\be=4.729}$&$\mr{L}_{\be=0.796}$
\\
\hline
(\ref{varSmu11})                    &0.99&0.84&0.68&0.55&0.96&0.83
\\
(\ref{varSmu12Smu21}), $c_1=-1/12$  &0.99&0.99&0.99&0.99&0.96&0.83
\\
(\ref{varSmu12Smu21}), $c_1=-0.331$ &0.99&0.99&0.99&0.99&0.96&0.83
\\
(\ref{varSmu12Smu21}), $c_1=-1.409$ &0.99&0.99&0.99&0.99&0.96&0.83
\\
(\ref{varSmu22}), $c_2=-1/48$       &0.99&0.84&0.68&0.55&0.99&0.99
\\
(\ref{varSmu22}), $c_2=-0.538$      &0.99&0.84&0.68&0.55&0.99&0.99
\\
\hline
\end{tabular}
\vspace{-5mm}
\end{center}
\caption{\sl Acceptance probabilities in the ``direct'' overrelaxation step if
the proposed flip $(\ref{propoverwils}, \ref{propoverimpr})$ is w.r.t.\ the
same staple as used in the gauge action $($diagonal entries$)$ or another
one. These figures are specific for the ``overlap'' projection procedure
{\rm \cite{Kiskis:2003rd}} as detailed in the appendix. The rates have been
determined from $100$ sweeps on a $8^4$ lattice, after thermalization.}
\label{tab:overrelaxrates}
\end{table}


\section{Smearing details}


The smearing prescriptions considered in this article will be dubbed ``APE'',
``HYP'' and ``EXP'', and they are defined as follows.

The APE procedure~\cite{ape} replaces (in smearing step $n$) the existing set
of links $\{U^{(n-1)}\}$ by
\begin{equation}
U_\mu^{(n)}(x)=P_{SU(3)}
\Big\{\,
(1\!-\!\al)U_\mu^{(n-1)}(x)+{\al\ovr6}S_\mu^{(n-1)}(x)\,
\Big\}
\label{defape}
\end{equation}
where $S_\mu^{(n-1)}(x)$ denotes the $1\!\times\!1$ staple (\ref{defS11}) built
from $\{U^{(n-1)}\}$ links.
The parameter $\al$ governs the relative weight of the fluctuation and the
projection $P_{SU(3)}$ may be applied optionally.

The HYP procedure has been introduced in \cite{hyp} and defines the replacement
via
\begin{eqnarray}
\bar V_{\mu,\nu\si}(x)&=&P_{SU(3)}
\Big\{\,
(1\!-\!\al_1)U_\mu^{(n-1)}(x)+{\al_1\ovr2}\sum_{\pm\rh\neq(\mu,\nu,\si)}
U^{(n-1)}_{\rh}(x)\,
U^{(n-1)}_{\mu}(x\!+\!\hat\rh)\,
U^{(n-1)}_{\rh}(x\!+\!\hat\mu)\dag\,
\Big\}
\nonumber\\
\til V_{\mu,\nu}(x)&=&P_{SU(3)}
\Big\{\,
(1\!-\!\al_2)U_\mu^{(n-1)}(x)+{\al_2\ovr4}\;\sum_{\pm\si\neq(\mu,\nu)}
\bar V_{\si,\mu\nu}(x)\,
\bar V_{\mu,\nu\si}(x\!+\!\hat\si)\,
\bar V_{\si,\mu\nu}(x\!+\!\hat\mu)\dag\,
\Big\}
\nonumber\\
U_\mu^{(n)}(x)&=&P_{SU(3)}
\Big\{\,
(1\!-\!\al_3)U_\mu^{(n-1)}(x)+{\al_3\ovr6}\;\;\sum_{\pm\nu\neq(\mu)}
\til V_{\nu,\mu}(x)\,
\til V_{\mu,\nu}(x\!+\!\hat\nu)\,
\til V_{\nu,\mu}(x\!+\!\hat\mu)\dag\,
\Big\}
\label{defhyp}
\end{eqnarray}
where I have taken the liberty to interchange $\al_1\leftrightarrow\al_3$
compared to \cite{hyp}, i.e.\ our $\al_1$ denotes the fluctuation weight in the
first step.
Again, $\al_1,...,\al_3$ will be considered free parameters and the projection
will be treated as an option.

The EXP procedure has been introduced in \cite{exp} (there the resulting links
are called ``stout links'') and is defined through (no summation over $\mu$)
\begin{equation}
U_\mu^{(n)}(x)=
\exp\Big(\,{\al\ovr2}
\Big\{
[S^{(n-1)}_\mu(x)U^{(n-1)}_\mu(x)\dag\!-\!U^{(n-1)}_\mu(x)S^{(n-1)}_\mu(x)\dag]
-{1\ovr3}\trc[...]
\Big\}
\,\Big)\,
U_\mu^{(n-1)}(x)
\label{defexp}
\end{equation}
where, in principle, any staple could be used, but we shall restrict ourselves
to the simplest choice, $S_\mu(x)=S_\mu^{1\times1}(x)$.
Upon expanding the exponential, one realizes that this is the resummed version
of the ansatz (8) in Ref.~\cite{Orginos:1999cr}.
The key feature of such a construct is that the resulting links $U^{(n)}$ are
differentiable w.r.t\ the input links $U^{(n-1)}$; hence the recipe
(\ref{defexp}) may be used in a HMC \cite{Orginos:1999cr,exp}, which is not the
case for (\ref{defape}, \ref{defhyp}) unless the projection is abandoned.


\section{Relationship with reduced HQ self energy}


The naive view is that substituting the ``thin'' links in a Wilson loop with
4D-smeared ``fat'' links replaces the static quark by a quasi-static one which
is allowed to rattle inside the 3D-cube perpendicular to the HQ line.
However, as discussed in \cite{DellaMorte},  at least for the time-like links
in the Wilson loop $W(R,T)$ it creates a legitimate -- and in fact better -- HQ
action.

\bigskip

\begin{figure}
\begin{center}
\includegraphics[width=11.2cm,angle=270]{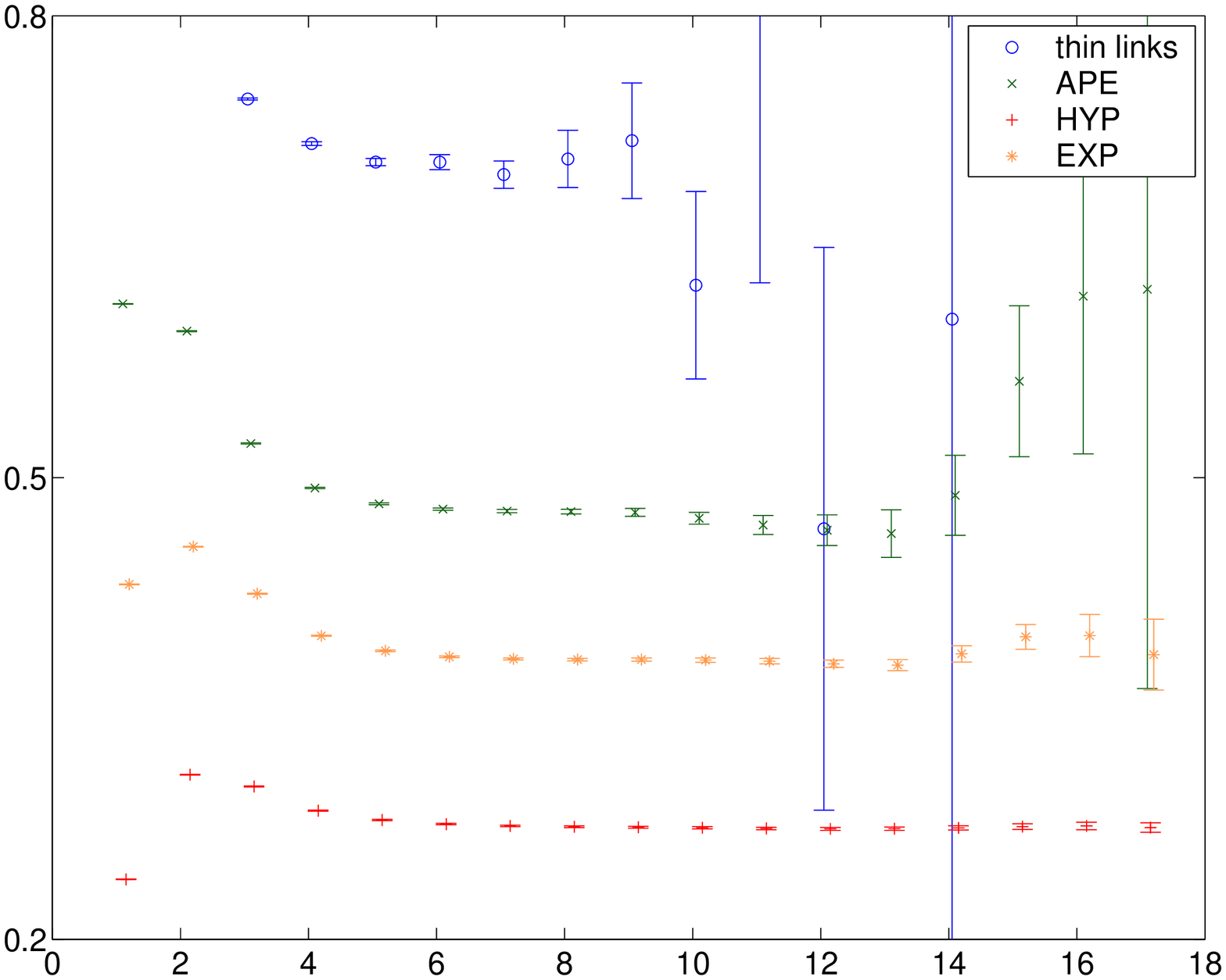}
\includegraphics[width=11.2cm,angle=270]{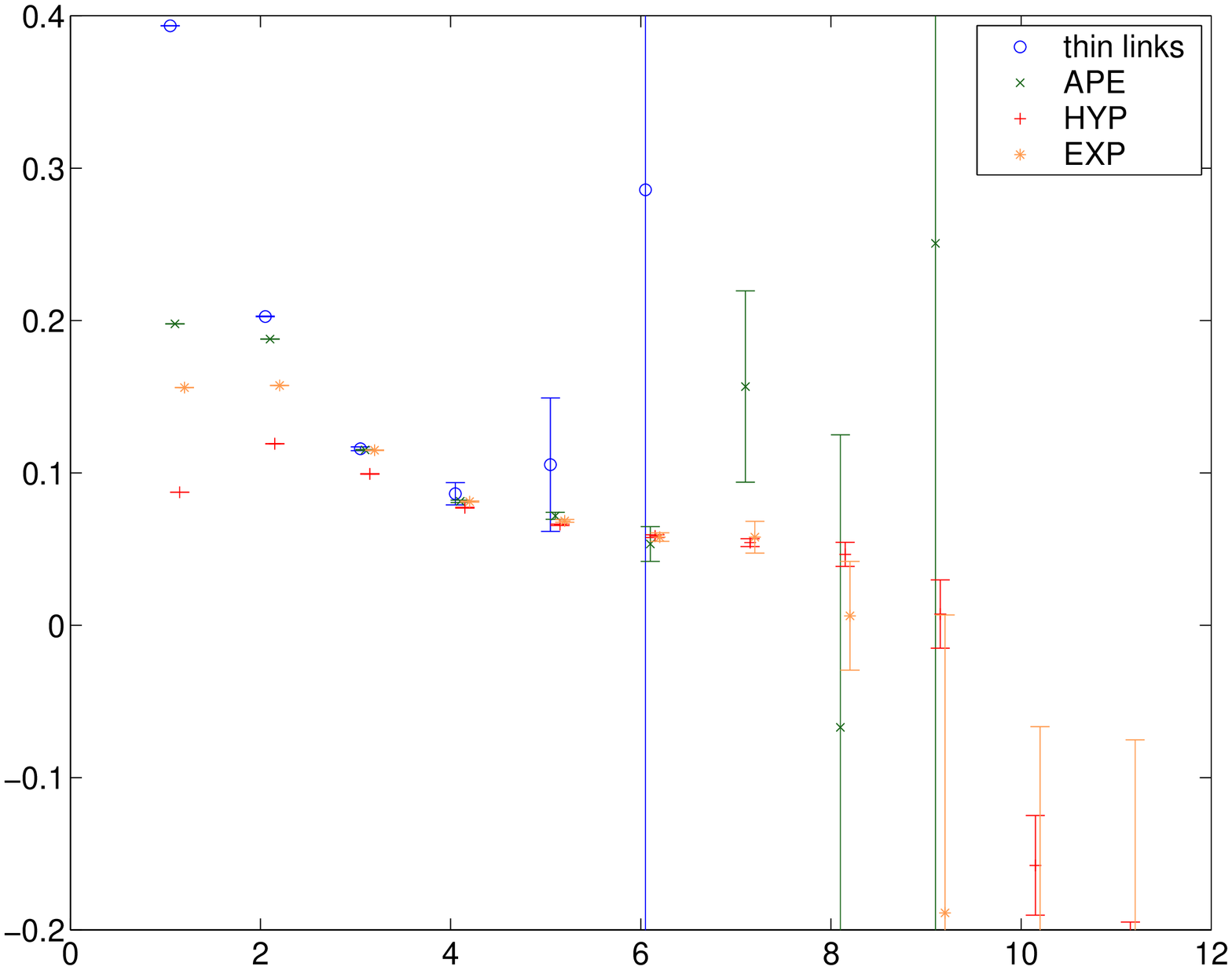}
\end{center}
\vspace{-8mm}
\caption{\sl Top: $M_\mr{eff}(T)\!=\!{1\ovr2}\log(W(R,T\!-\!1)/W(R,T\!+\!1))$
vs.\ $T$ at $R\!=\!3$ after one step with projection and $\al_\mr{APE}\!=\!0.5,
\al_\mr{HYP}\!=\!(0.3,0.6,0.75), \al_\mr{EXP}\!=\!0.2$. The thick links at the
initial and final time-slice result in a loss of overall positivity. For the
relation of plateau-height and -length cf.\ Fig.\,\ref{fig:signal_noise}.
Bottom: $-\!\log(\ch(x))$ vs.\ $x$ without smearing and after one step as
above. The data stem from $500$ configurations on a $12^3\!\times\!18$ lattice
with the Wilson gauge action at $\be\!=\!6.0$.}
\label{fig:meff_creutz}
\end{figure}

If the spatial links are subject to 3D smearing, the operator insertion takes
place at a specific time and the interpretation in the transfer matrix
formalism goes through.
If the spatial links are subject to 4D smearing, positivity may be lost (cf.\
the short-distance behavior in Fig.~\ref{fig:meff_creutz}, top), but one
can still learn something about the smearing process.
If the interpretation \cite{Lepage:1991ui} that smearing reduces the divergent
self energy ($\de\!m\propto\!1/a$) of the heavy quark is correct, one
expects that in the ansatz for the expectation value of a Wilson loop at
smearing level $n$
\begin{equation}
W^{(n)}(R,T)\simeq\mr{const}\;e^{-\mu^{(n)}(R+T)-\si^{(n)} RT}
\label{wilsansatz}
\end{equation}
the perimeter coefficient gets reduced while the squared string tension stays
invariant, i.e.\
\begin{eqnarray}
\mu^{(n)}&<&\mu^{(n-1)}
\label{expect1}
\\
\si^{(n)}&=&\si^{(0)}
\label{expect2}
\;.
\end{eqnarray}

A simple way to test the hypothesis (\ref{expect1}) is to look at the
$T$-dependence of an elongated ($R\!\ll\!T$) Wilson loop with a fixed $R$ small
enough that the perimeter part dominates in (\ref{wilsansatz}).
Such an object may be seen as the 2-point correlator of a ``frozen''
$Q\overline Q$-state (missing its bound-state dynamics), hence a change of its
effective mass will only reflect a change of the HQ self energies.
Fig.~\ref{fig:meff_creutz} (top) shows
$M_\mr{eff}(T)\!=\!{1\ovr2}\log(W(R,T\!-\!1)/W(R,T\!+\!1))$ with fixed
$R\!=\!3$, without smearing and after one step of APE/HYP/EXP smearing,
respectively.
Obviously, the effective mass plateau is lowered [more so under HYP than under
EXP or APE] and there is a correspondence between this self energy reduction
and the factor by which the error-bar is decreased and hence the plateau
elongated -- we shall come back to this point below.

To test the hypothesis (\ref{expect2}) it is useful to look at the symmetric
diagonal Creutz ratio \cite{creutzratio}
\begin{eqnarray}
\ch(x)&\equiv&\sqrt{\ch_+(x,x)\ch_-(x,x)}
\label{defcreutz}
\\
\ch_+(R,T)&=&{W(R,T)W(R\!+\!1,T\!+\!1)\ovr W(R\!+\!1,T)W(R,T\!+\!1)}
\simeq e^{-\si}
\nonumber\\
\ch_-(R,T)&=&{W(R,T)W(R\!-\!1,T\!-\!1)\ovr W(R\!-\!1,T)W(R,T\!-\!1)}
\simeq e^{-\si}
\nonumber\end{eqnarray}
which (for large $x$) is independent of $\mu$.
In Fig.~\ref{fig:meff_creutz} (bottom) $-\!\log(\ch(x))$ vs.\ $x$ is shown,
both for the original loop and after one APE/HYP/EXP-step.
The squared string tension is invariant under smearing, but the error-bar
changes.
HYP reduces it more aggressively than APE and EXP (for the parameters used in
Fig.~\ref{fig:meff_creutz}, the parameter dependence is addressed below).

In summary, Fig.~\ref{fig:meff_creutz} suggests that the reduced self energy
of the quasi-static quark (i.e.\ its divergent piece $\de m\!\sim\!1/a$) is
responsible for the improved signal-to-noise ratio after smearing.

\begin{figure}[t]
\begin{center}
\includegraphics[width=11.2cm,angle=270]{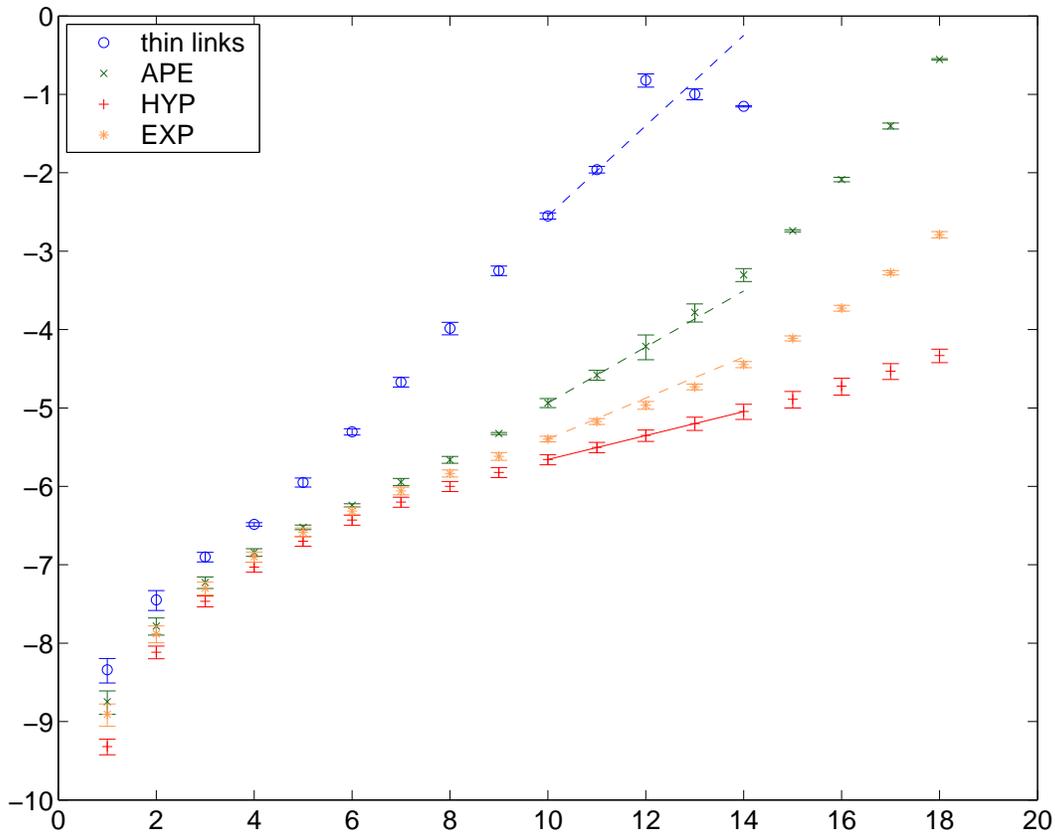}
\end{center}
\vspace{-6mm}
\caption{\sl Log of noise over signal in the Wilson loop shown in
{\rm Fig.~\ref{fig:meff_creutz} (top)}, without smearing and after one step of
{\rm APE/HYP/EXP}-smearing. The full line is a fit to the {\rm HYP}-data with
range $[10,14]$; the slopes of the dashed lines represent the prediction
$(\ref{signaltonoise})$ based on the $\Delta E_{Q\overline Q}$ taken from
{\rm Fig.~\ref{fig:meff_creutz} (top)} at $T\!=\!7$. The error on the error
is defined as as $((j_2\!-\!j_4)^2+(j_4\!-\!j_6)^2)^{1/2}$ with $j_n$ the
jackknife error at binlength $n$. Configurations and parameters are those of
{\rm Fig.~\ref{fig:meff_creutz}}.}
\label{fig:signal_noise}
\end{figure}

In fact, this discussion can be made more precise by virtue of an argument due
to Lepage~\cite{Lepage:1991ui}.
The key idea is rather than determining the variance of a correlator
$C(x)=\<O\dag(x)O(0)\>$ a posteriori from its fluctuation one could estimate
it in the MC via $\<O\dag(x)O(0)\til O\dag(0)\til O(x)\>$.
Here, a different symbol has been chosen for the ``copy'' creation and
annihilation operators to exclude additional unwanted Wick contractions, i.e.\
the idea is that they couple to different ``flavors'' (which, apart from the
flavor number, have identical properties).
In our case this noise-squared estimator consists of two Wilson loops, with
opposite orientation, on top of each other.
Asymptotically, it is dominated by two flux-tubes of length zero, i.e.\ by a
state with mass $2V(0)\!=\!0$ for unitary links.
Thus, since the original correlator falls of with an effective mass
$E_{Q\overline Q}$ and its noise is asymptotically constant, the Lepage
argument \cite{Lepage:1991ui} predicts
\begin{equation}
{\mr{signal}\ovr\mr{noise}}(t)\;=\;\mr{const}\;e^{-E_{Q\overline Q}\,t}
\label{signaltonoise}
\end{equation}
and Fig.~\ref{fig:signal_noise} shows that this expectation is indeed well
satisfied.
It is clear that the specific recipes and parameters in
Figs.\,\ref{fig:meff_creutz},\ref{fig:signal_noise} are irrelevant -- the
Lepage argument is a general one.


\section{Matching boxes via the local force}


Our goal is to compare various smearing recipes on lattices generated with
either the Wilson gauge action or an improved one.
In order to give a fair comparison, these lattices should be matched, i.e.\
have the same spacing $a$ in units of a physical scale.
Ideally, this is done by computing the HQ potential, but rather than moving on
to compute various $R\!\times\!T$ ``thin-link'' Wilson loops, I propose
(specifically for the present investigation) two shortcuts.

The first one is to avoid the $T\!\to\!\infty$ extrapolation and to determine
the force $F(R,T)\!=\!{1\ovr2}(\pa_R\!+\!\pa_R^*)M_\mr{eff}(R,T)$ with
$M_\mr{eff}(R,T)\!=\!{1\ovr2}\log(W(R,T\!-\!1)/W(R,T\!+\!1))$ at one fixed
entry, and I choose $R\!=\!T\!=\!4$.
Since there is no $T\!\to\!\infty$ extrapolation, one can stay in a relatively
small volume and the finite-volume effect will be the same for all final
(tuned) lattices.

The second shortcut is even less innocent.
In order to get precise values for $F(4,4)$, one is tempted to use smeared
links.
From the observation of Ref.\ \cite{hyp} that (one step of) smearing affects
(for $R\!>\!2$) mainly the additive constant in the HQ potential, one expects
$F(4,4)$ to be unchanged, since its arguments are large enough.
This is indeed the plan; albeit with the proviso that all three smearing
recipes will be used to have a cross check.

Hence, the problem with measuring only $F(4,4)$ is that the fixed time
separation does not allow to select a definite state (ideally the groundstate),
and the mixture of states that one has at $T\!=\!4$ may vary with the gauge
action and the type of link used at $T\!=\!0$.
Therefore, having several ``thick'' links as creation and absorption operators
helps, since they couple differently to excited states and one may hope that
the problem is detected, if it is numerically sizable.

\bigskip

\begin{table}
\begin{center}
\begin{tabular}{|cl|cccc|}
\hline
&&original&APE&HYP&EXP\\
\hline
W   &$\be\!=\!6.00$&0.0576 (60)&0.0550(10)&0.0513(06)&0.0550(07)\\
\hline
S   &$\be\!=\!4.40$&0.0692(115)&0.0594(21)&0.0555(13)&0.0589(16)\\
    &$\be\!=\!4.50$&0.0494 (75)&0.0462(15)&0.0452(09)&0.0474(11)\\
    &$\be\!=\!4.60$&0.0343 (51)&0.0388(10)&0.0376(07)&0.0398(07)\\
\hline
I   &$\be\!=\!2.65$&0.0557 (72)&0.0552(16)&0.0527(11)&0.0559(12)\\
    &$\be\!=\!2.75$&0.0396 (43)&0.0444(10)&0.0429(07)&0.0450(08)\\
    &$\be\!=\!2.85$&0.0371 (35)&0.0384(10)&0.0371(08)&0.0391(09)\\
\hline
DBW2&$\be\!=\!1.01$&0.0605 (53)&0.0604(15)&0.0602(12)&0.0628(13)\\
    &$\be\!=\!1.04$&0.0466 (34)&0.0504(10)&0.0504(10)&0.0528(11)\\
    &$\be\!=\!1.07$&0.0469 (31)&0.0491(12)&0.0478(09)&0.0502(10)\\
\hline
BKL &$\be\!=\!4.70$&0.0644(105)&0.0566(19)&0.0537(12)&0.0573(14)\\
    &$\be\!=\!4.75$&0.0384 (97)&0.0527(19)&0.0500(11)&0.0532(13)\\
    &$\be\!=\!4.80$&0.0645 (75)&0.0486(14)&0.0439(08)&0.0469(10)\\
\hline
L   &$\be\!=\!0.79$&0.0562 (63)&0.0551(17)&0.0541(11)&0.0562(12)\\
    &$\be\!=\!0.80$&0.0529 (53)&0.0505(13)&0.0490(09)&0.0512(10)\\
    &$\be\!=\!0.81$&0.0430 (48)&0.0482(12)&0.0477(08)&0.0502(09)\\
\hline
\end{tabular}
\vspace{-5mm}
\end{center}
\caption{\sl Details of the tuning process to get approximately matched
lattices. Given is the force $F(R,T)\!=\!{1\ovr4}\log(W(R\!+\!1,T\!-\!1)/
W(R\!+\!1,T\!+\!1)\!\cdot\!W(R\!-\!1,T\!+\!1)/W(R\!-\!1,T\!-\!1))$ on a $8^4$
lattice for $R\!=\!T\!=\!4$. The columns {\rm APE/HYP/EXP} give the result
after one step with the same parameters as in {\rm Fig.~\ref{fig:meff_creutz}}.
The errors are from a jackknife with binlength 2, using 800 {\rm (W)} or 200
{\rm (all other actions)} configurations per run, after discarding 10
configurations for thermalization.}
\label{tab:tuning_protocol}
\end{table}

Table~\ref{tab:tuning_protocol} gives details of the tuning process aimed at
finding $\be$-values which (approximately) match a quenched
$\be_\mr{W}\!=\!6.0$ with the Wilson gauge action.
The sections ``I'' and ``BKL'' refer to the tree-level choice in
(\ref{choice_c1}, \ref{choice_c2}), i.e.\ $\til c_1\!=\!c_1$ and
$\til c_2\!=\!c_2$ throughout.

\begin{table}
\begin{center}
\begin{tabular}{|c|cccc|}
\hline
&original&APE&HYP&EXP\\
\hline
S    &4.459 (44)&4.427(30)&{\bf 4.439}(17)&4.429(22)\\
I    &2.555(298)&2.632(42)&{\bf 2.652}(31)&2.643(37)\\
DBW2 &0.999(276)&1.026(28)&{\bf 1.047}(14)&1.041(14)\\
BKL  &   ---    &4.720(13)&{\bf 4.729}(12)&4.726(12)\\
L    &0.790 (14)&0.788(06)&{\bf 0.796}(06)&0.791(09)\\
\hline
\end{tabular}
\vspace{-5mm}
\end{center}
\caption{\sl $\be$-values needed with the
{\rm Symanzik/Iwasaki/DBW2/Beinlich-Karsch-Laermann/ Langfeld} actions to have
an approximately matched lattice w.r.t.\ $\be\!=\!6.0$ with the Wilson gauge
action, from interpolating data of {\rm Table~\ref{tab:tuning_protocol}}. The
{\rm HYP}-values will be used subsequently.}
\label{tab:tuning_results}
\end{table}

By linearly interpolating, for each action, the simulated couplings one finds
the matched (according to the local force at $R\!=\!T\!=\!4$ criterion)
$\be$-values.
Results are given in Table~\ref{tab:tuning_results}, without smearing in the
first column (where the error is largest) and with one APE/HYP/EXP-step in
subsequent columns.
The unsmeared matched $\be$-values are compatible with any smeared counterpart,
but also the agreement among the smeared ones is satisfactory.
We rate this as a sign that our interpolated matched couplings are essentially
correct.
Obviously, this does not invalidate the theoretical concerns discussed above,
but apparently our statistical precision is not good enough to really suffer
from this.

\bigskip

In \cite{Necco:2003vh} Necco gives an interpolation formula for $\log(r_0/a)$
versus $\be$ for the Iwasaki- and the DBW2-action, analogous to the one in
\cite{Guagnelli:1998ud} for the Wilson gauge action.
Solving them, one finds that a lattice with $\be_\mr{W}\!=\!6.0$ is matched
via $\be_\mr{I}\!=\!2.635$ and $\be_\mr{DBW2}\!=\!1.034$ when employing one of
the RG-motivated actions, respectively.
My $\be_\mr{I}\!=\!2.652(31)$ and $\be_\mr{DBW2}\!=\!1.047(14)$ are consistent.
In spite of the limitations mentioned above one may thus feel confident that
the $\be$-values listed in Table~\ref{tab:tuning_results} yield lattices which
are reasonably (say: better than to 10\%) matched.


\section{Effect on square Wilson loops and the Creutz ratio}


Whether an improved gauge action is worth the investment is usually decided on
the basis of the cut-off effects involved.
Another practically relevant criterion might be whether such an
action is amenable to noise-reduction techniques like the APE/HYP/EXP-smearing
discussed above.
Obviously, physical observables should depend as little as possible on the
details of the smearing process, e.g.\ the parameters employed.

\bigskip

Following Ref.\ \cite{exp}, we shall first have a look at a few Wilson loops,
comparing the effect of changing the parameter(s) and/or the iteration level.

To this end I have prepared 100 lattices of size $12^4$ for each gauge action,
using the $\be$-values of the HYP column in Table~\ref{tab:tuning_results}.
After this is done, one applies the smearing recipes discussed in section 3,
varying the parameters in a reasonable range.
For HYP I choose to concentrate on the direction of the ``canonical''
$\al$-value introduced in \cite{hyp}, i.e.\ I consider six $\al$-values,
\begin{equation}
\begin{array}{ll}
\mbox{APE}:&\al\in\{0.20,0.40,0.60,0.80,1.00,1.20\}\\
\mbox{HYP}:&\al\in\{0.25,0.50,0.75,1.00,1.25,1.50\}\cdot\al_\mr{HYP}^\mr{std}
\quad\mbox{with}\;\;\al_\mr{HYP}^\mr{std}=(0.3,0.6,0.75)\\
\mbox{EXP}:&\al\in\{0.05,0.10,0.20,0.35,0.55,0.80\}
\end{array}
\label{parameterset}
\end{equation}
where I deviate from Ref.~\cite{hyp} in my notation
$\al_\mr{HYP}\!=\!(\al_\mr{step\,1},\al_\mr{step\,2},\al_\mr{step\,3})$.
I choose to apply $2^n$ smoothing steps between successive measurements,
resulting in the smearing levels
\begin{equation}
n_\mr{tot}\in\{1,3,7,15\}
\;.
\end{equation}

\begin{figure}
\begin{center}
\scriptsize
\psfrag{al_1}{$\alpha^{(1)}$}
\psfrag{al_2}{$\alpha^{(2)}$}
\psfrag{al_3}{$\alpha^{(3)}$}
\psfrag{al_4}{$\alpha^{(4)}$}
\psfrag{al_5}{$\alpha^{(5)}$}
\psfrag{al_6}{$\alpha^{(6)}$}
\includegraphics[width=84mm,angle=0]{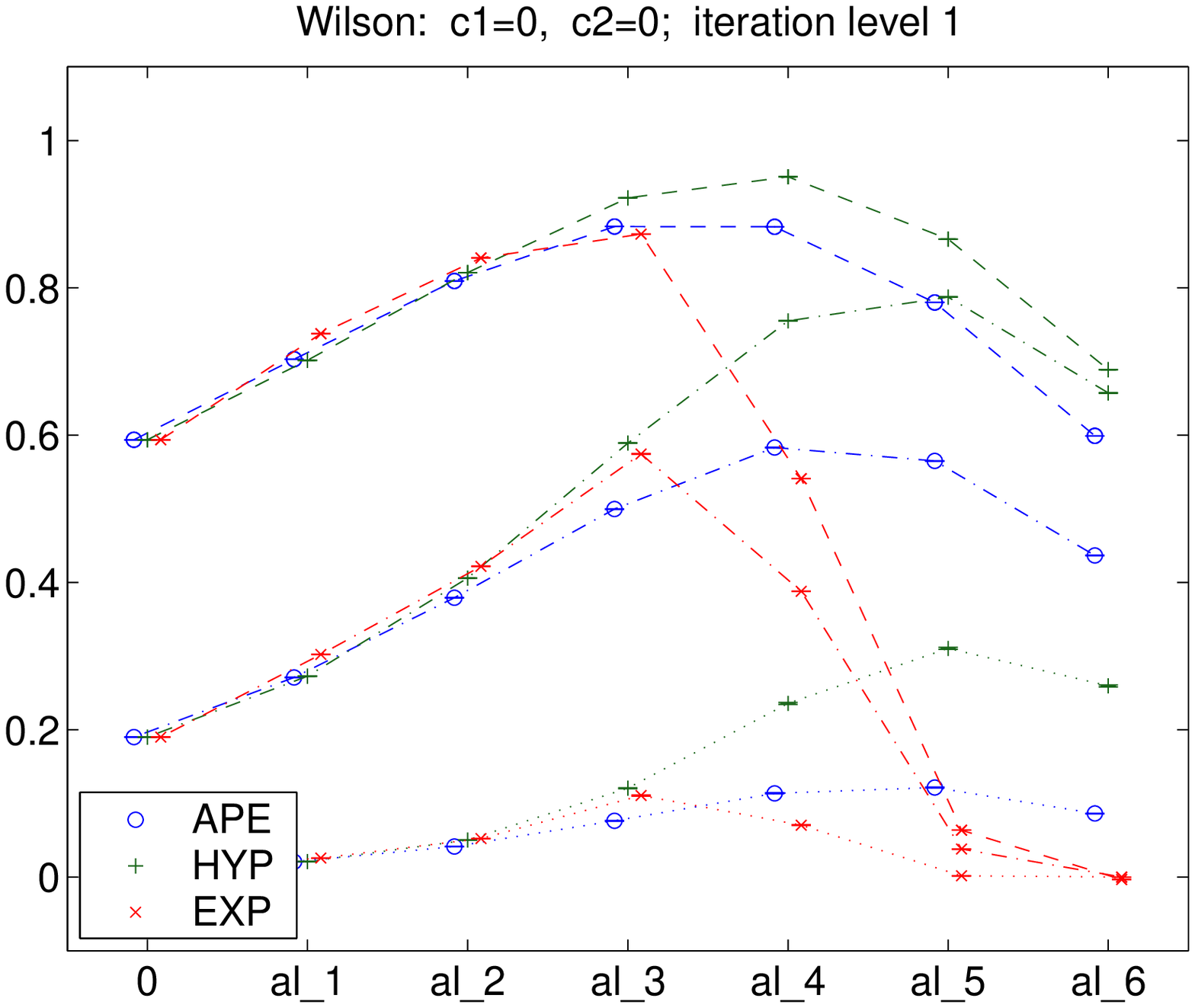}
\includegraphics[width=84mm,angle=0]{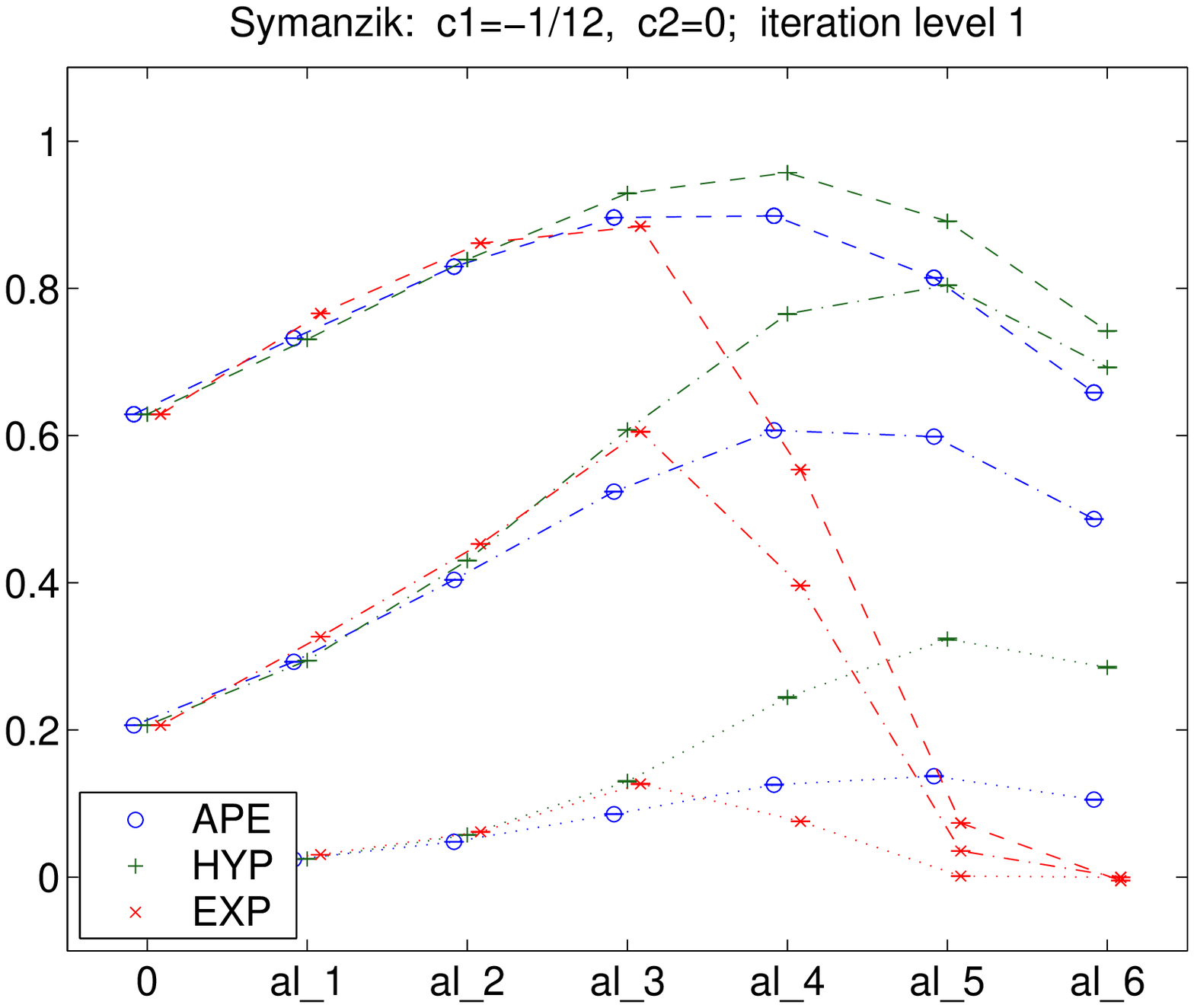}\\[4mm]
\includegraphics[width=84mm,angle=0]{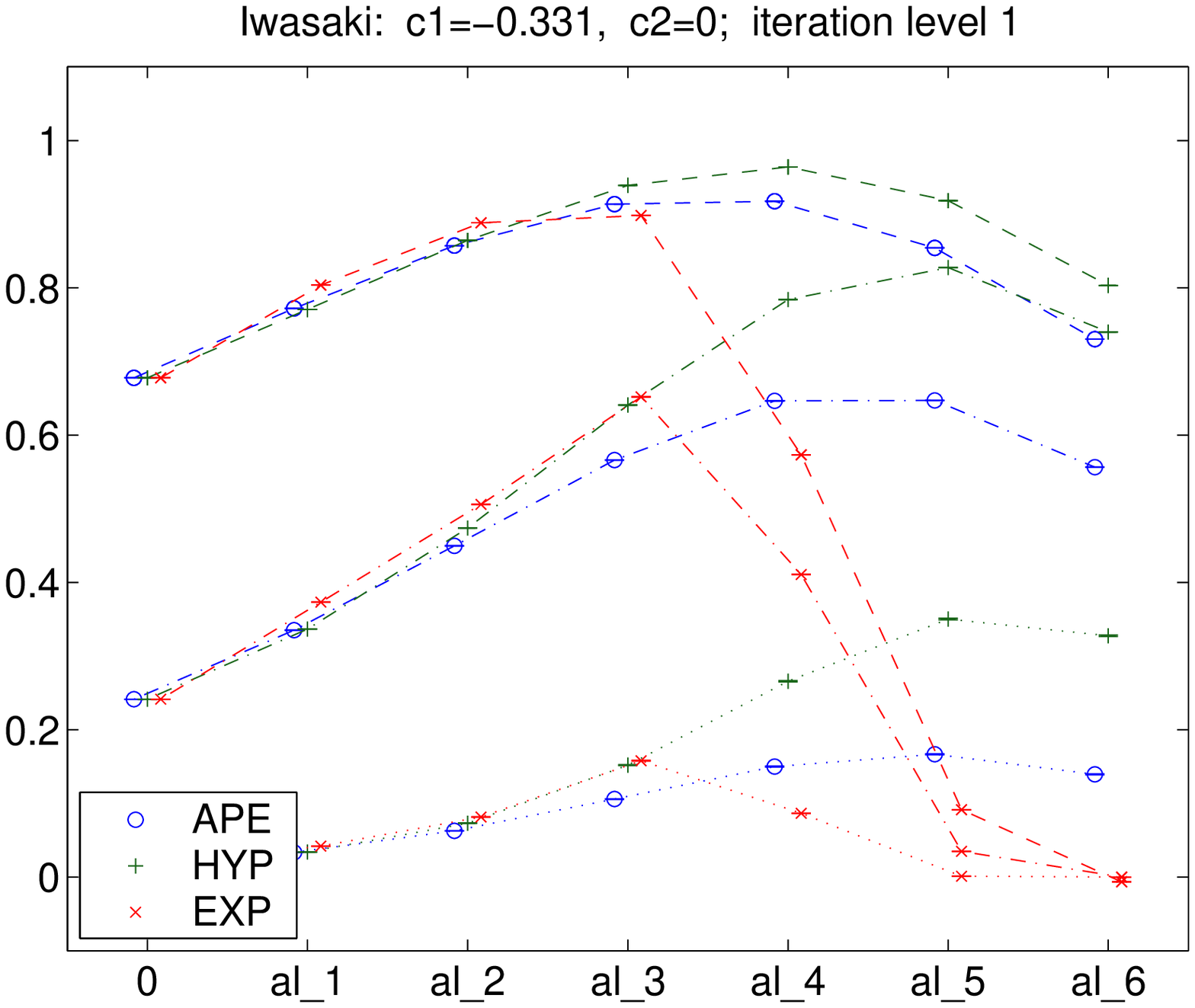}
\includegraphics[width=84mm,angle=0]{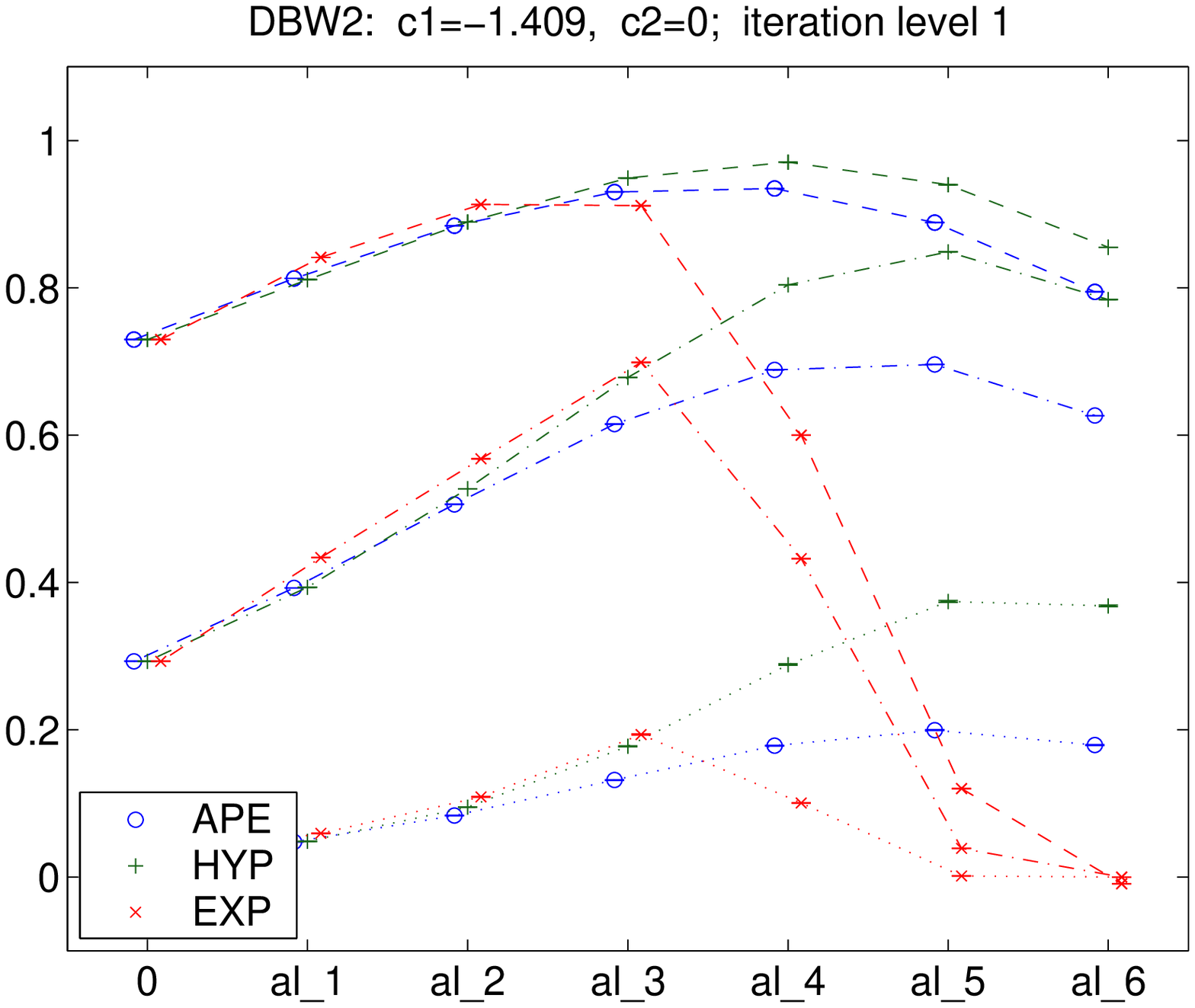}\\[4mm]
\includegraphics[width=84mm,angle=0]{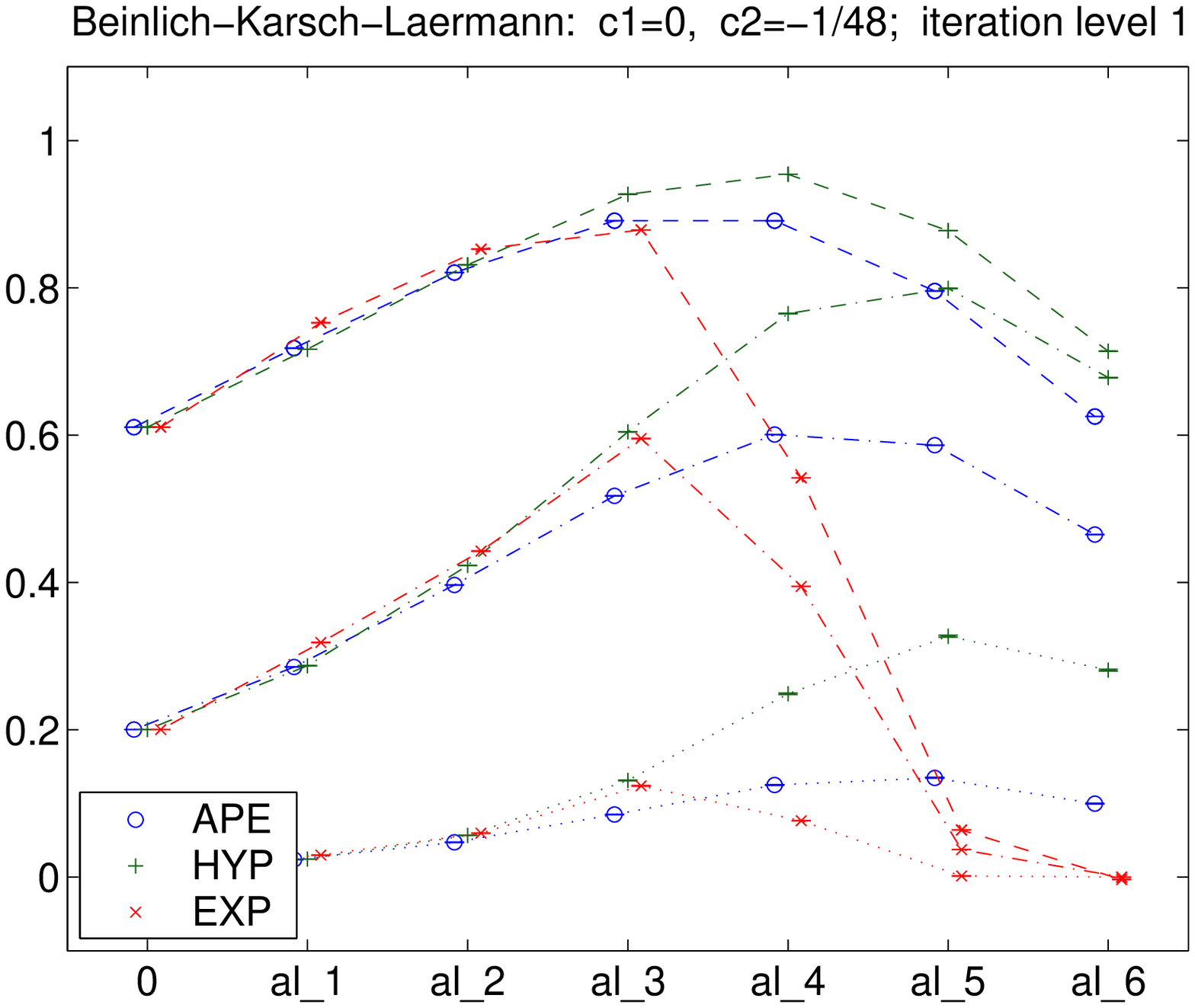}
\includegraphics[width=84mm,angle=0]{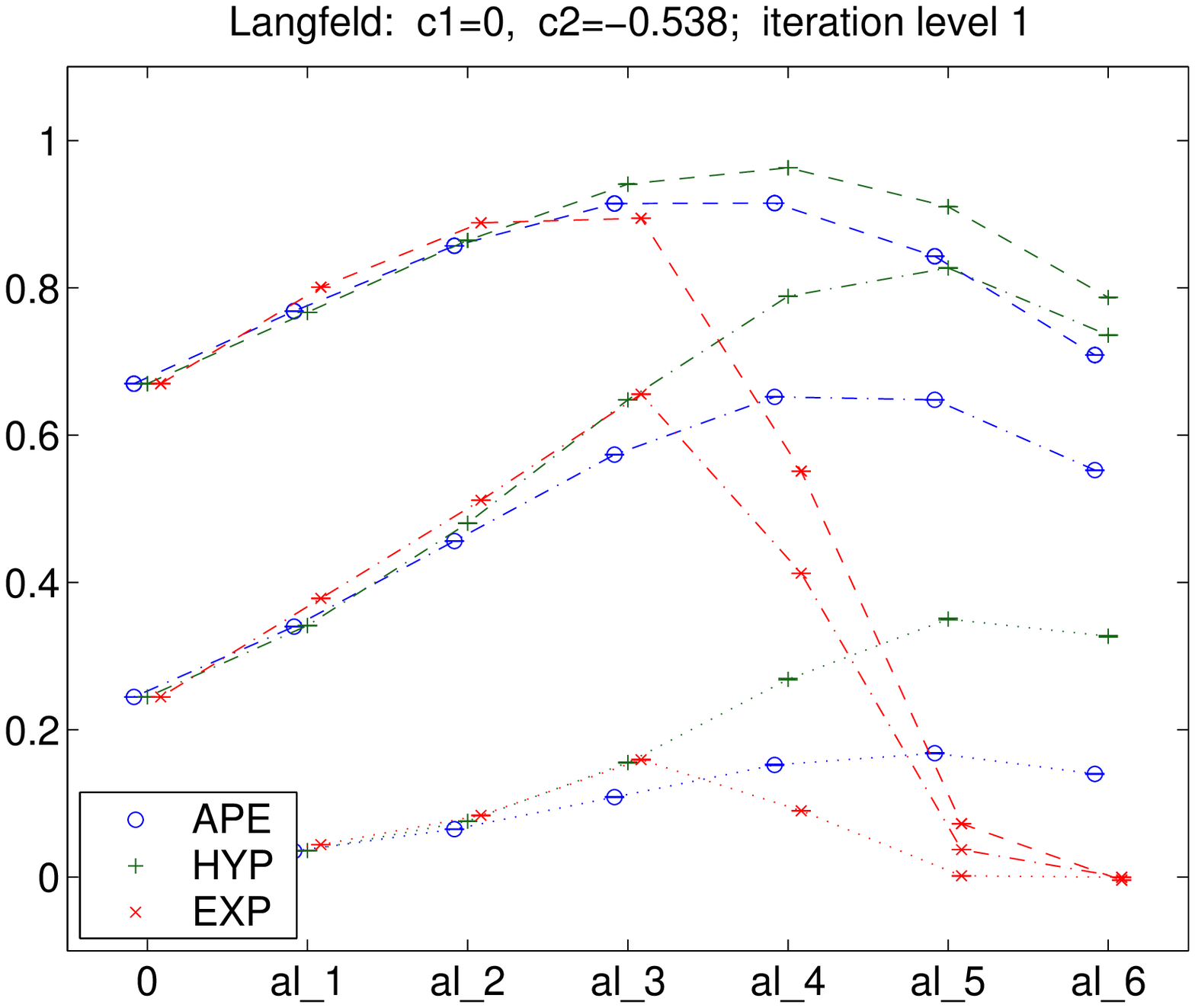}
\end{center}
\vspace{-6mm}
\caption{\sl For each gauge action $W_{1\times1}, W_{2\times2}, W_{4\times4}$
is plotted versus the parameter number in {\rm (\ref{parameterset})}, 0
indicates the unsmeared starting point. Throughout 1 smearing step ({\rm APE},
{\rm HYP} with projection) is applied. Best $W_{1\times1}$-parameters are
$\al_\mr{APE}\!\simeq\!0.7, \al_\mr{HYP}\!\simeq\!(0.3,0.6,0.75),
\al_\mr{EXP}\!\simeq\!0.2$.}
\label{fig:smear_lev1}
\end{figure}

\begin{figure}
\begin{center}
\scriptsize
\psfrag{al_1}{$\alpha^{(1)}$}
\psfrag{al_2}{$\alpha^{(2)}$}
\psfrag{al_3}{$\alpha^{(3)}$}
\psfrag{al_4}{$\alpha^{(4)}$}
\psfrag{al_5}{$\alpha^{(5)}$}
\psfrag{al_6}{$\alpha^{(6)}$}
\includegraphics[width=84mm,angle=0]{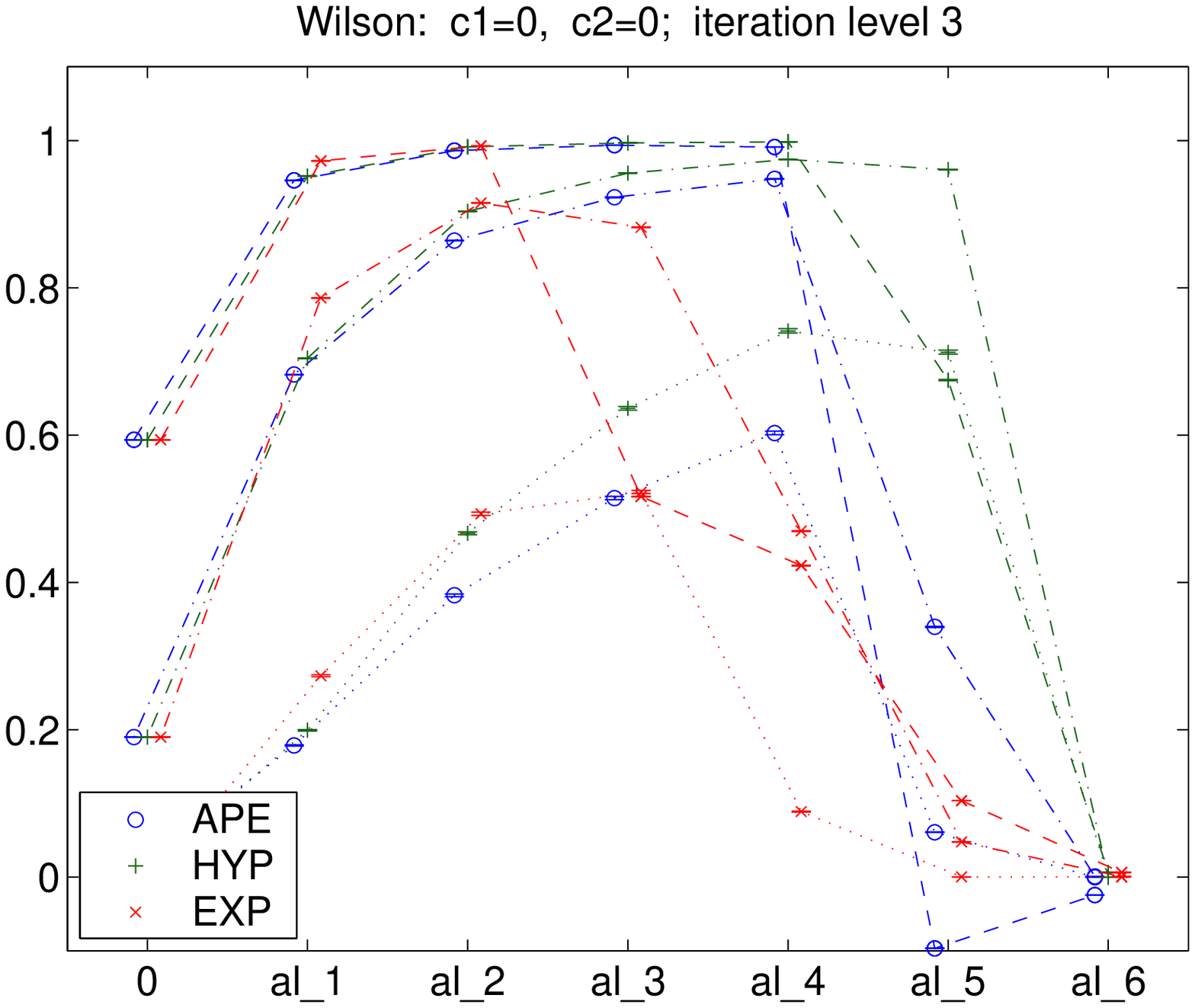}
\includegraphics[width=84mm,angle=0]{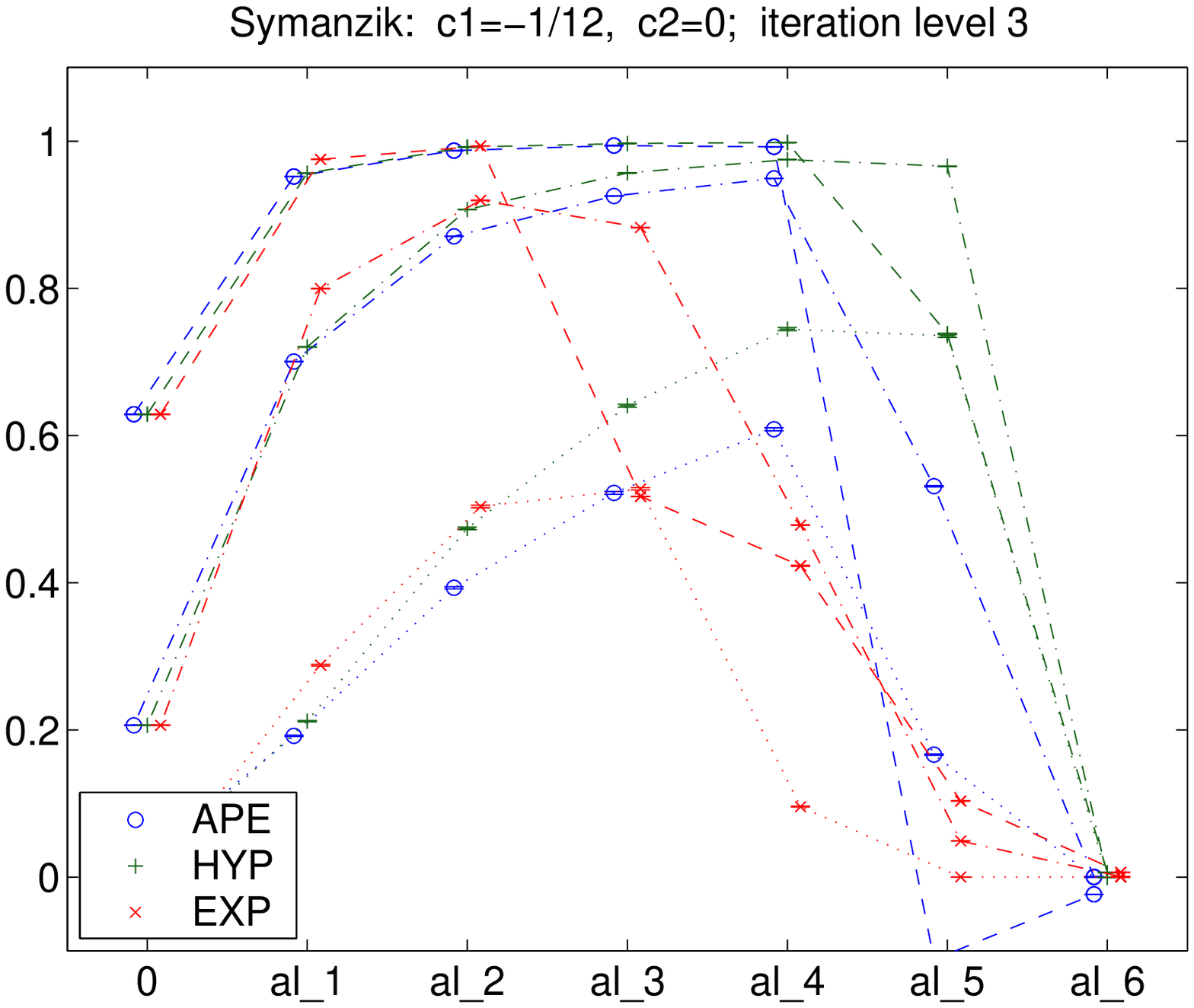}\\[4mm]
\includegraphics[width=84mm,angle=0]{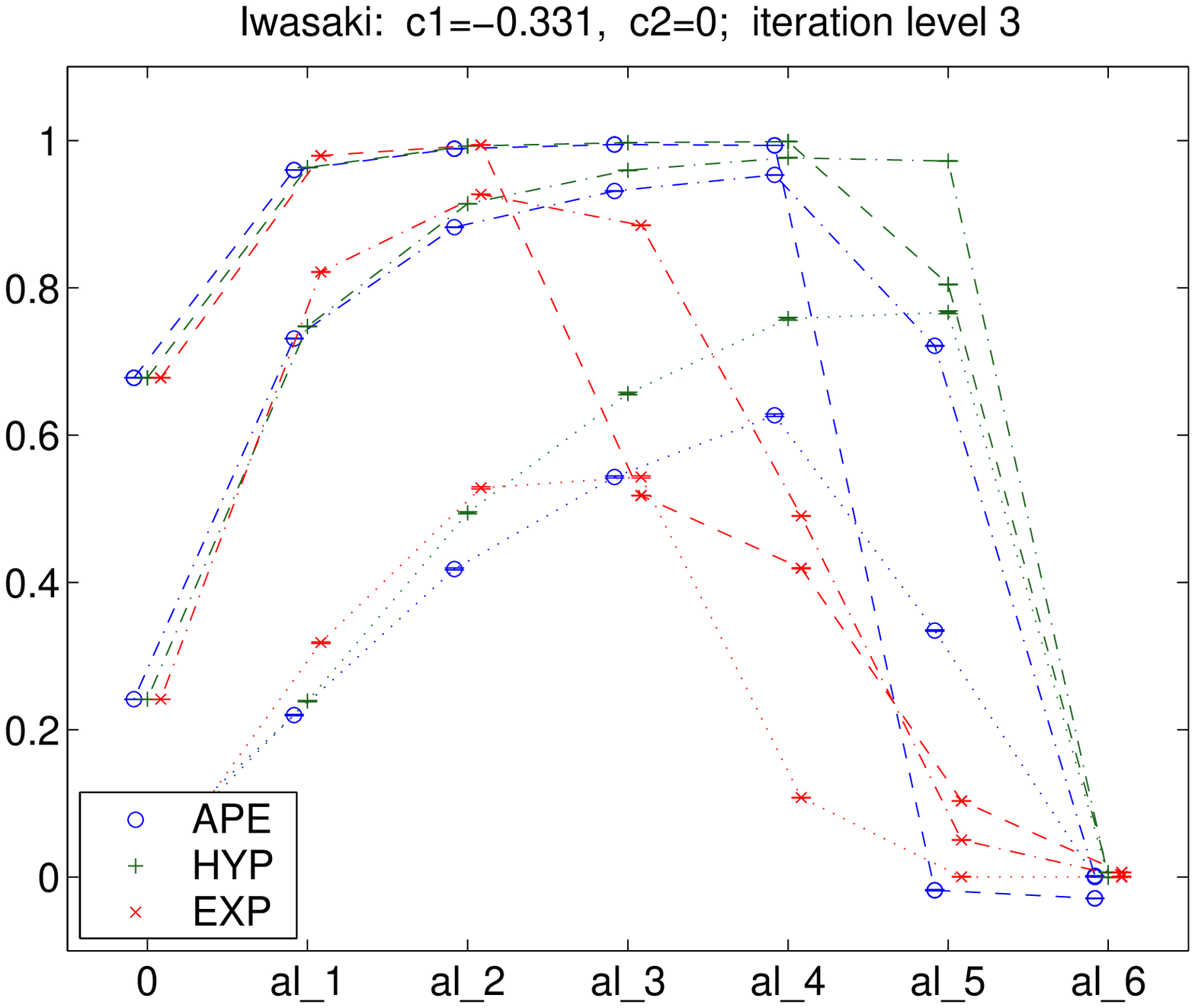}
\includegraphics[width=84mm,angle=0]{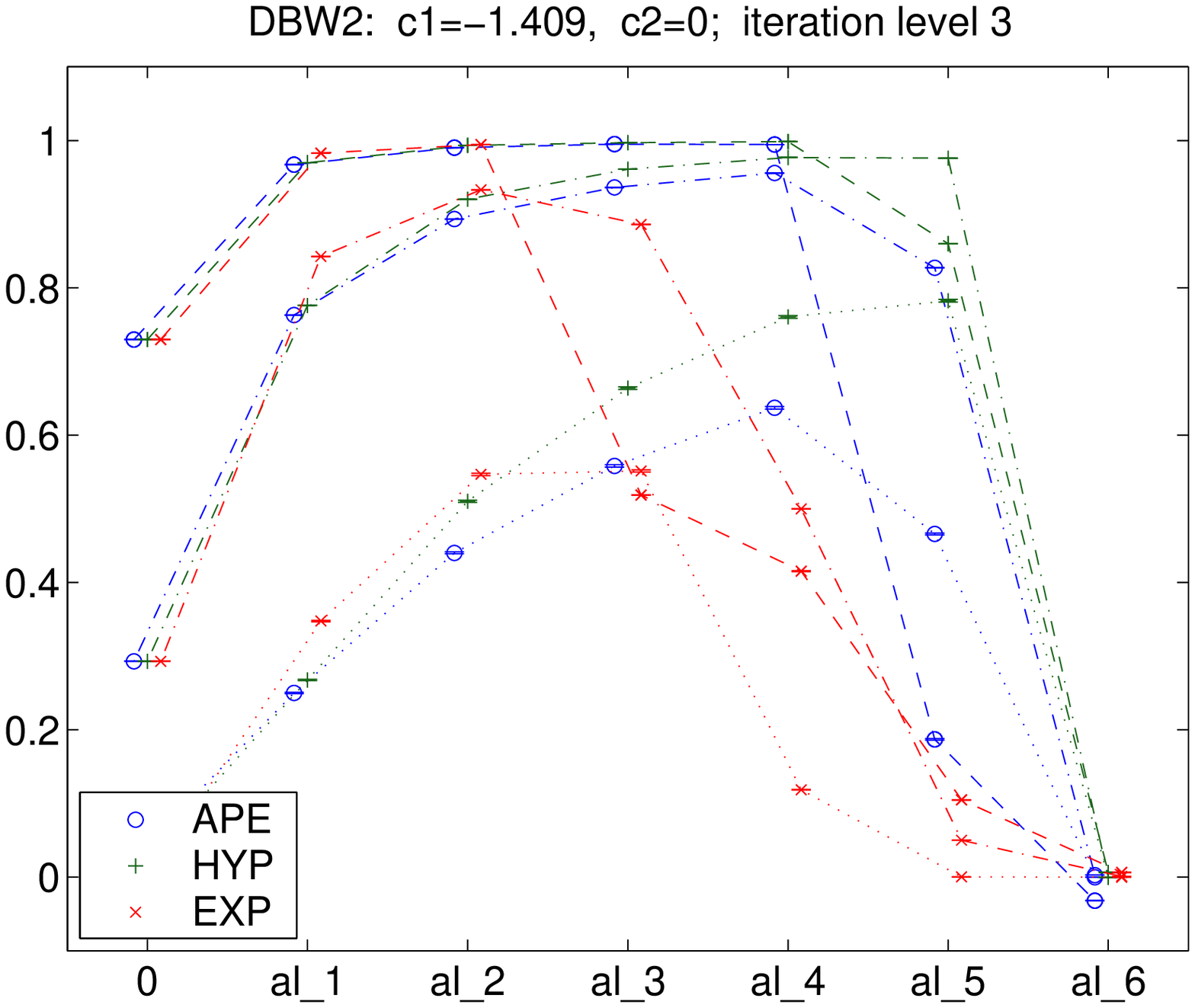}\\[4mm]
\includegraphics[width=84mm,angle=0]{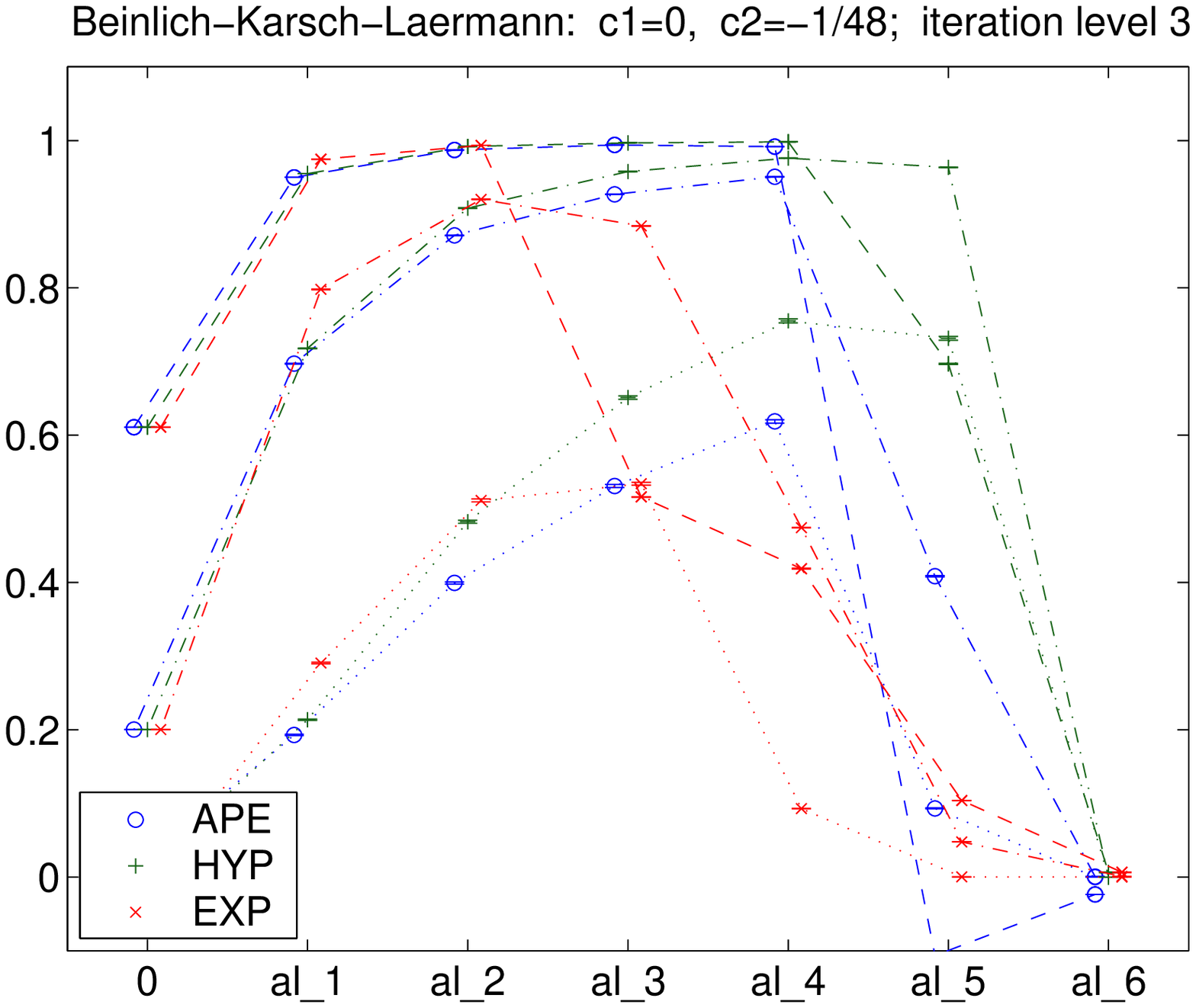}
\includegraphics[width=84mm,angle=0]{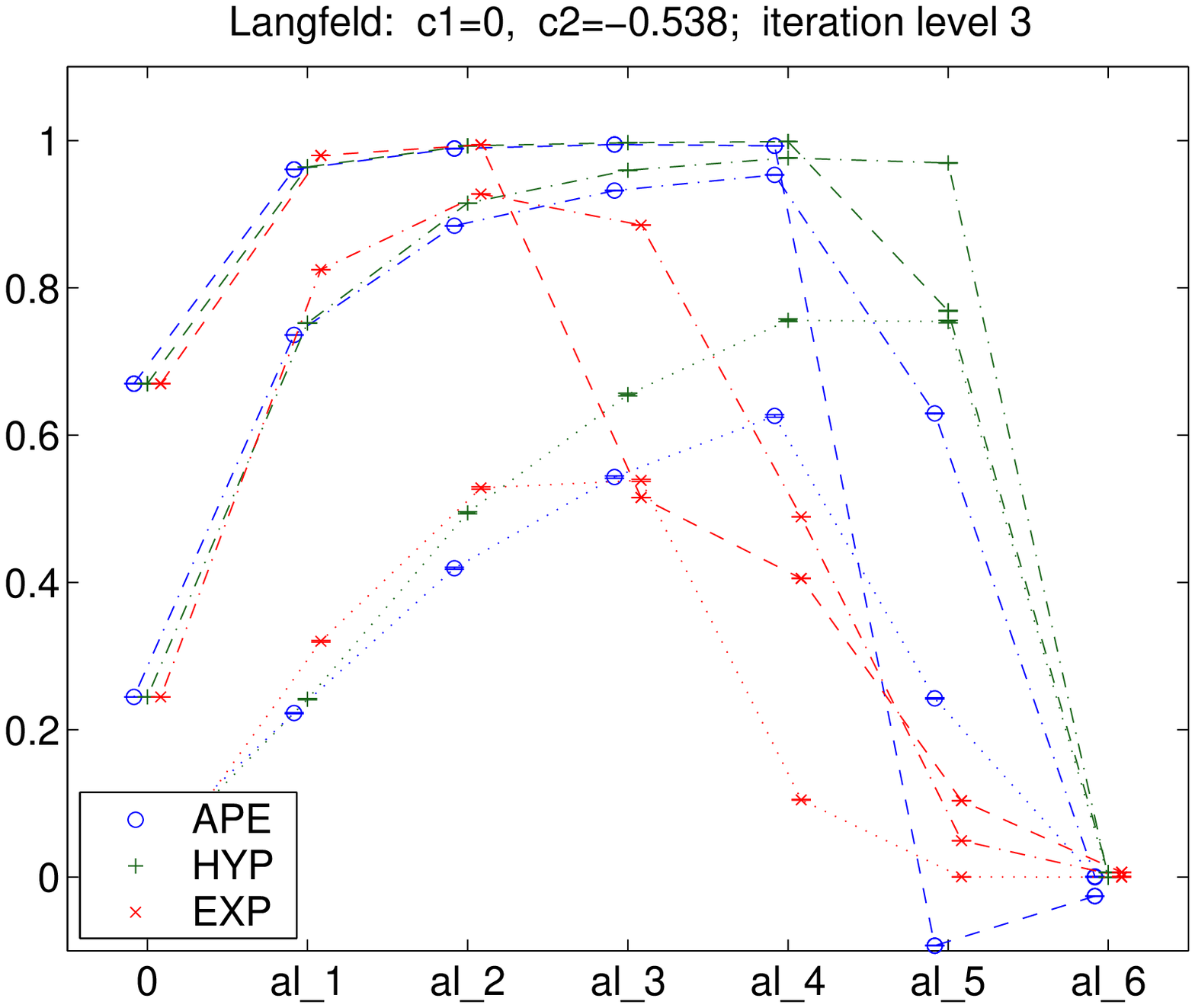}
\end{center}
\vspace{-6mm}
\caption{\sl For each gauge action $W_{1\times1}, W_{2\times2}, W_{4\times4}$
is plotted versus the parameter number in {\rm (\ref{parameterset})}, 0
indicates the unsmeared starting point. Throughout 7 smearing steps ({\rm APE},
{\rm HYP} with projection) are used. Best $W_{1\times1}$-parameters are
$\al_\mr{APE}\!\simeq\!0.6, \al_\mr{HYP}\!\simeq\!(0.3,0.6,0.75),
\al_\mr{EXP}\!\simeq\!0.1$.}
\label{fig:smear_lev3}
\end{figure}

\begin{figure}
\begin{center}
\scriptsize
\psfrag{al_1}{$\alpha^{(1)}$}
\psfrag{al_2}{$\alpha^{(2)}$}
\psfrag{al_3}{$\alpha^{(3)}$}
\psfrag{al_4}{$\alpha^{(4)}$}
\psfrag{al_5}{$\alpha^{(5)}$}
\psfrag{al_6}{$\alpha^{(6)}$}
\includegraphics[width=84mm,angle=0]{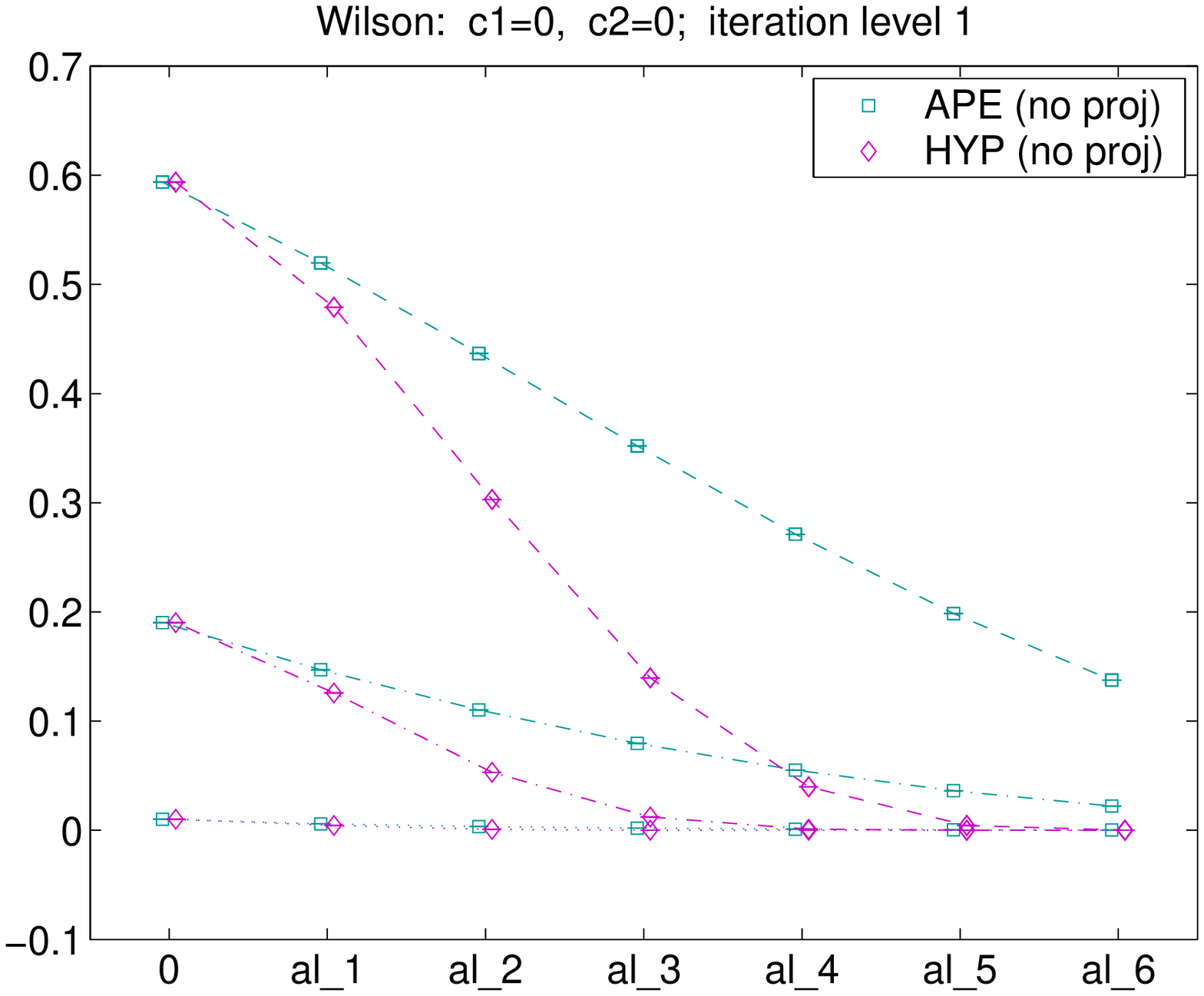}
\includegraphics[width=84mm,angle=0]{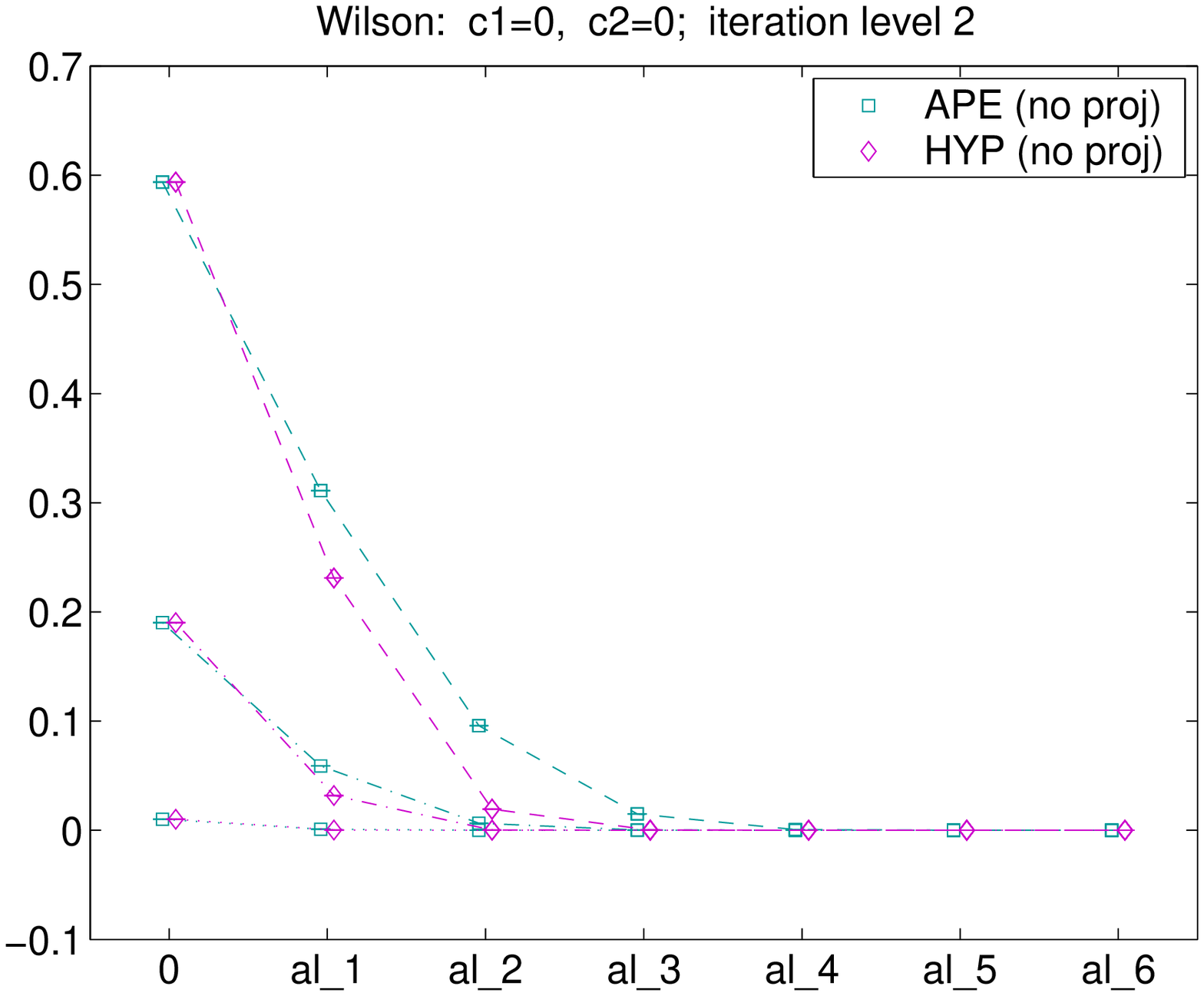}
\end{center}
\vspace{-6mm}
\caption{\sl $W_{1\times1}, W_{2\times2}, W_{4\times4}$ versus the smearing
parameter in the {\rm APE} and {\rm HYP} procedure without projection, using
1 (left) or 3 (right) iterations, on Wilson backgrounds. Note the difference
to the first graph in Fig.\,\ref{fig:smear_lev1}.}
\label{fig:smear_noproj}
\end{figure}

\begin{figure}
\begin{center}
\includegraphics[width=84mm,angle=0]{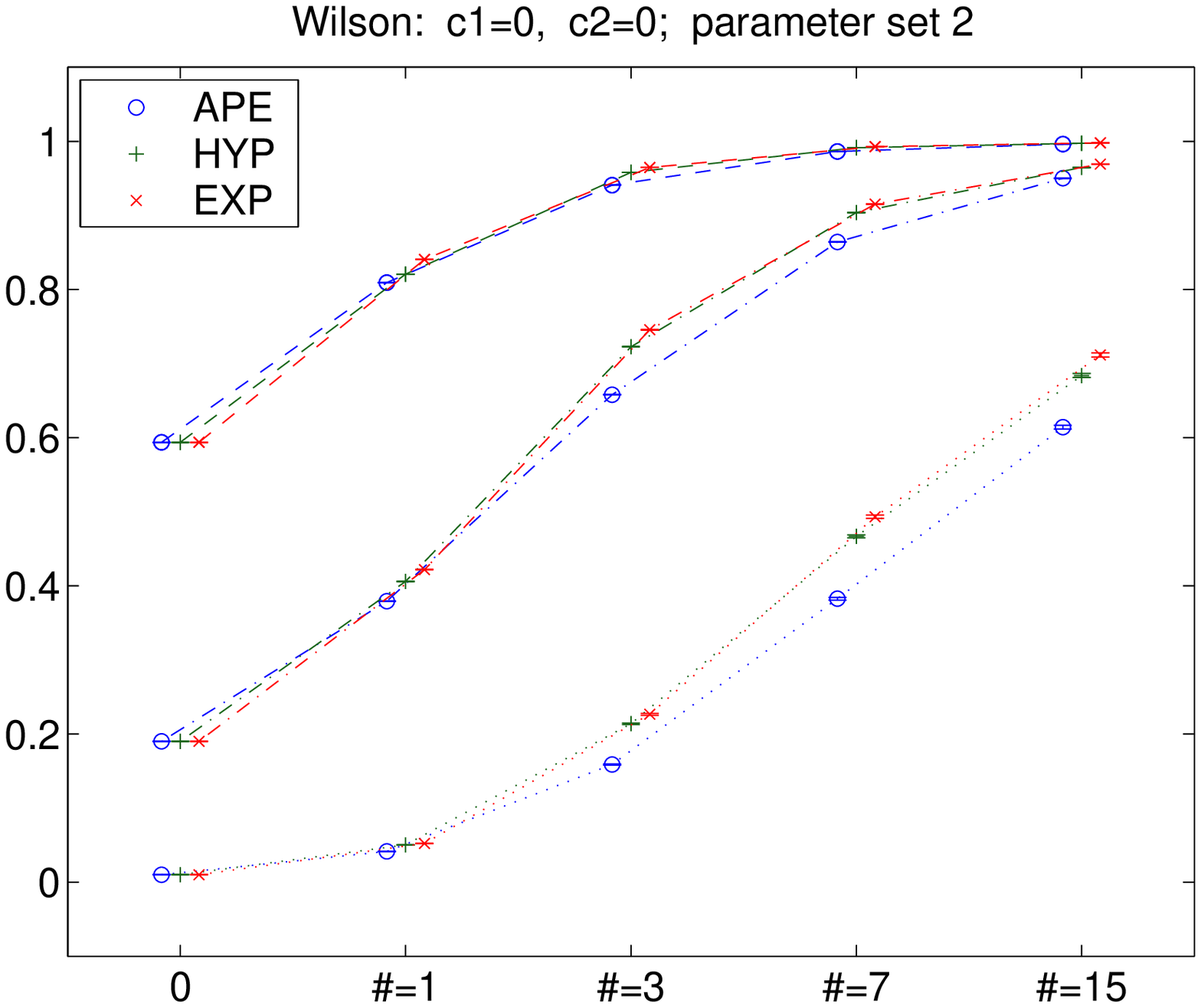}
\includegraphics[width=84mm,angle=0]{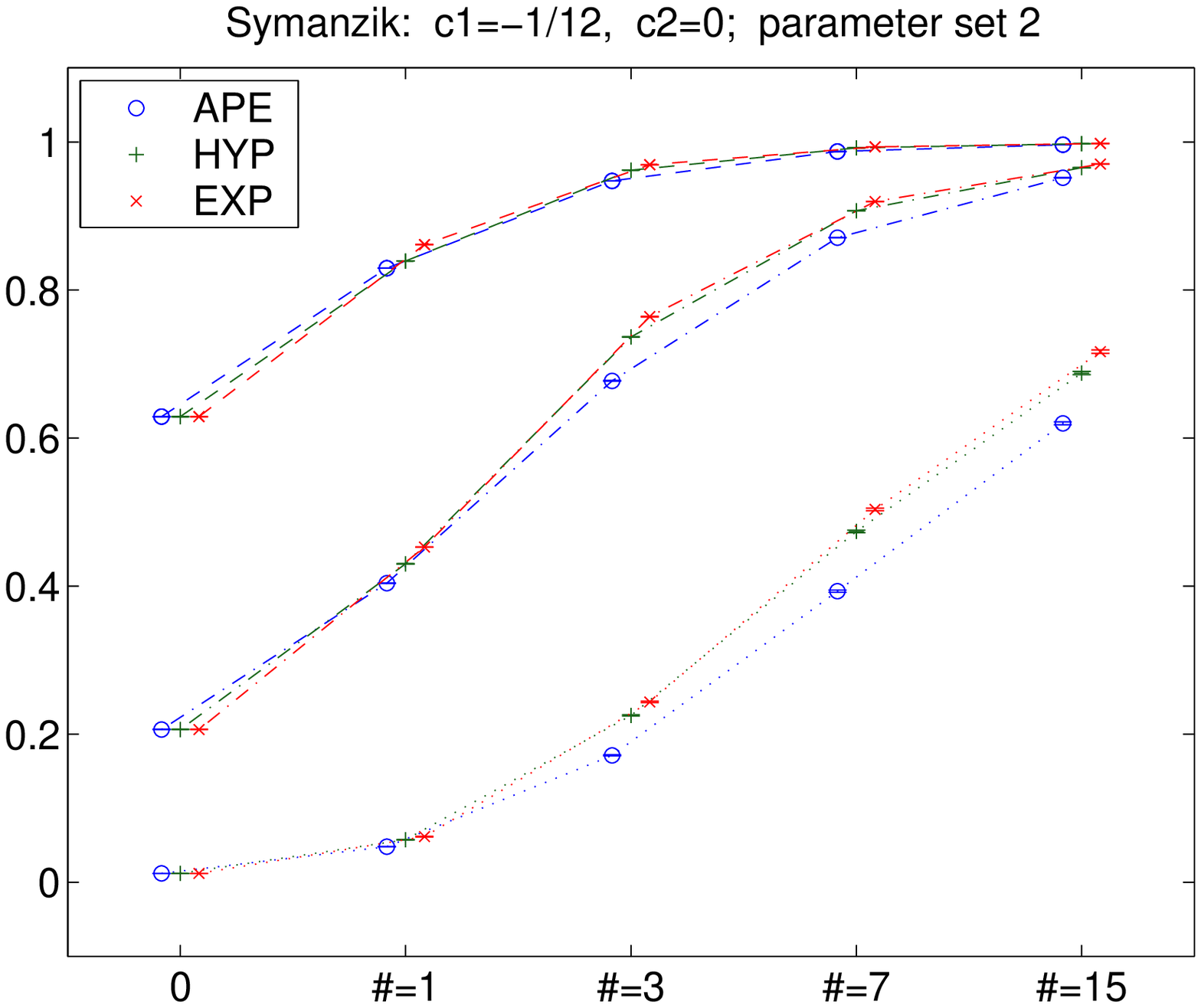}\\[4mm]
\includegraphics[width=84mm,angle=0]{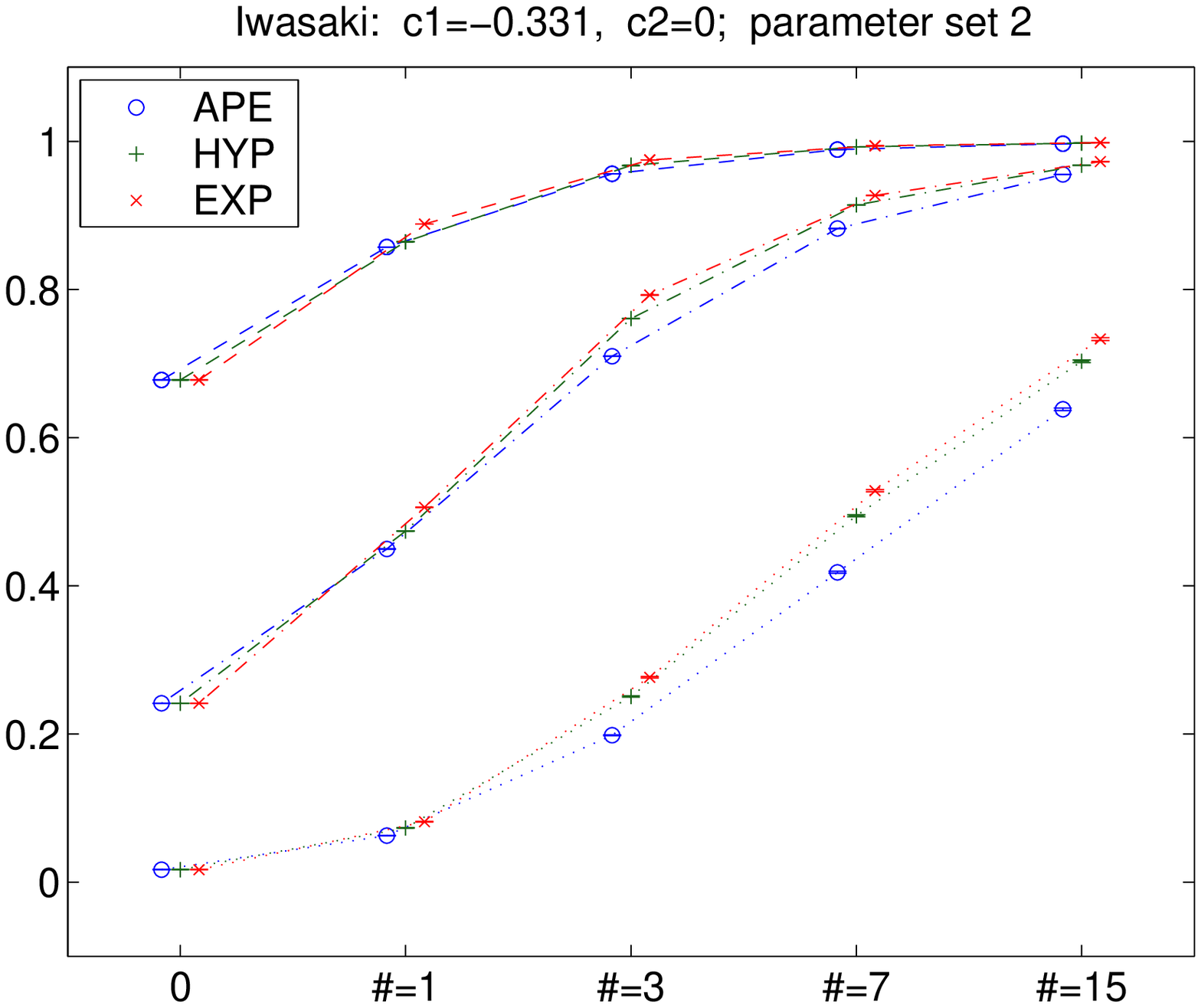}
\includegraphics[width=84mm,angle=0]{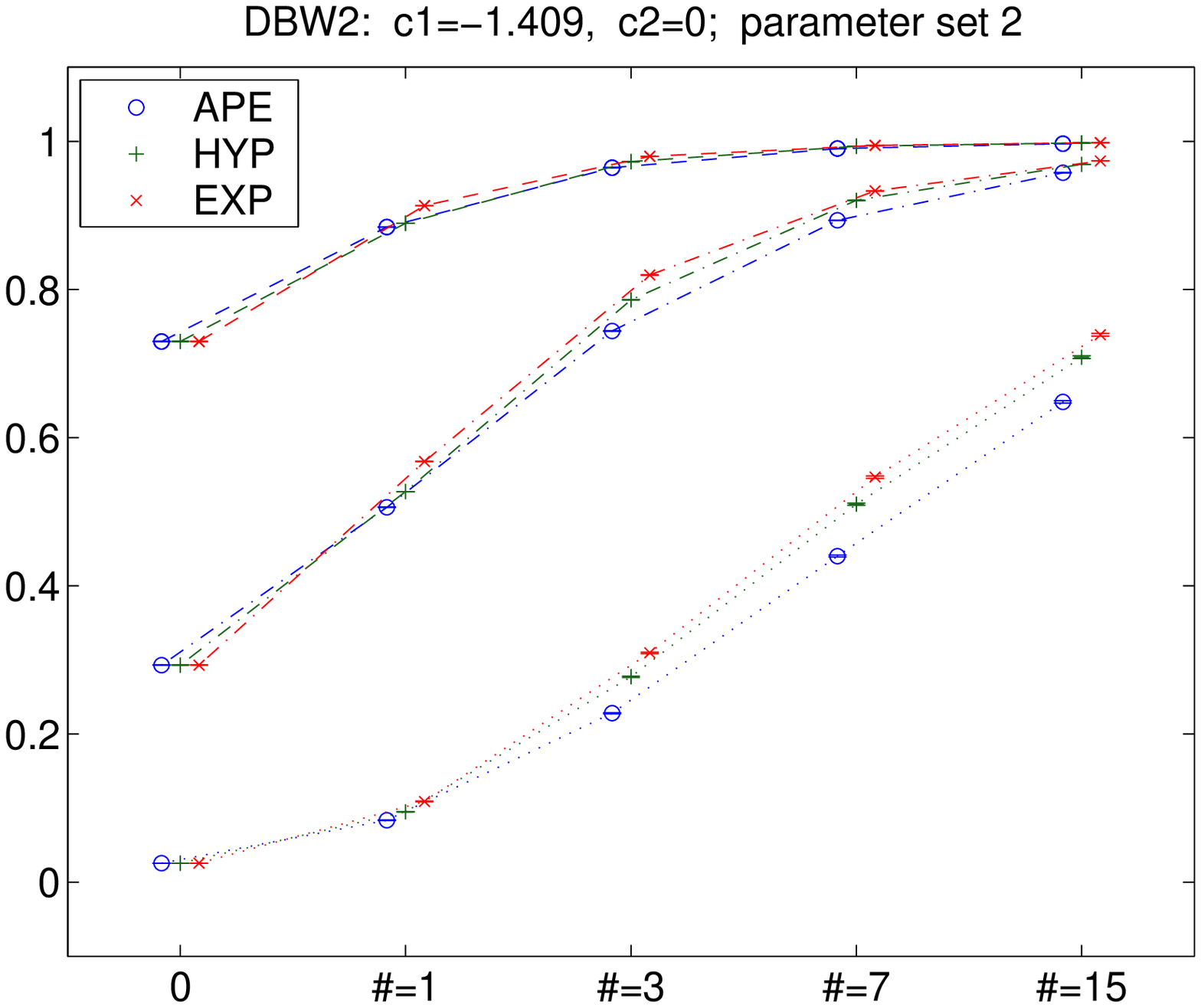}\\[4mm]
\includegraphics[width=84mm,angle=0]{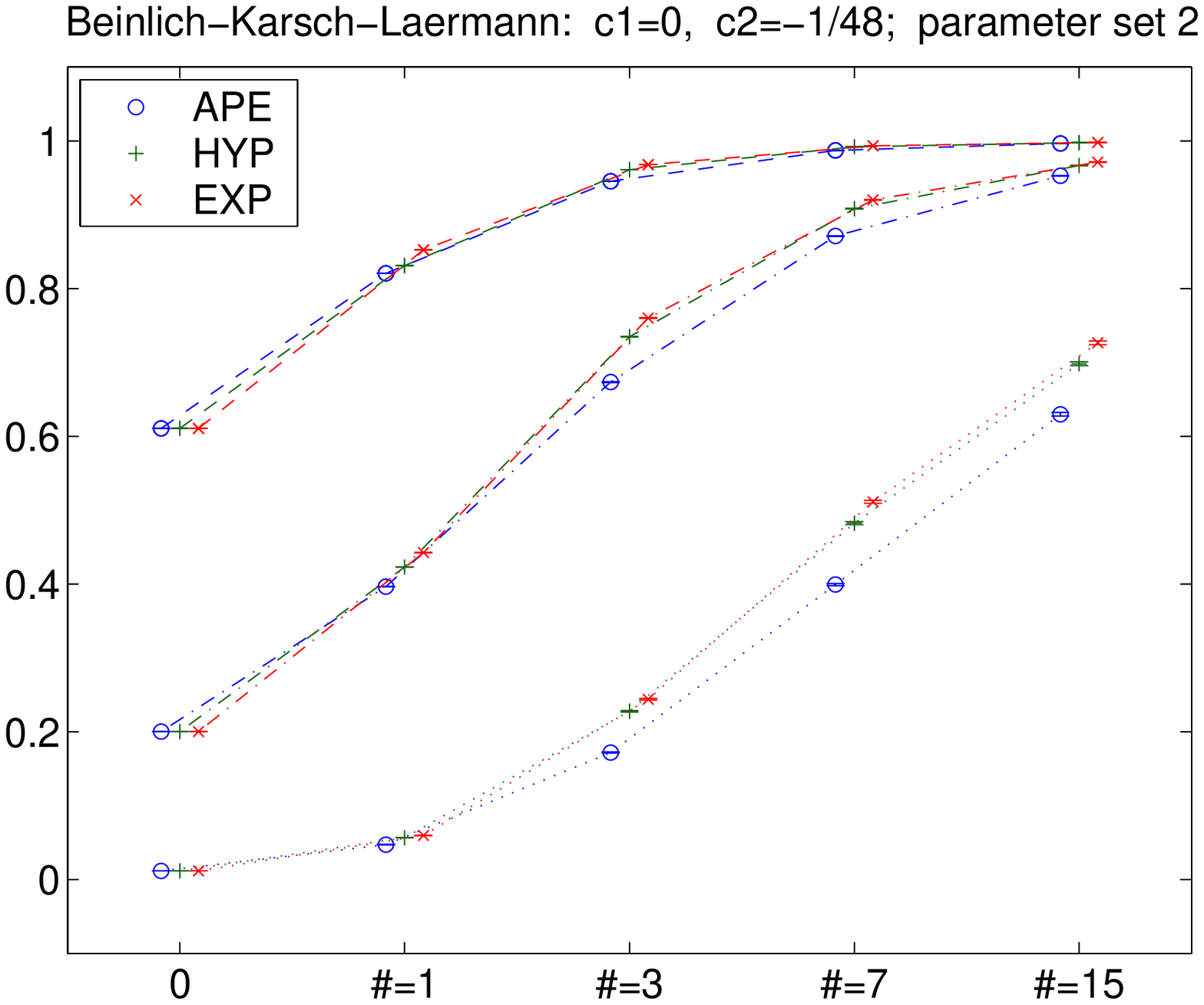}
\includegraphics[width=84mm,angle=0]{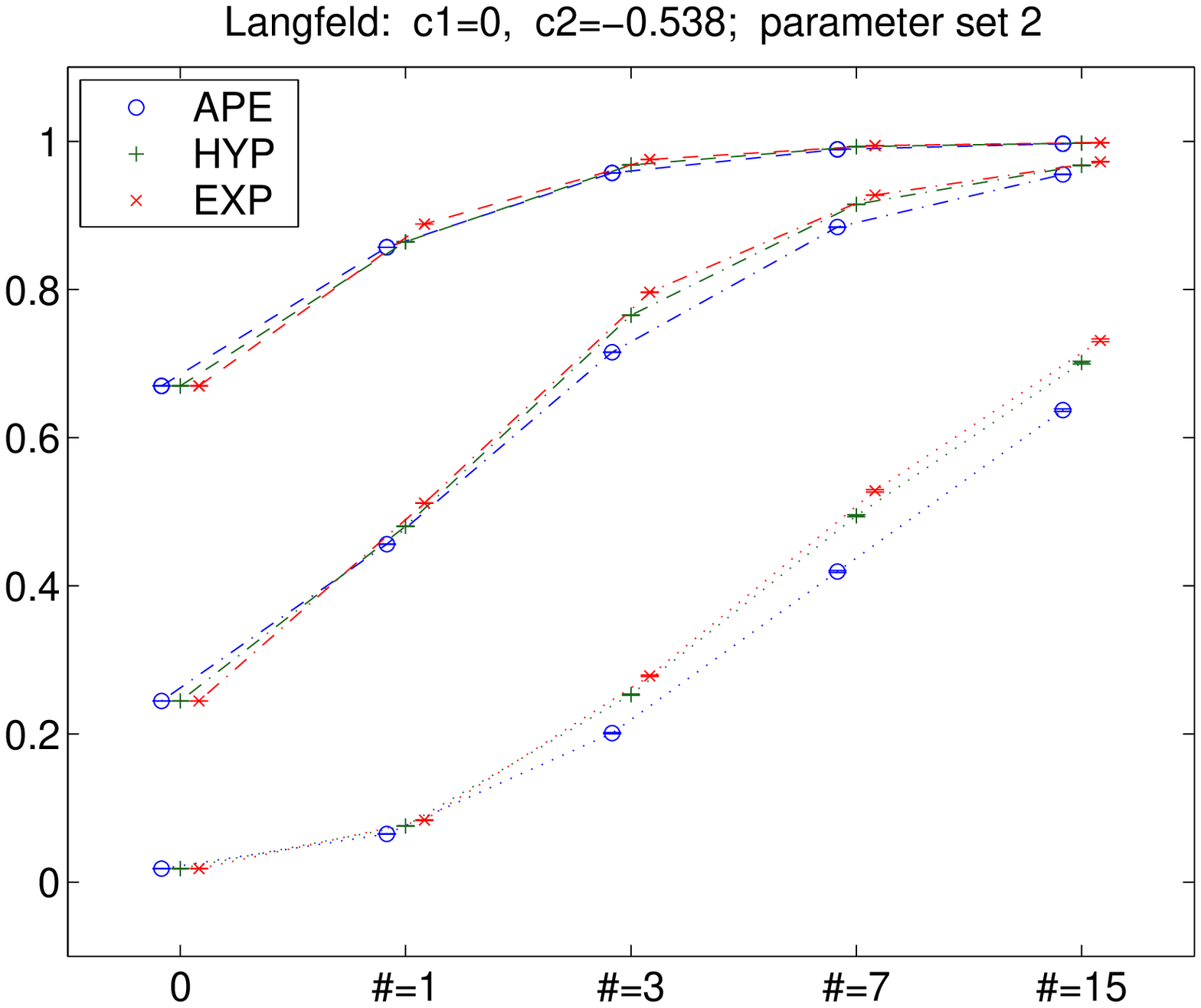}
\end{center}
\vspace{-6mm}
\caption{\sl For each gauge action $W_{1\times1}, W_{2\times2}, W_{4\times4}$
is plotted versus the number of smearing steps, 0 indicates the unsmeared
starting point. Throughout the second parameter in {\rm (\ref{parameterset})}
({\rm APE}, {\rm HYP} with projection) is used: For small parameters many
smearing steps seem favorable.}
\label{fig:smear_par2}
\end{figure}

\begin{figure}
\begin{center}
\includegraphics[width=84mm,angle=0]{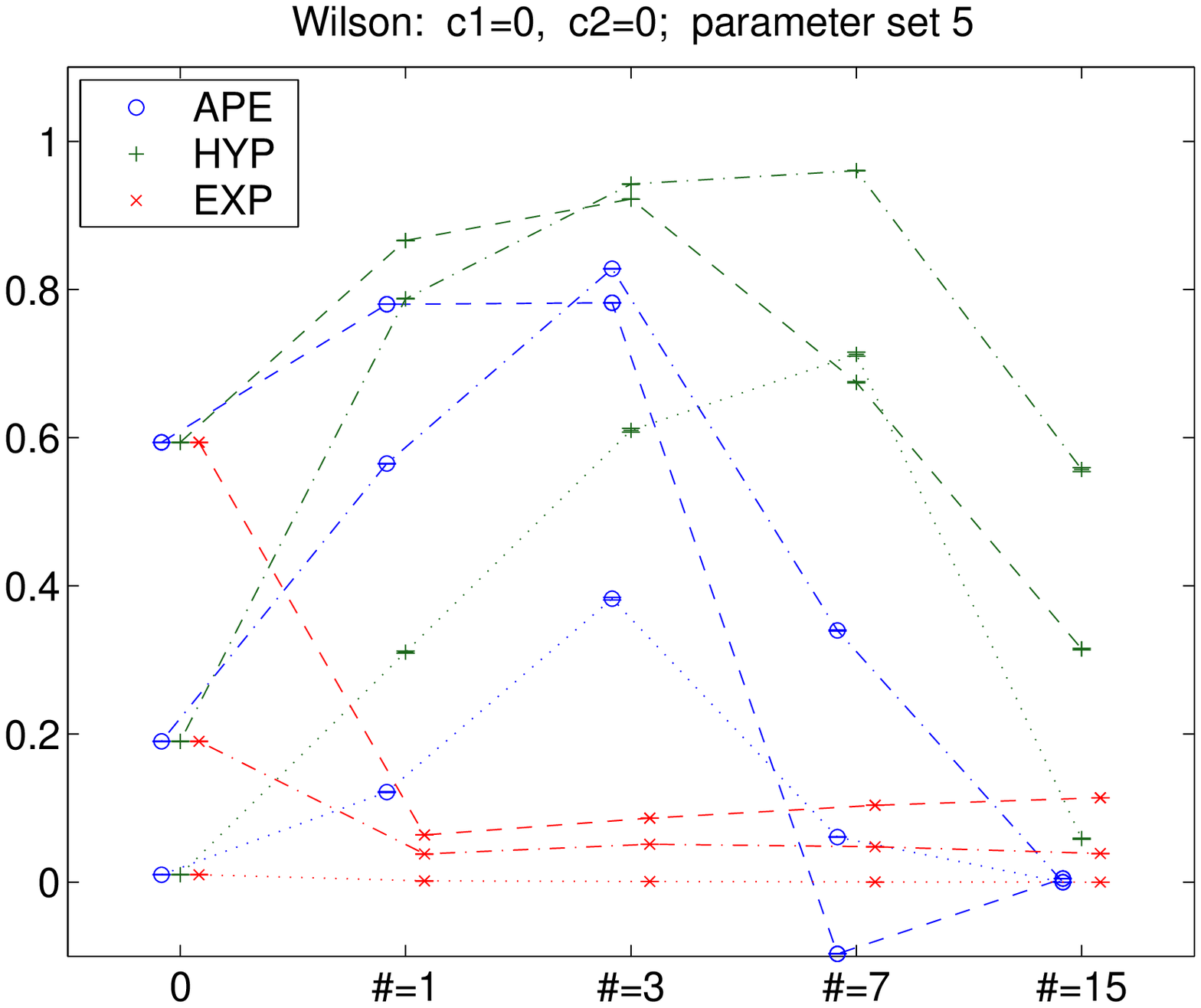}
\includegraphics[width=84mm,angle=0]{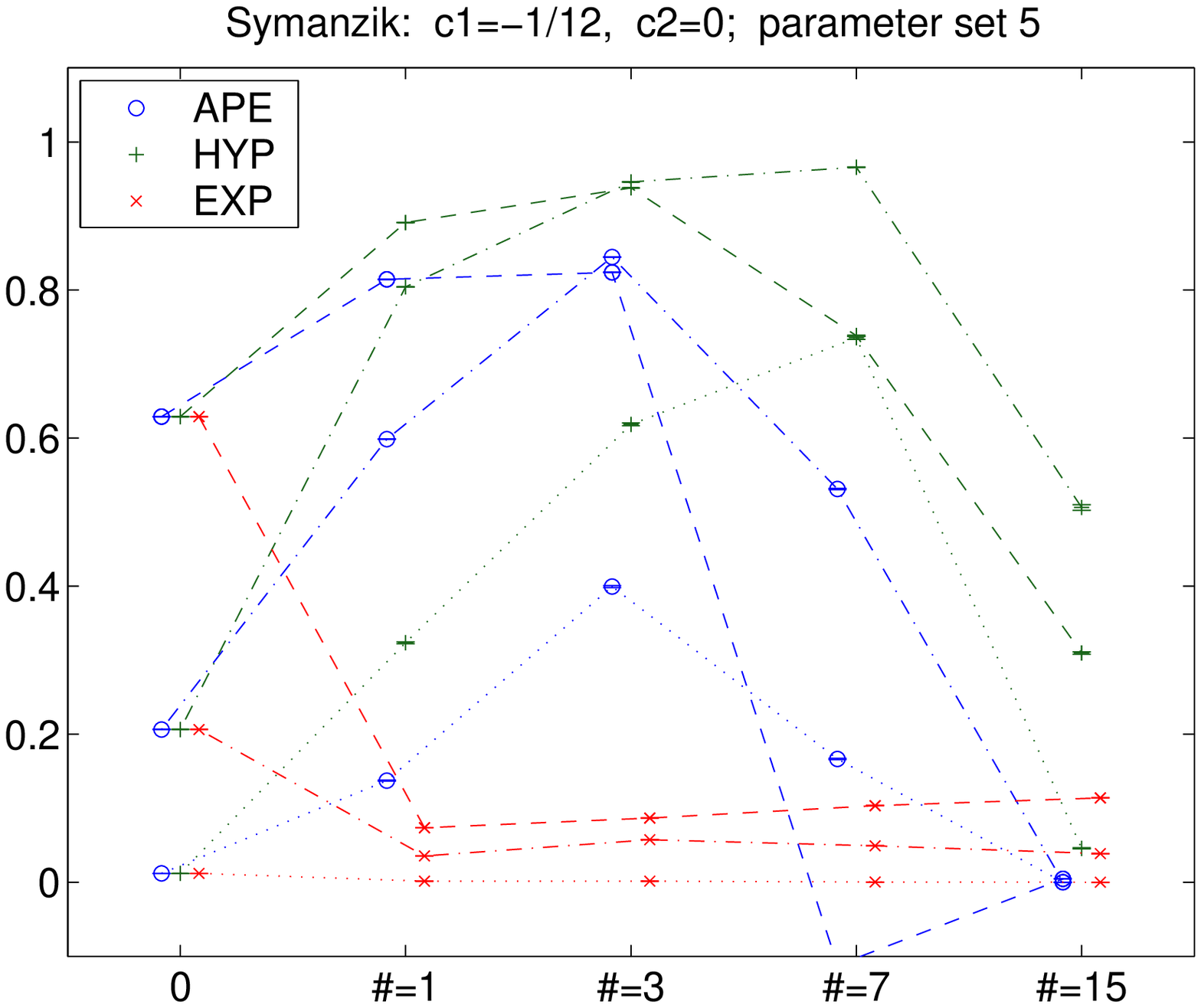}\\[4mm]
\includegraphics[width=84mm,angle=0]{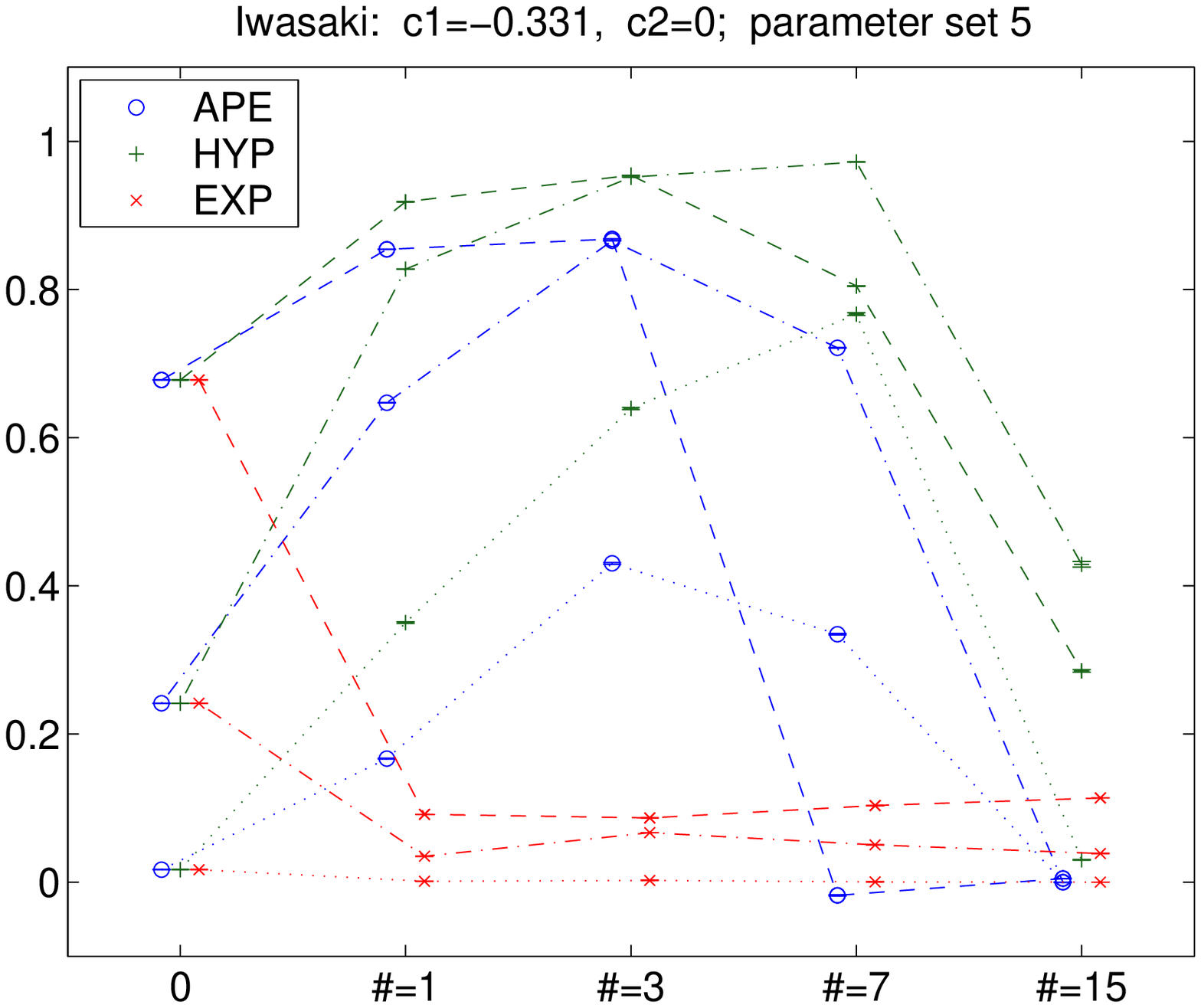}
\includegraphics[width=84mm,angle=0]{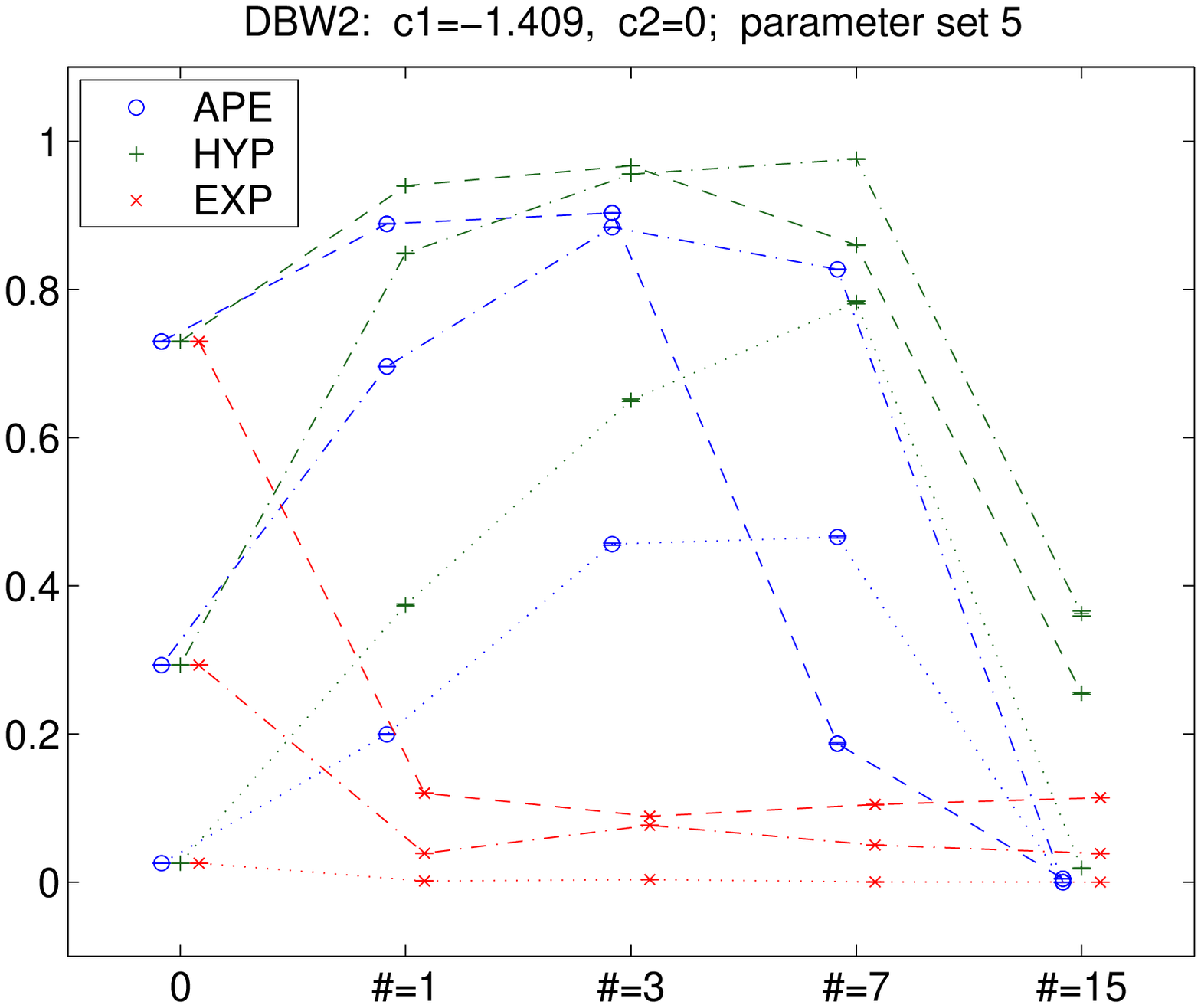}\\[4mm]
\includegraphics[width=84mm,angle=0]{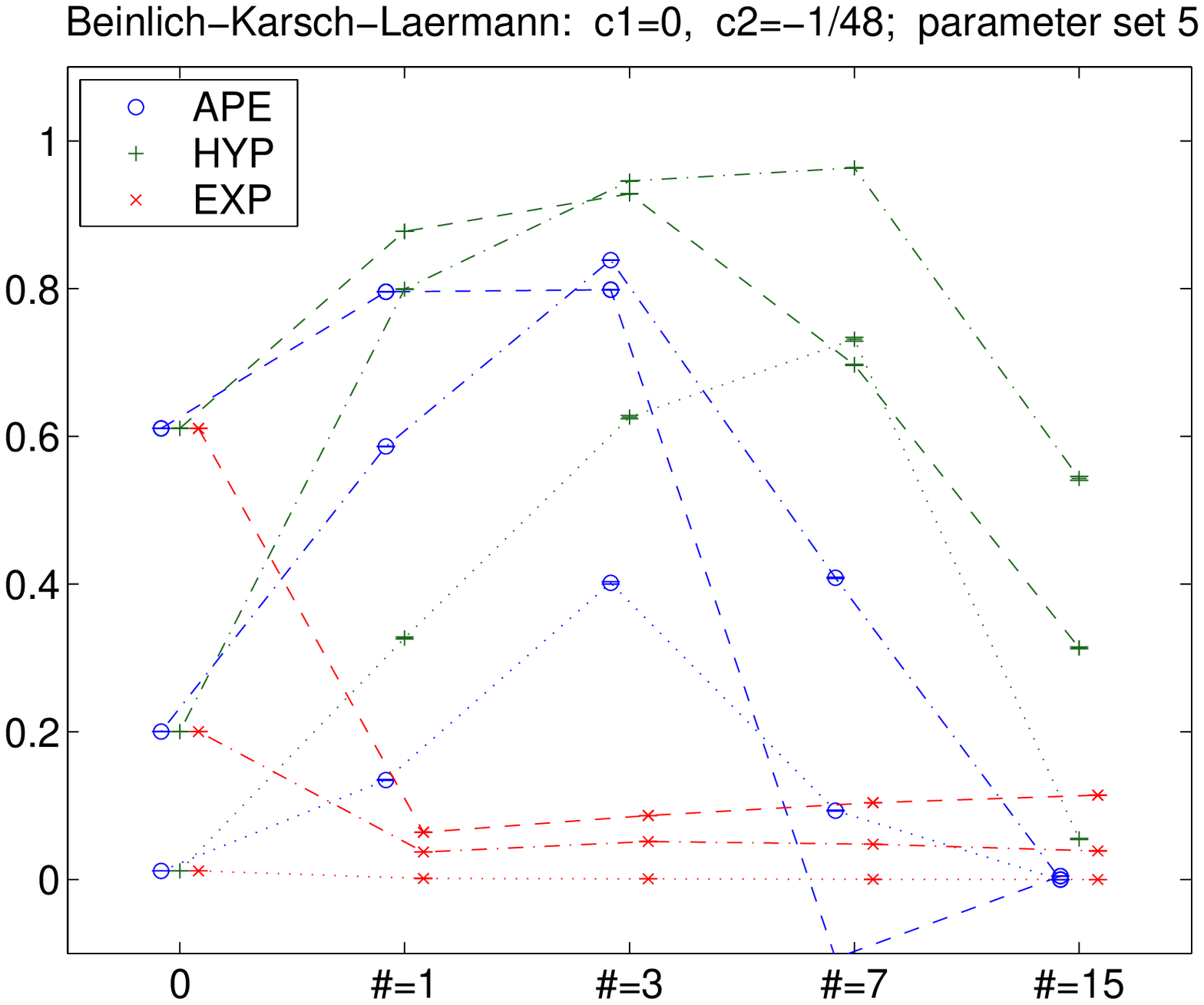}
\includegraphics[width=84mm,angle=0]{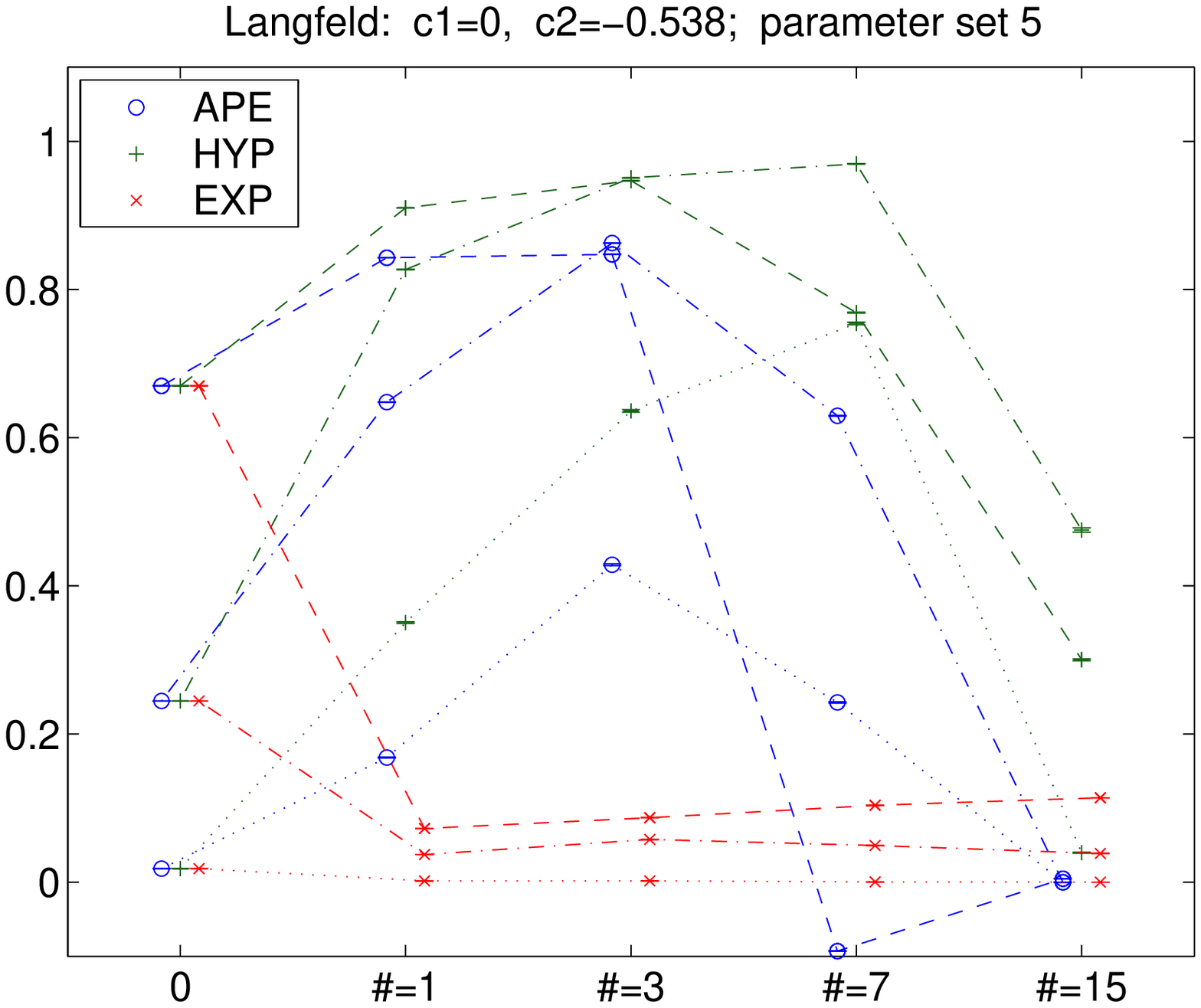}
\end{center}
\vspace{-6mm}
\caption{\sl For each gauge action $W_{1\times1}, W_{2\times2}, W_{4\times4}$
is plotted versus the number of smearing steps, 0 indicates the unsmeared
starting point. Throughout the fifth parameter in {\rm (\ref{parameterset})}
({\rm APE}, {\rm HYP} with projection) is used: For large parameters the
result deteriorates after a few steps.}
\label{fig:smear_par5}
\end{figure}

Fig.~\ref{fig:smear_lev1} shows, for each action, the Wilson
loops $W_{1\times1}, W_{2\times2}, W_{4\times4}$ after 1 smearing step (with
projection back to the gauge group) as a function of the smearing parameter.
The plaquette gets increased for small $\al$ reaching a maximum
around $\al_\mr{APE}\!\simeq\!0.7, \al_\mr{HYP}\!\simeq\!\al_\mr{HYP}^\mr{std},
\al_\mr{EXP}\!\simeq\!0.2$; for larger parameters it decreases again.
This holds for the larger Wilson loops, too, albeit the maximum may be shifted
towards larger $\al$ (for APE and HYP).
What all recipes have in common is that excessive parameters drive all loops
to 0 (disorder) rather than to 1 (order).

Fig.~\ref{fig:smear_lev3} is the same after 7 smearing steps.
For small $\al$ all Wilson loops are driven more efficiently to 1 than before,
but beyond a certain ``critical'' parameter the descent to disorder is more
pronounced, too.
For APE and HYP this culminating $\al$ is barely changed compared to the
previous figure, for EXP it is smaller.

Fig.~\ref{fig:smear_noproj} demonstrates a dramatic change in the behavior of
any Wilson loop under smearing, if the projection in the APE or HYP recipe is
abandoned.
The graphs have been obtained with Wilson glue, but other actions yield the
same type of monotonic decrease in $\al$.

Fig.~\ref{fig:smear_par2} shows the Wilson loops $W_{1\times1}, W_{2\times2},
W_{4\times4}$ (with projection in the APE and HYP cases) versus the iteration
level for the second set of parameters in (\ref{parameterset}).
With such small $\al$-values one always observes a benevolent pattern.

Fig.~\ref{fig:smear_par5} shows that with somewhat larger parameters
(throughout the fifth one out of the set (\ref{parameterset})) iterating
the smearing is not really a good idea.
Note that these parameters are to the right of the maximum in
Fig.~\ref{fig:smear_lev1}, for ``sub-critical'' $\al$-values no such
breakdown has been seen.

It is worth emphasizing how little the smearing ``profile'' depends on the
gauge action, for matched ensembles.
A maximum value of $P\!=\!\<W_{1\times1}\>$ might be used as a criterion to
``optimize'' the APE/HYP/EXP-parameters (the rather similar values with 1
and 7 iterations are found in the captions).
Of course, such a criterion is arbitrary, but at least the result is
\emph{universal}, i.e.\ (to a good extent) independent of the gauge action
used (for fixed physical spacing).

\bigskip

More interesting than the behavior of the Wilson loop itself is how the mean
and the error of a physical observable derived from it evolve under smearing,
i.e.\ whether different recipes yield consistent results.
For that aim let us consider the squared string
tension $\si\!=\!-\log(\ch)$ defined via the symmetric Creutz ratio
(\ref{defcreutz}), using the same gauge configurations as before.

\begin{figure}
\begin{center}
\scriptsize
\psfrag{al_1}{$\alpha^{(1)}$}
\psfrag{al_2}{$\alpha^{(2)}$}
\psfrag{al_3}{$\alpha^{(3)}$}
\psfrag{al_4}{$\alpha^{(4)}$}
\psfrag{al_5}{$\alpha^{(5)}$}
\psfrag{al_6}{$\alpha^{(6)}$}
\includegraphics[width=84mm,angle=0]{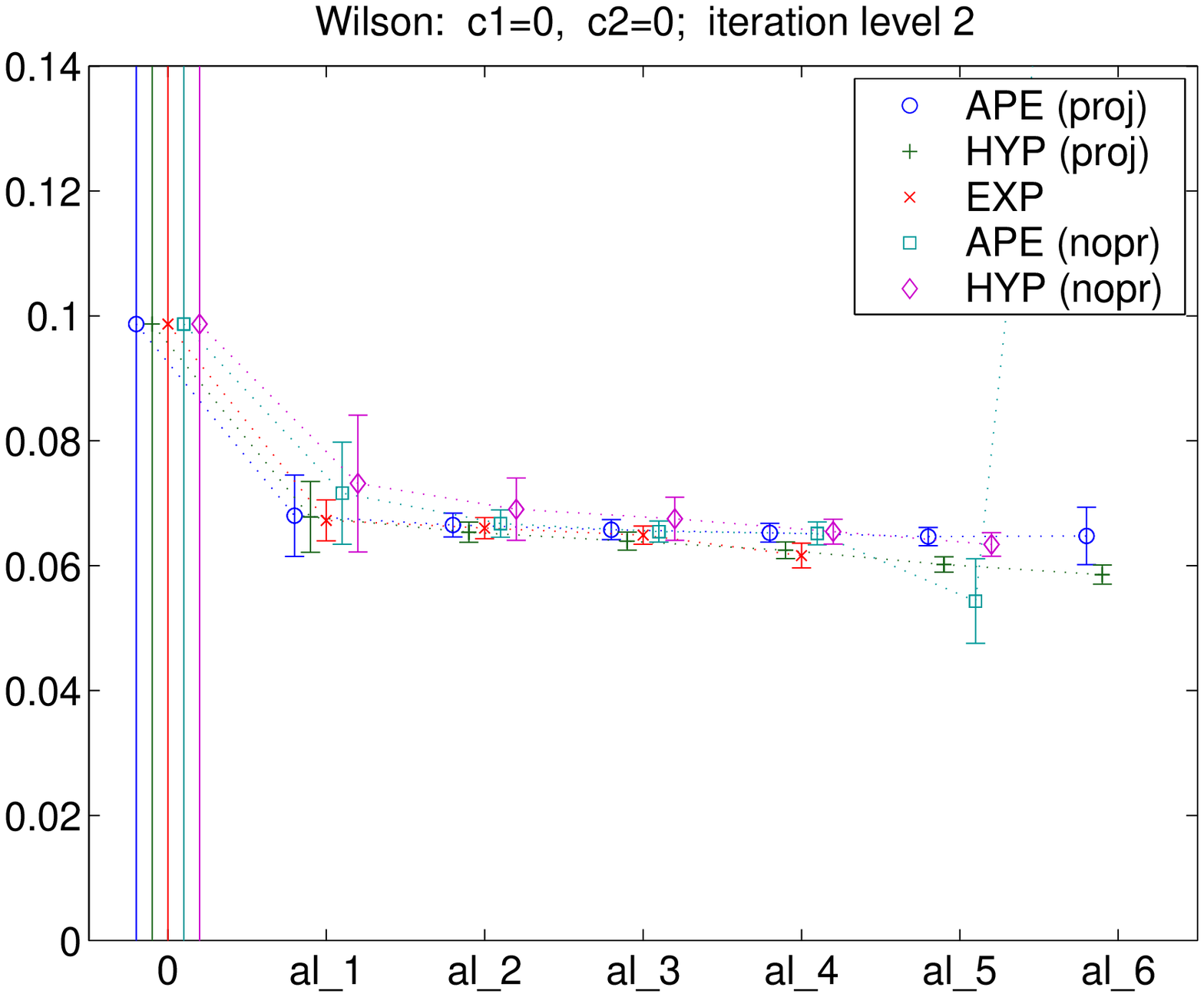}
\includegraphics[width=84mm,angle=0]{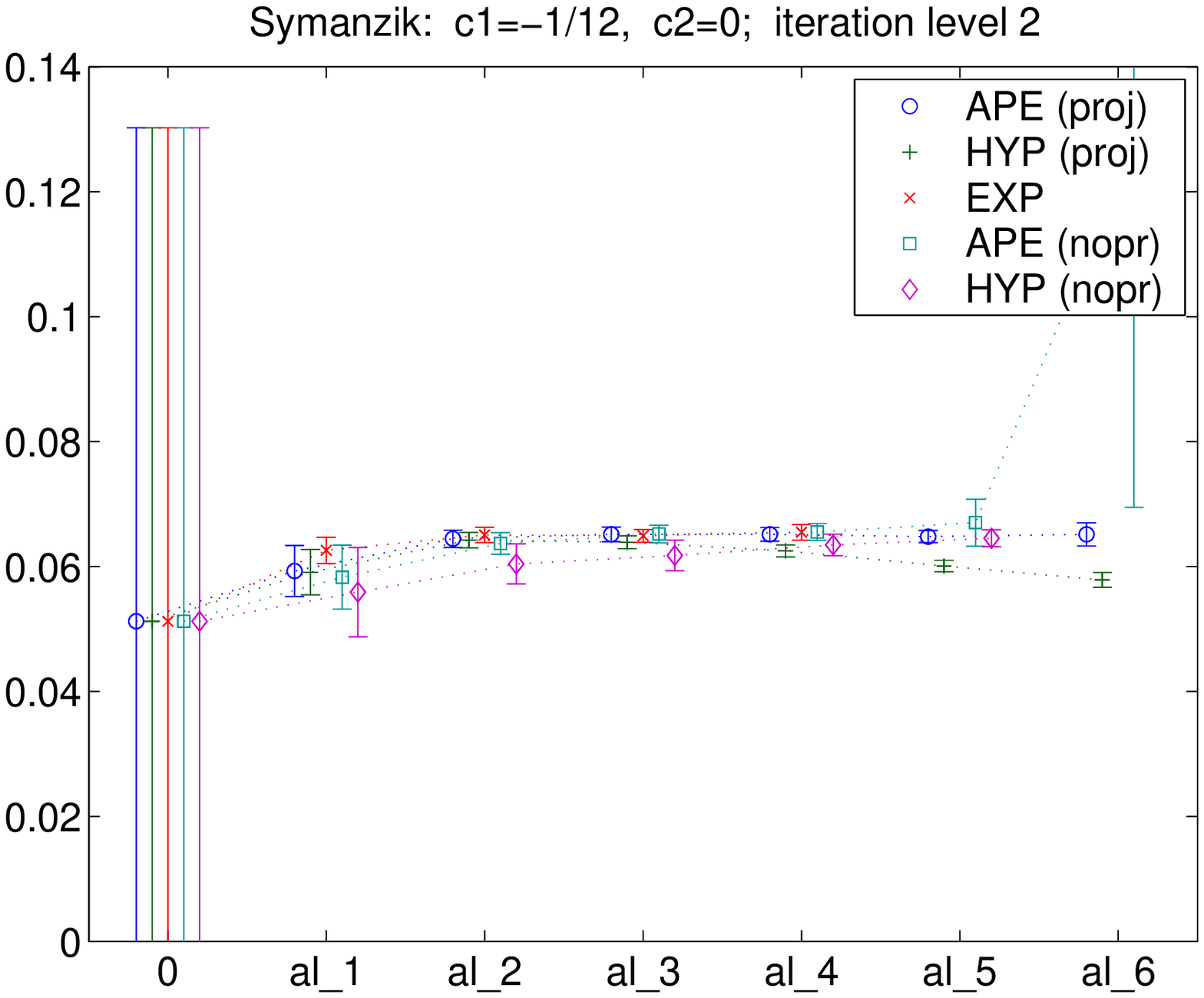}\\[4mm]
\includegraphics[width=84mm,angle=0]{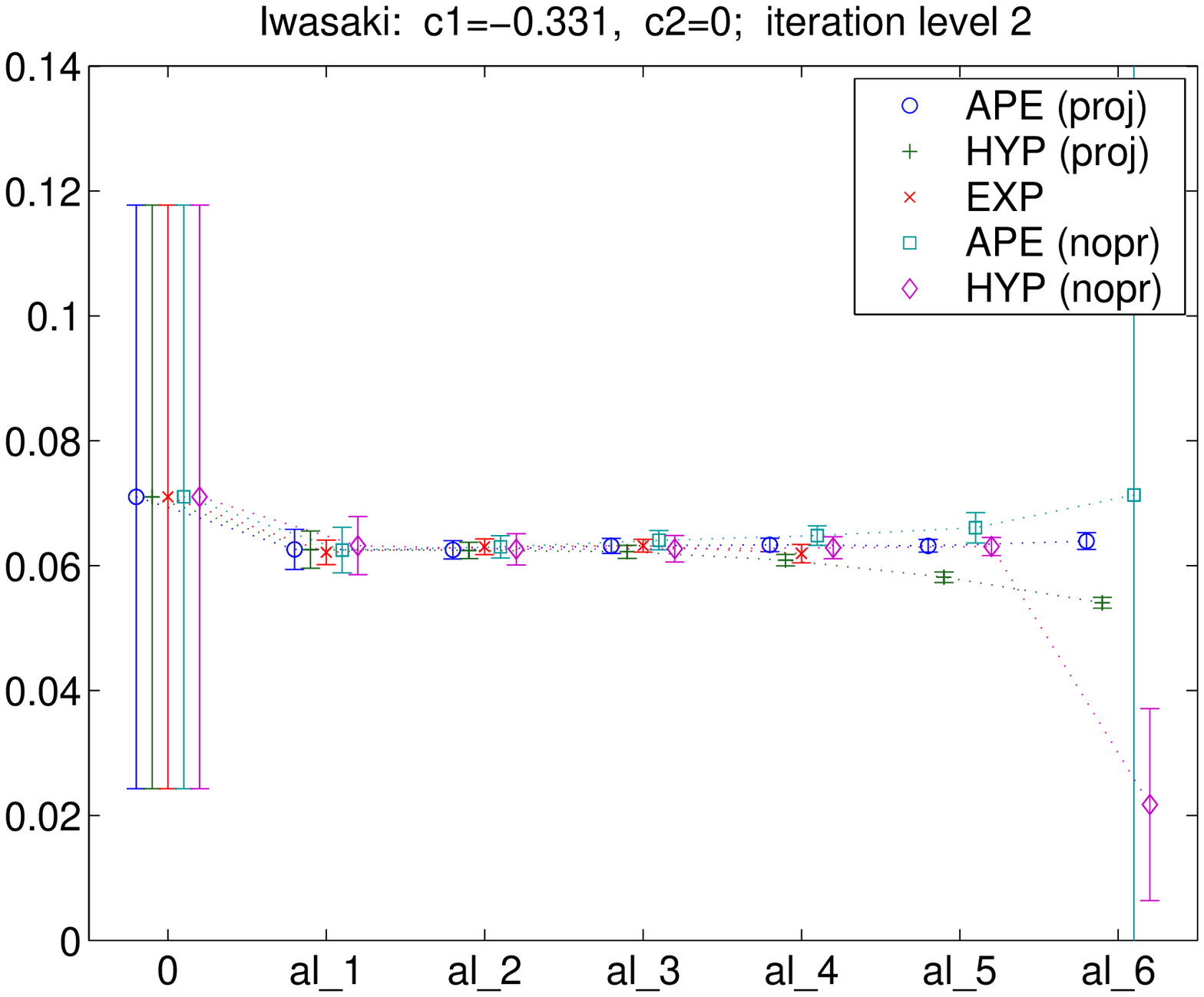}
\includegraphics[width=84mm,angle=0]{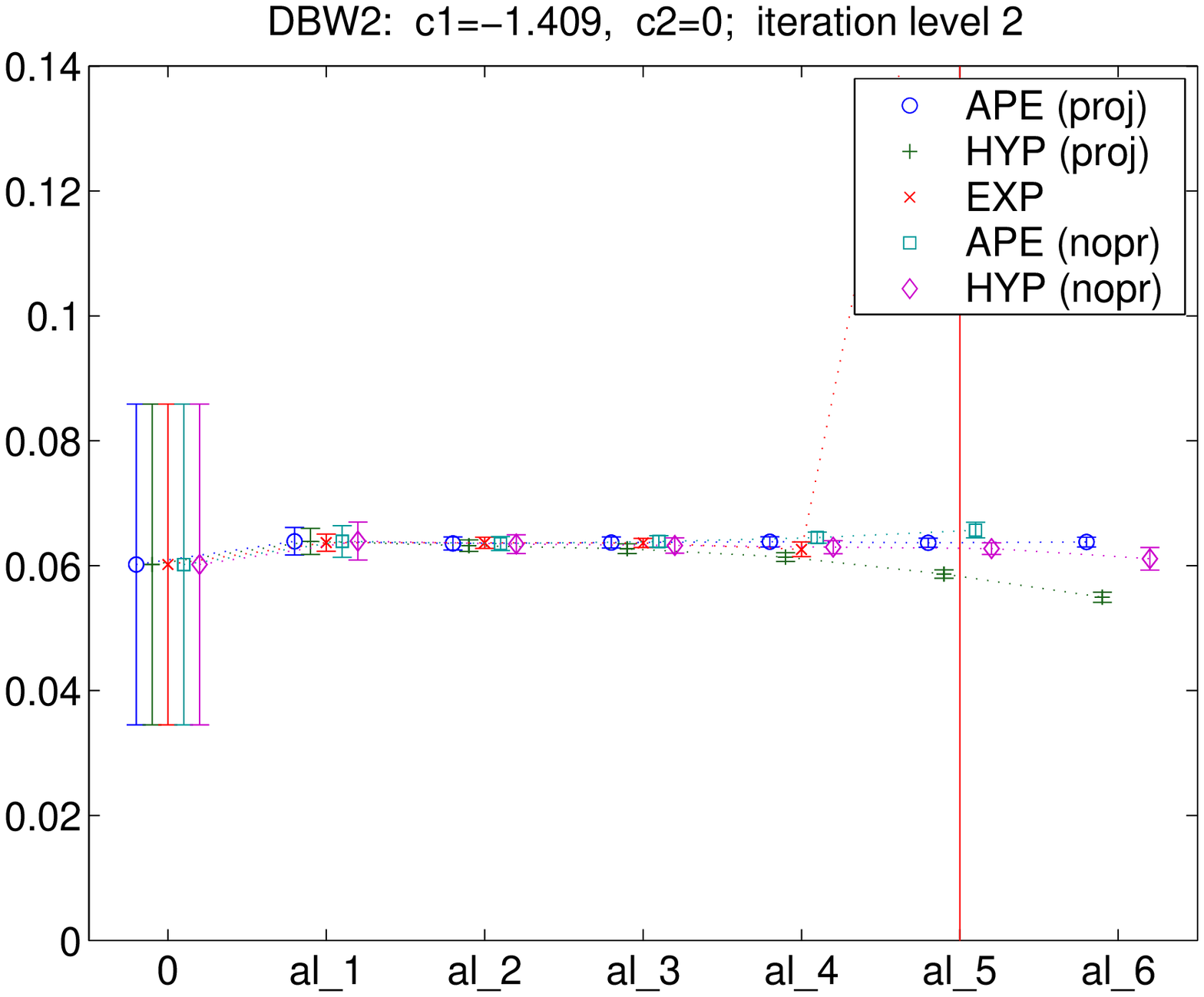}\\[4mm]
\includegraphics[width=84mm,angle=0]{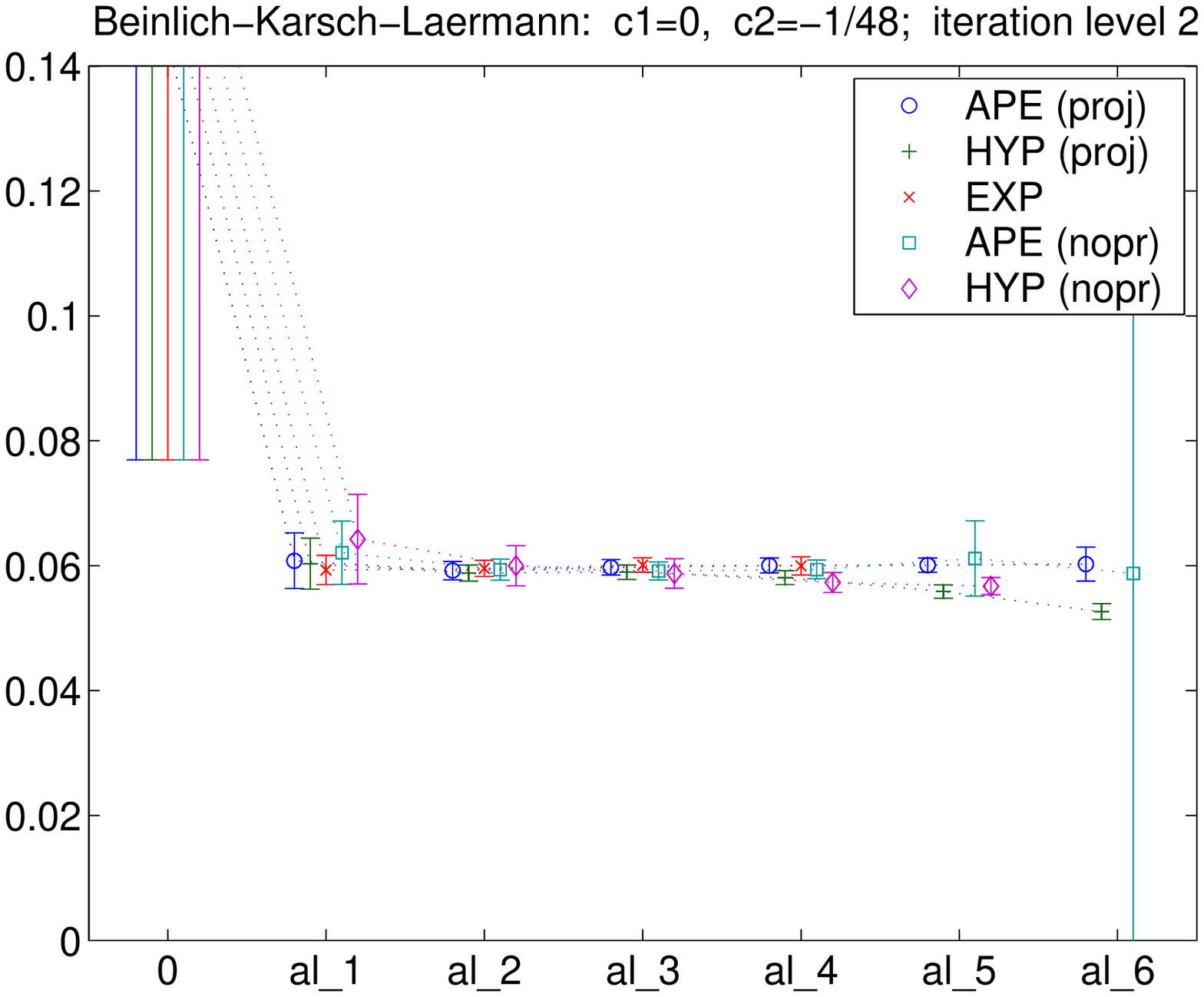}
\includegraphics[width=84mm,angle=0]{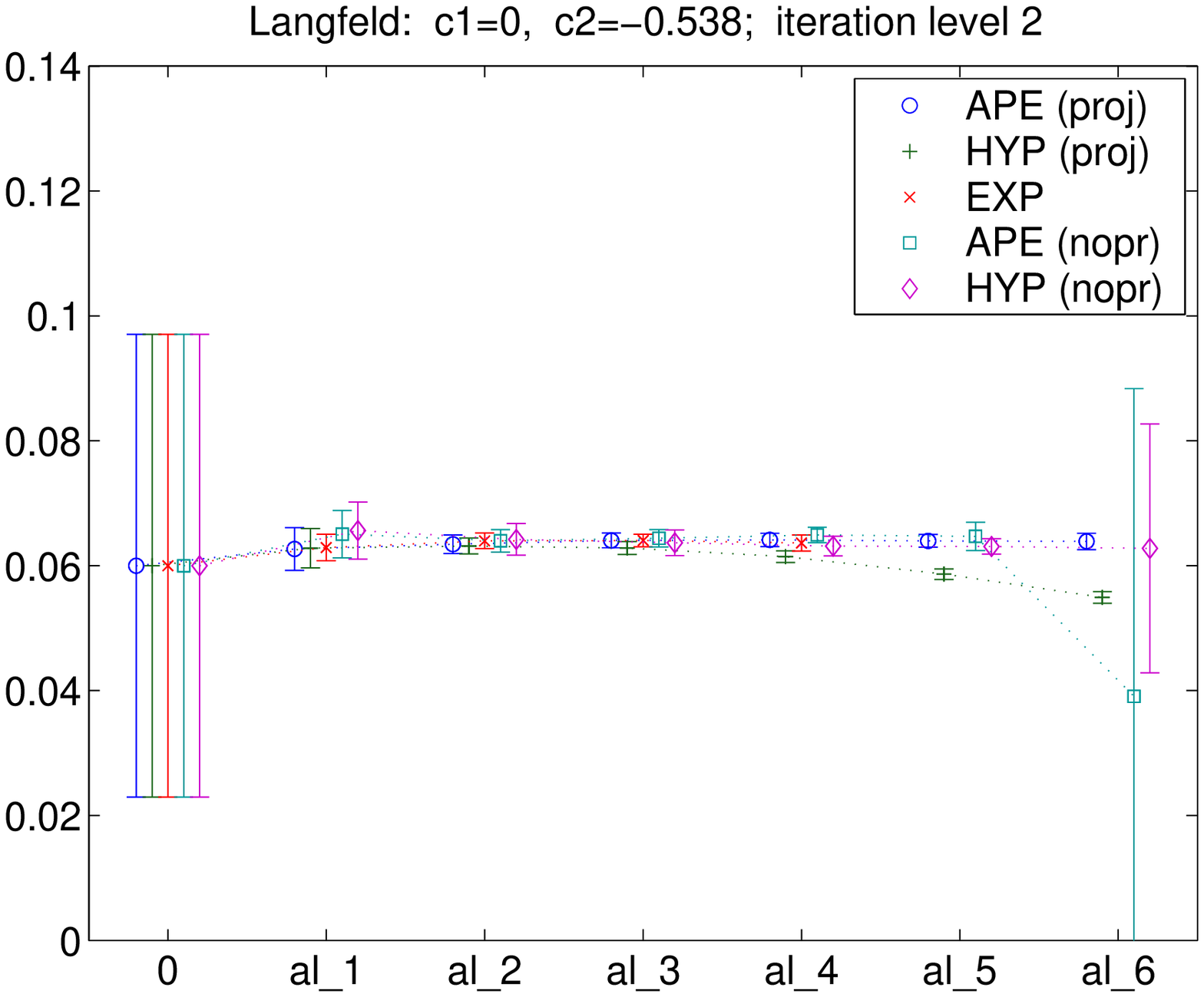}
\end{center}
\vspace{-6mm}
\caption{\sl For each gauge action the squared string tension
$\si\!=\!-\log(\ch(5))$ is plotted versus the parameter number in
{\rm (\ref{parameterset})}, 0 indicates the unsmeared starting point.
Throughout 3 smearing steps have been applied. Data with negative central
values have been removed.}
\label{fig:creutz_lev2}
\end{figure}

\begin{figure}
\begin{center}
\includegraphics[width=84mm,angle=0]{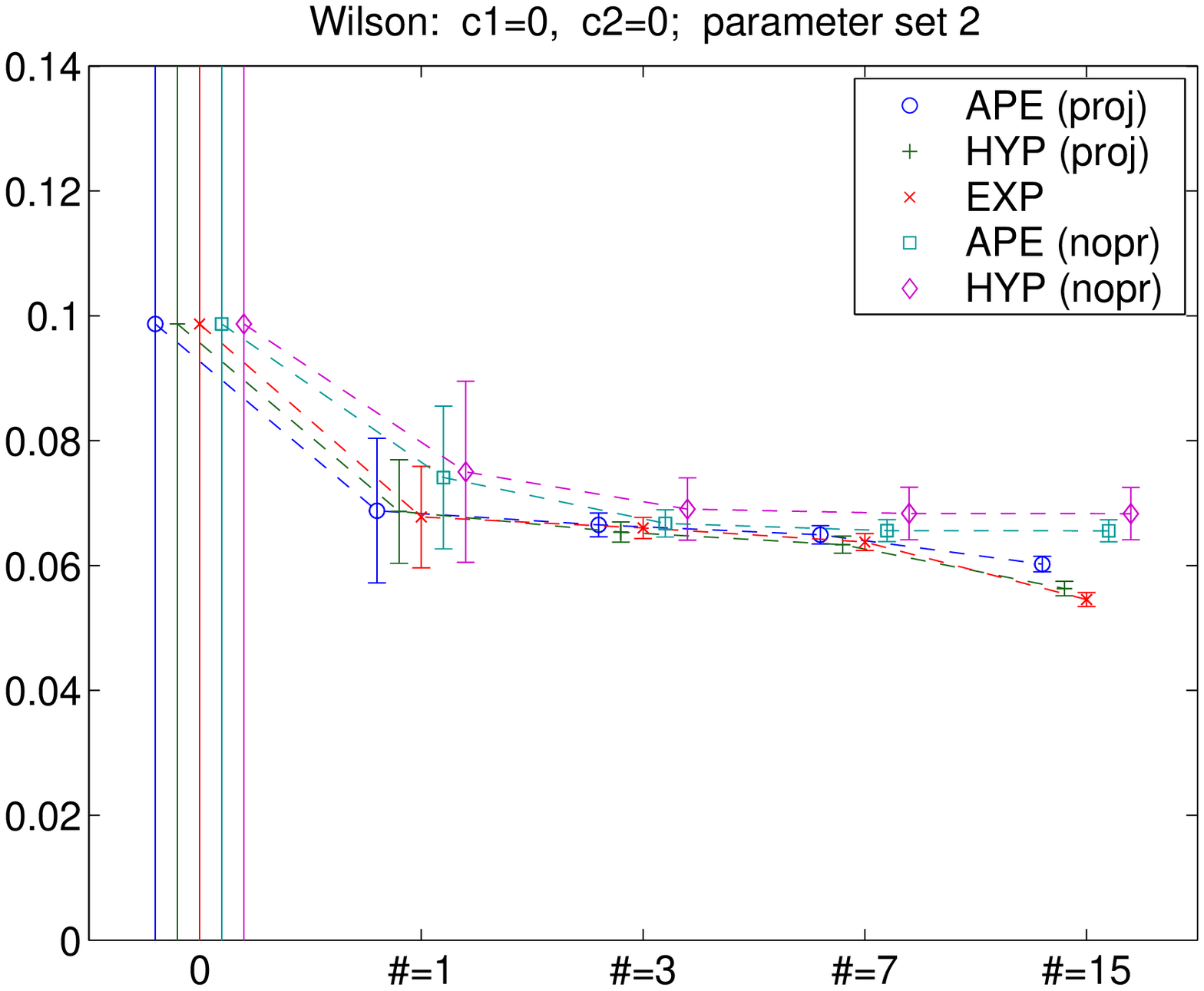}
\includegraphics[width=84mm,angle=0]{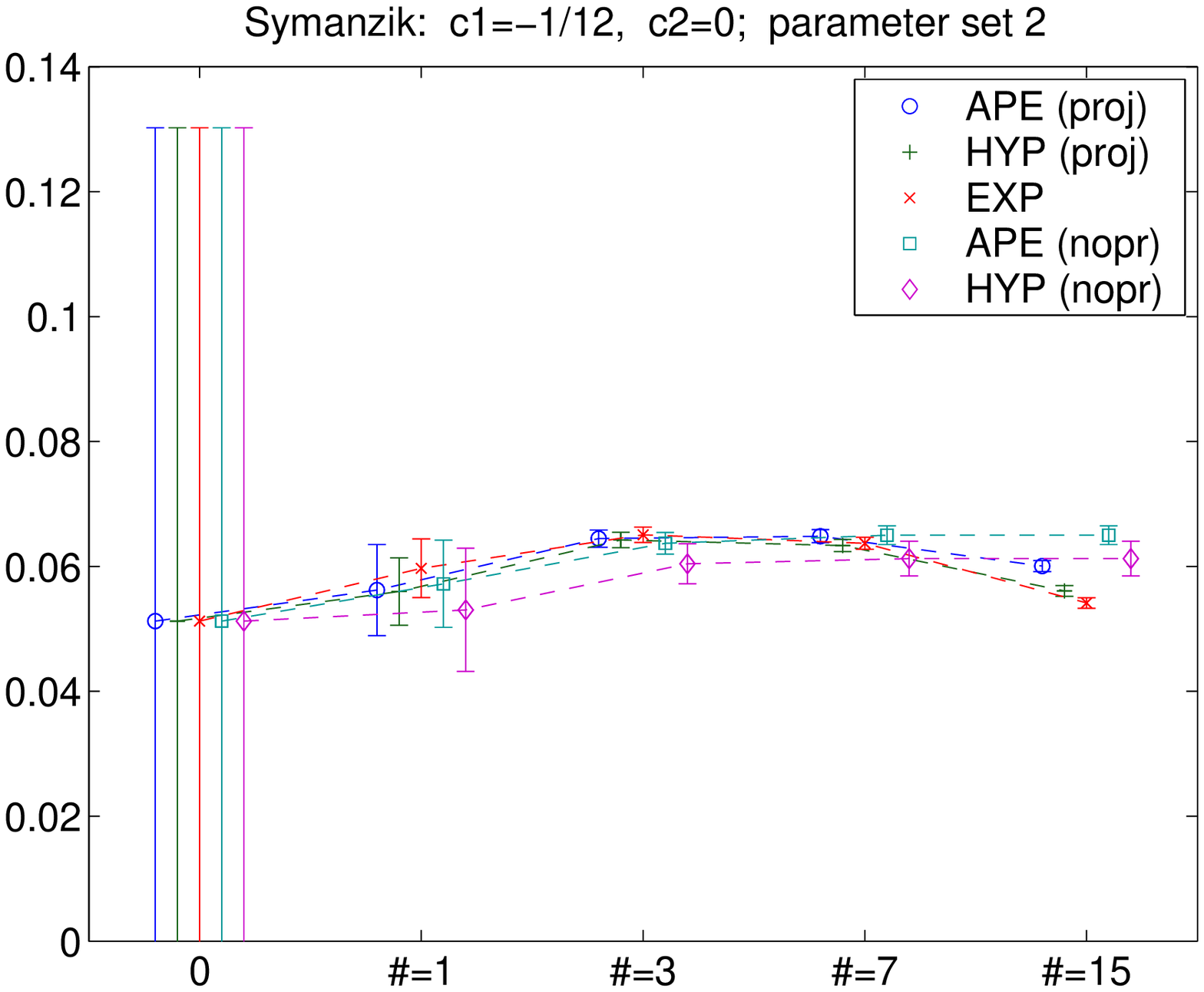}\\[4mm]
\includegraphics[width=84mm,angle=0]{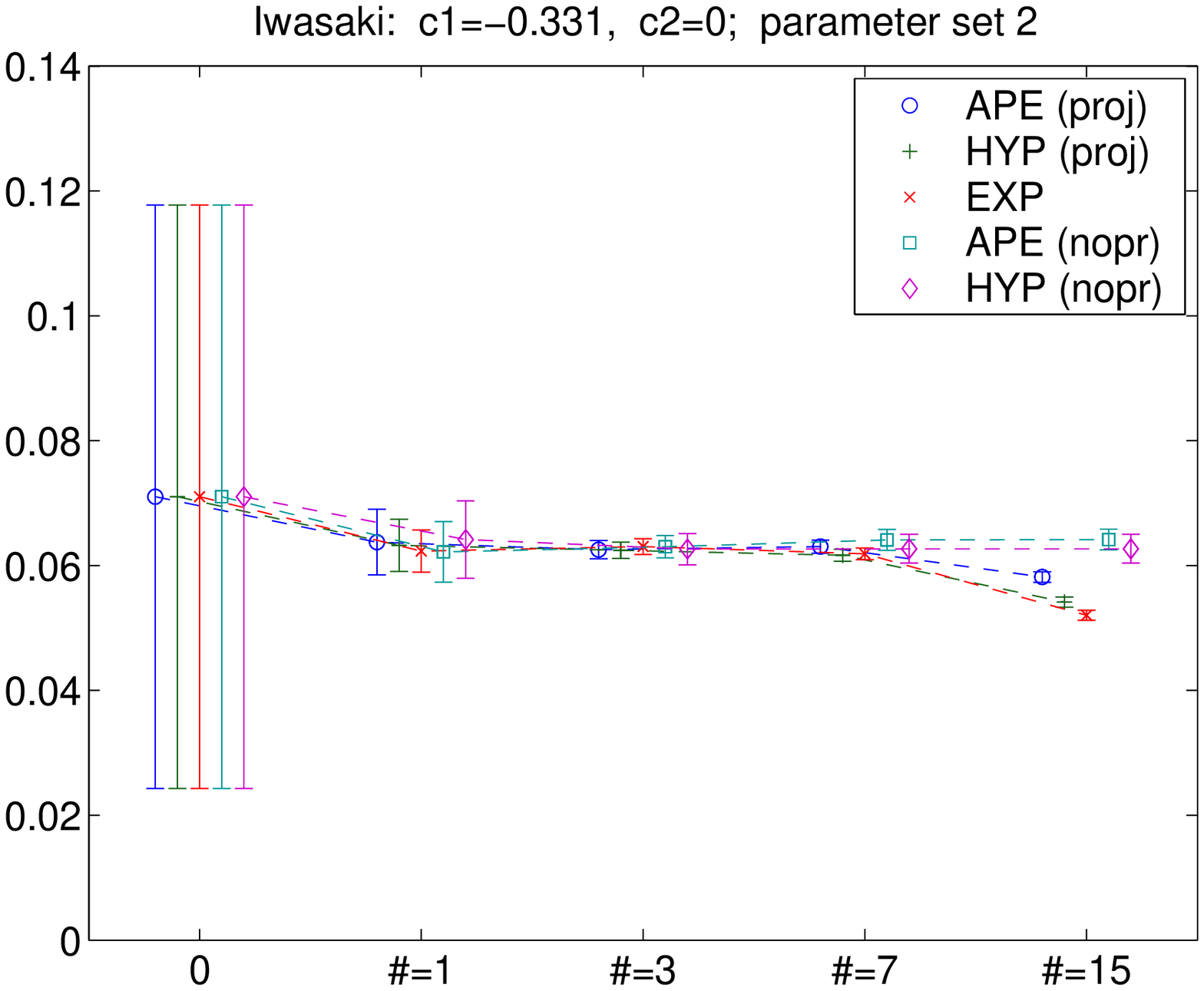}
\includegraphics[width=84mm,angle=0]{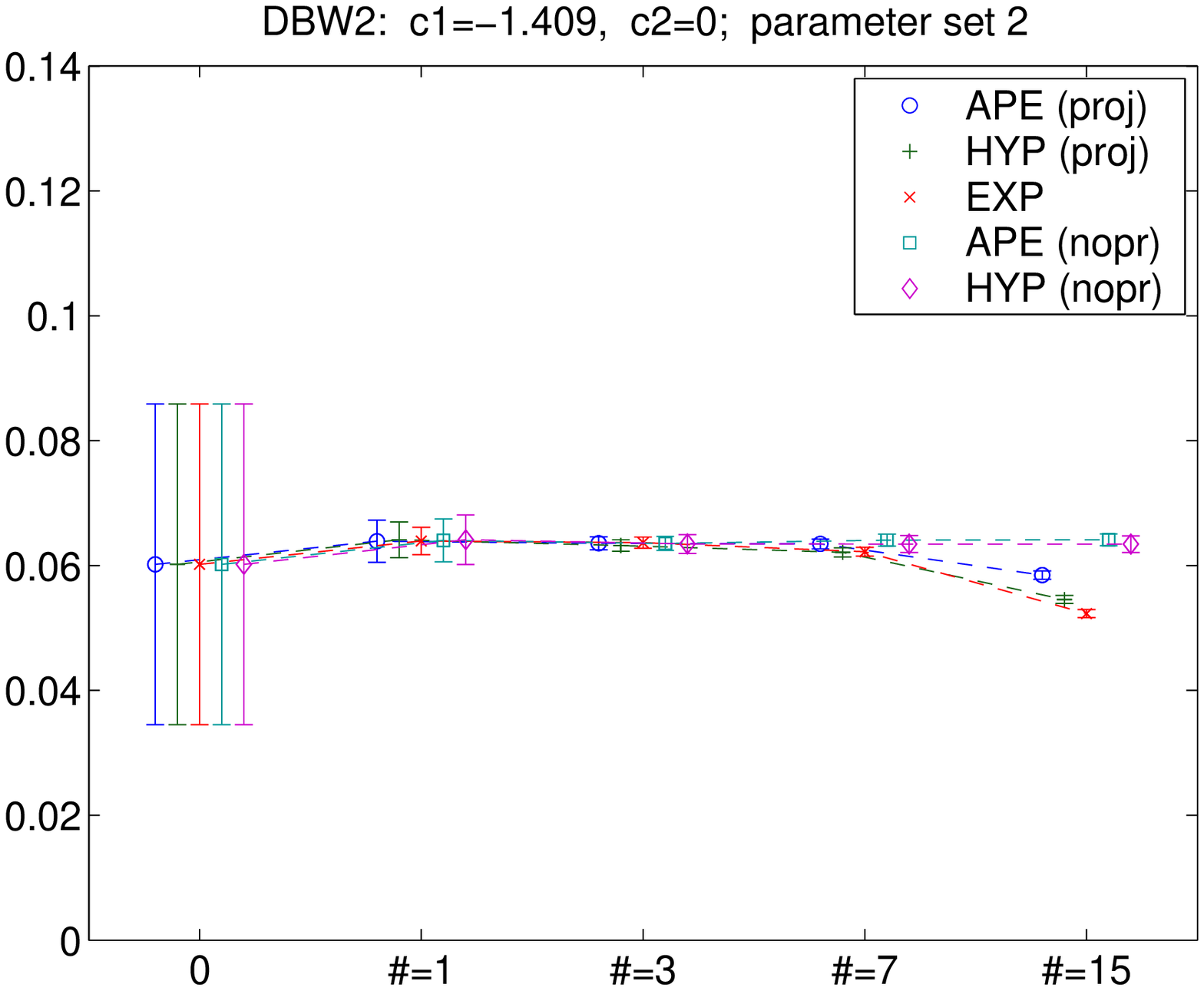}\\[4mm]
\includegraphics[width=84mm,angle=0]{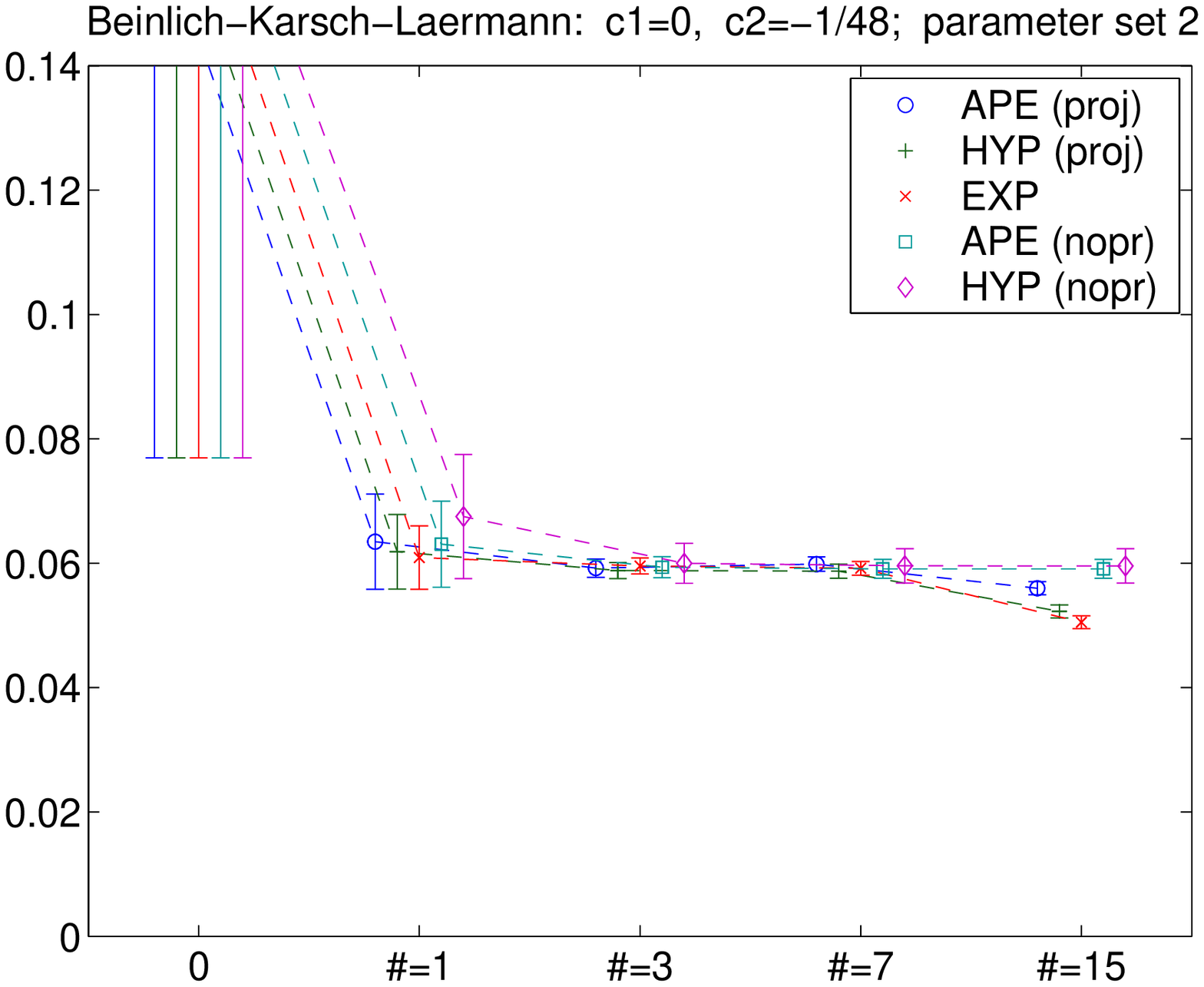}
\includegraphics[width=84mm,angle=0]{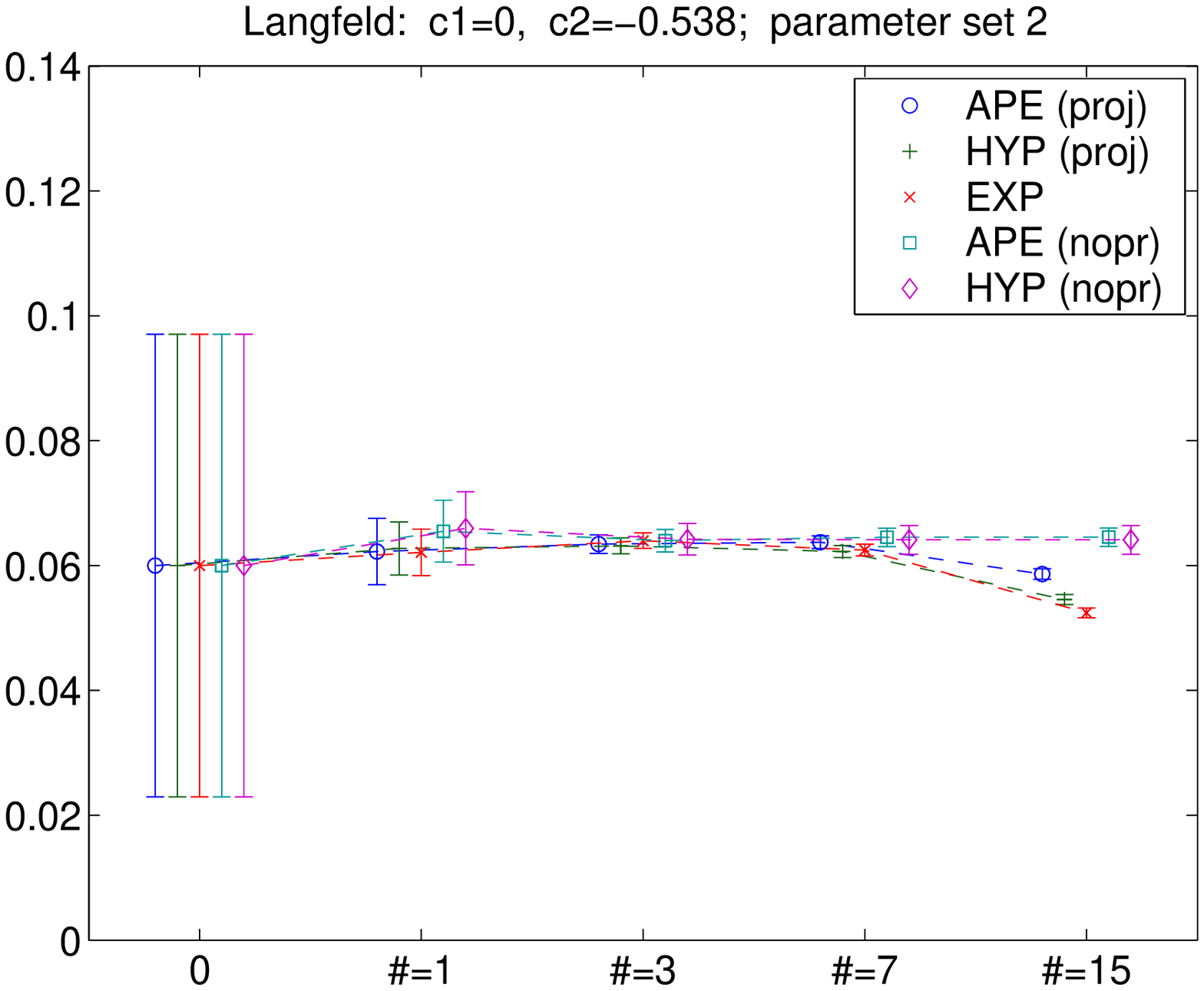}
\end{center}
\vspace{-6mm}
\caption{\sl For each gauge action the squared string tension
$\si\!=\!-\log(\ch(5))$ is plotted versus the number of smearing steps, 0
indicates the unsmeared starting point (where improved actions give a
smaller error than the Wilson action). Throughout the second parameter
in {\rm (\ref{parameterset})} is used.}
\label{fig:creutz_par2}
\end{figure}

Fig.~\ref{fig:creutz_lev2} plots $-\log(\ch(5))$ versus the parameter $\al$ in
(\ref{parameterset}) after 3 smearing steps.
Moderate parameters suffice to tremendously reduce the noise and and different
recipes prove consistent within errors.
However, as the smearing parameter is raised beyond a reasonable level
(typically something of the order of the ``critical'' parameter that achieved
the maximum in the plaquette), the \emph{error may grow} again.
When repeating this with a higher iteration level basically the same
behavior emerges, although the onset of systematic distortions with large
parameters gets more pronounced.
There are two more points that deserve a comment.
First, I find it astonishing that the projection-free APE and HYP recipes
manage to decrease the error in a physical observable, even though we have just
seen that they drive the plaquette to 0 rather than 1.
Does this indicate that the naive interpretation is wrong, and what one gets
is not disorder but some type of anti-ferromagnetic ordering ?
Obviously, this calls for further investigation, but it seems non-trivial to
address this issue in a gauge-invariant manner without invoking extreme
requests for CPU time.
Finally, comparing the unsmeared values, one notices that all improved actions
have an error bar which is smaller than the one with the Wilson action.
Hence an improved action may lead to a smaller error (with identical
statistics) in a gluonic observable, but it is clear that this is not
competitive with the effect of smearing.

Fig.~\ref{fig:creutz_par2} plots $-\log(\ch(5))$ versus the number of smearing
steps for the second parameter in the set (\ref{parameterset}).
Here, a ``plateau'' includes $1$ to $\sim\!7$ iterations -- further out
consistency is lost.
With a larger parameter a similar picture emerges, but the ``plateau'' ends
earlier.
Numerically, a value $\si\!\simeq\!0.06(1)$ corresponds to a string tension
$\sqrt{\si}\!\simeq\!490(40)\MeV$, if $a^{-1}\!=\!2\GeV$ is assumed.
Hence a statistical accuracy of 8\% can be achieved with 100 lattices of size
$12^4$.

In summary, these figures indicate a trade-off situation between a statistical
error on one side and a systematic error on the other.
Smearing reduces the former, but when applied in excess (either through too
large a parameter or too many iterations) it introduces clear artifacts.
It would be interesting to test how much of these can be attributed to the loss
of locality.
The good news is that a detailed optimization of the parameters is not needed
-- smearing proves beneficial as long as one can explicitly demonstrate that
one is in the \emph{two-fold plateau}; i.e.\ as long as a even larger
parameters and/or higher iteration levels do not alter the result.


\section{Effect on field-theoretic topological charge}


Knowing that smearing reduces the UV-fluctuations and given the
relation between UV-fluctua\-tions and large renormalization factors and/or
cut-off effects, one may ask whether smearing reduces the latter.
An observable from which one can learn something in this respect without having
to take the continuum limit is the ``naive'' field-theoretic topological charge
\begin{equation}
q_\mr{nai}[U]
={1\ovr 4\pi^2}\int\!dx\;\trc(F_{12}F_{34}-F_{13}F_{24}+F_{14}F_{23}) 
={1\ovr32\pi^2}\int\!dx\;\trc(\ep_{\mu\nu\si\rh}F_{\mu\nu}F_{\si\rh}) 
\;.
\label{defqnai}
\end{equation}

I have implemented (\ref{defqnai}) with $F_{\mu\nu}$ defined as the hermitean
part of $-\mr{i}U_{\mu\nu}=-\mr{i}I+aF_{\mu\nu}+ ...$ where $U_{\mu\nu}(x)$ is
the mean of the four plaquettes in the $(\mu,\nu)$-plane with starting point
$x$.
The goal is to see what happens if one replaces $U_{\mu\nu}$ by one which is
constructed from smeared links.

Using the same 100 lattices (for each action) of size $12^4$ as in the previous
section, I decided to stay with a single parameter fixed at near-optimal
values (according to the $W(1,1)$ criterion), 
$\al_\mr{APE}\!=\!0.5$, $\al_\mr{HYP}\!=\!(0.3, 0.6, 0.75)$,
$\al_\mr{EXP}\!=\!0.2$ and three smearing levels $n_\mr{tot}\in\{1,3,7\}$.

\begin{figure}
\begin{center}
\includegraphics[width=84mm,angle=0]{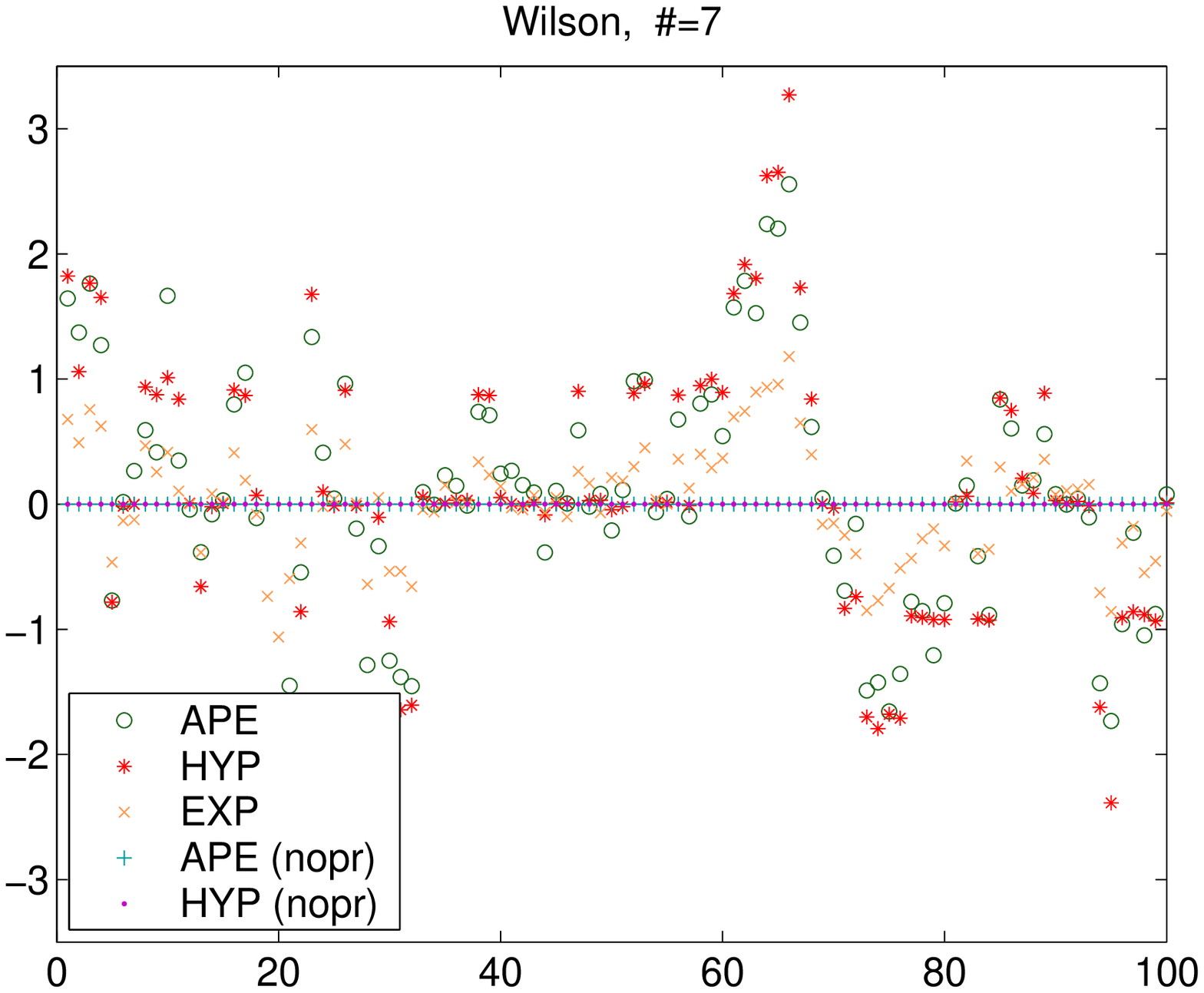}
\includegraphics[width=84mm,angle=0]{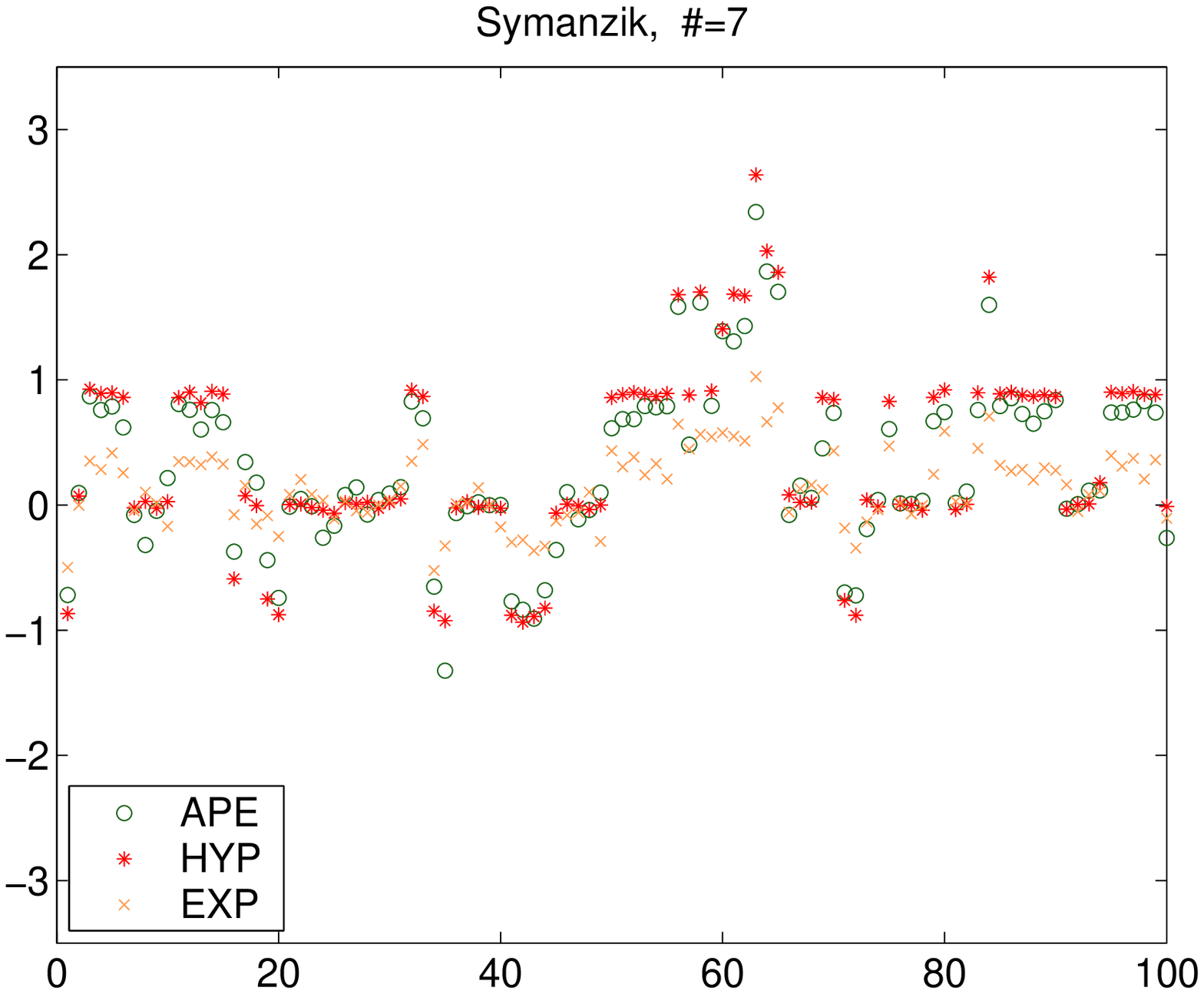}\\[4mm]
\includegraphics[width=84mm,angle=0]{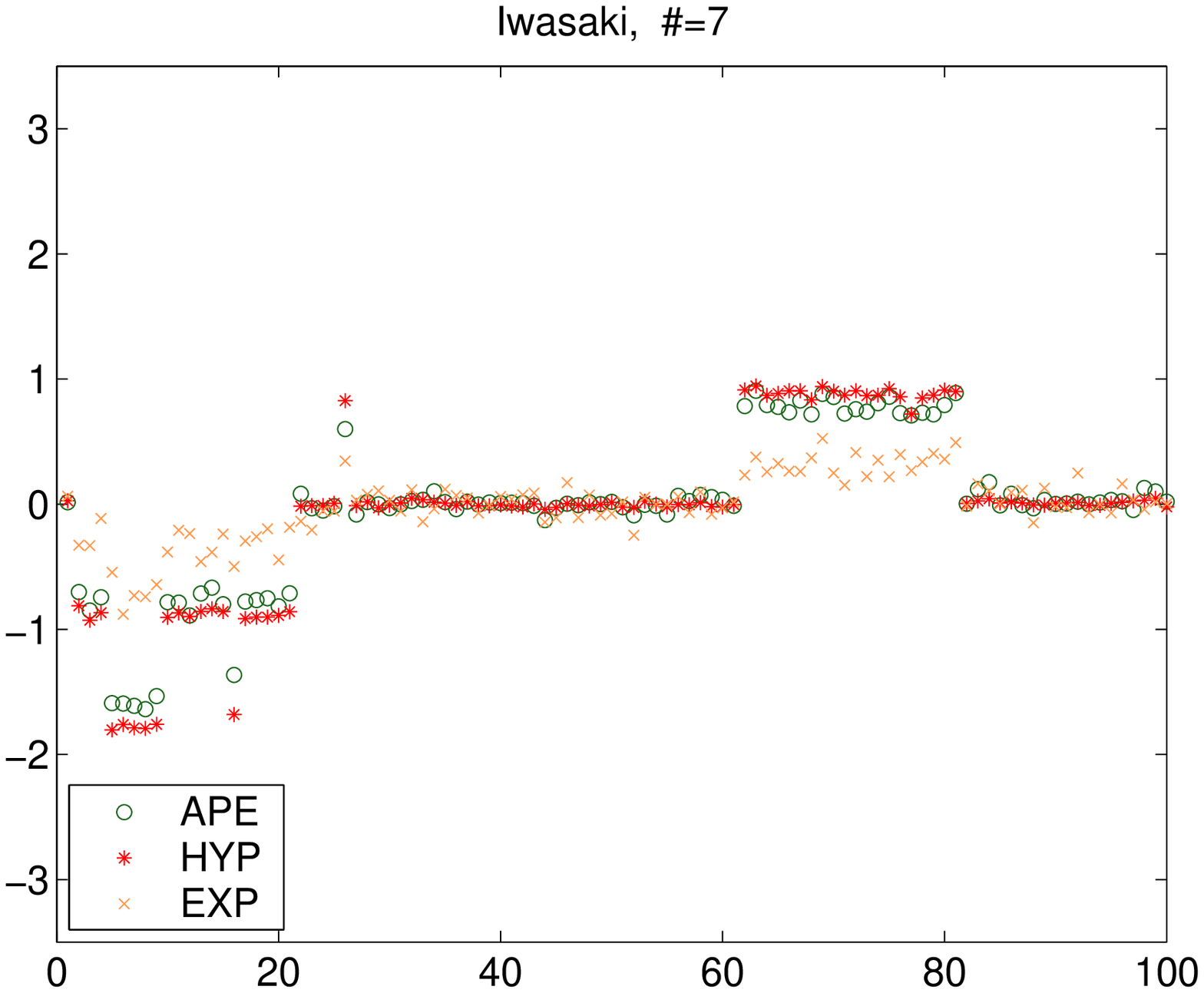}
\includegraphics[width=84mm,angle=0]{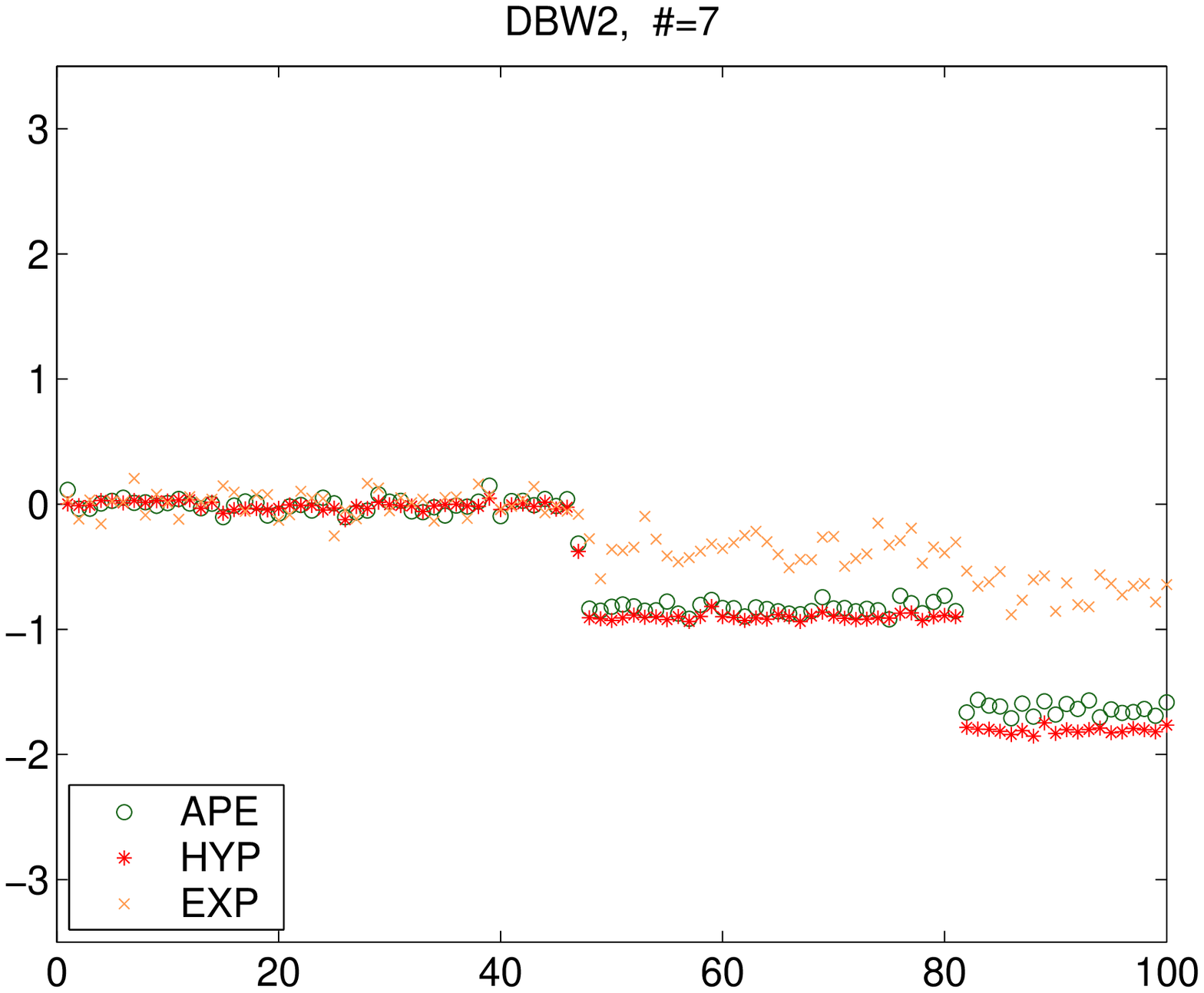}\\[4mm]
\includegraphics[width=84mm,angle=0]{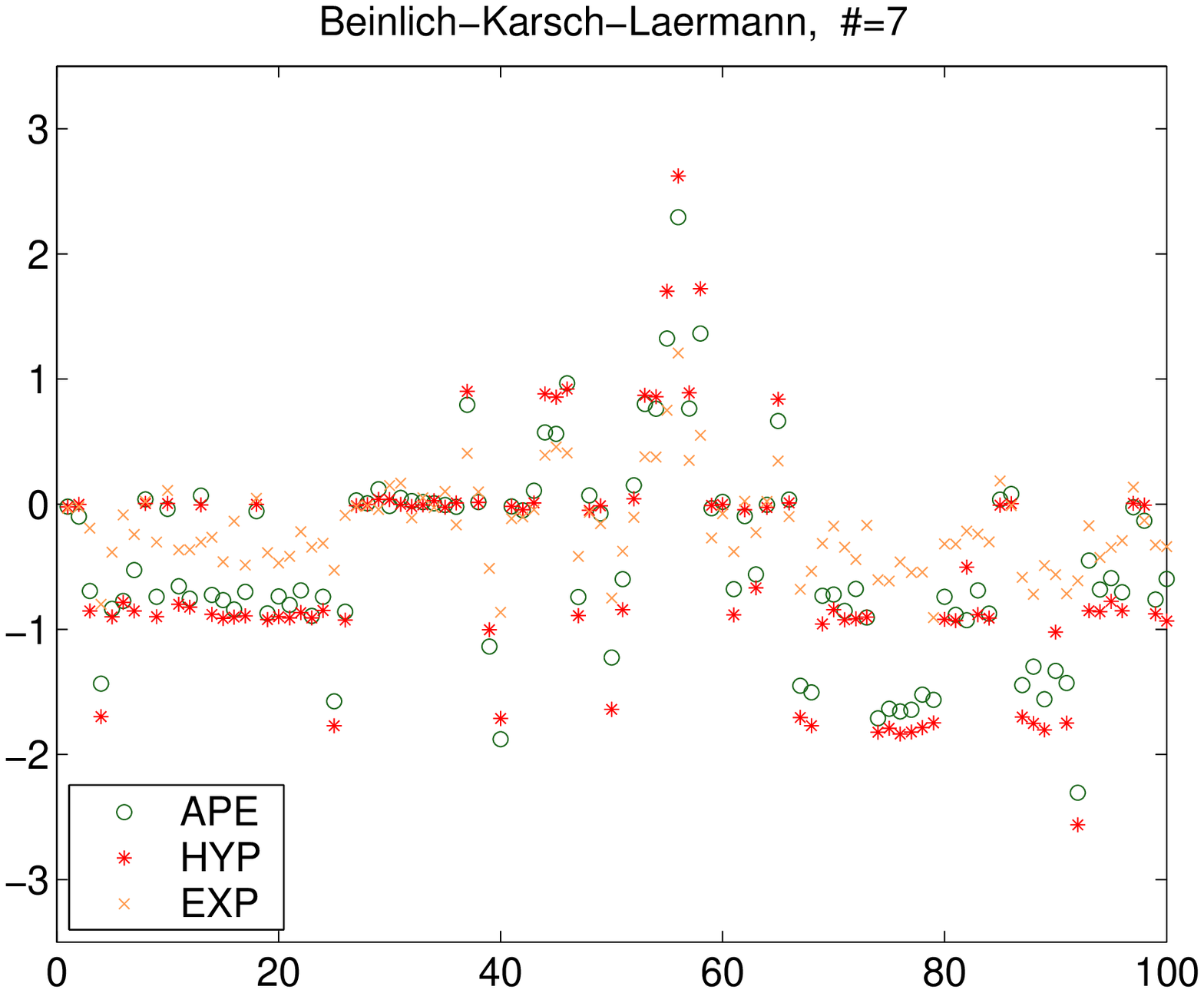}
\includegraphics[width=84mm,angle=0]{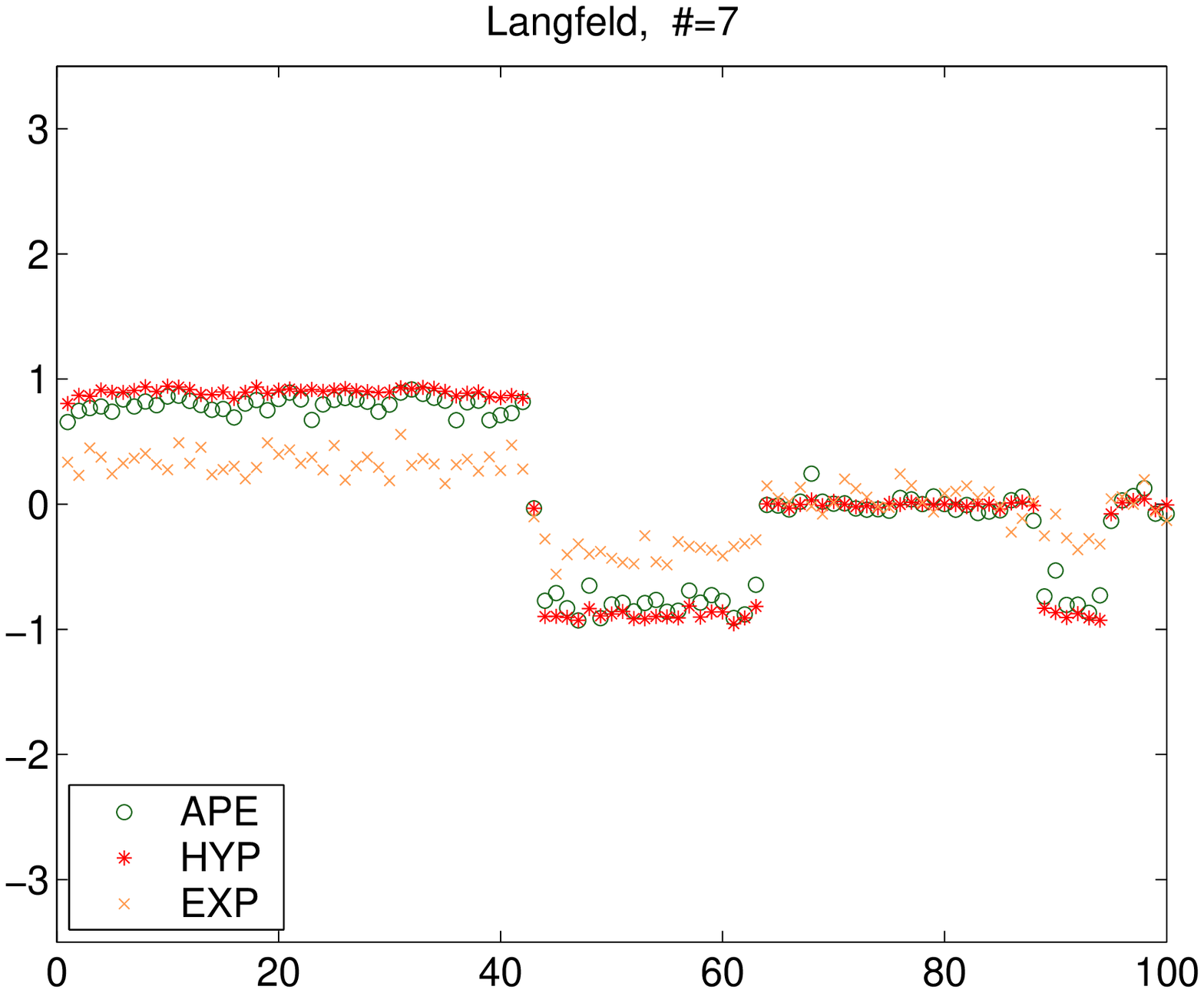}
\end{center}
\vspace{-6mm}
\caption{\sl Time history of the bare topological charge {\rm (\ref{defqnai})}
on the matched $12^4$ lattices after 7 smearing steps in the definition of
$F_{\mu\nu}$ (parameters as in Fig.\,\ref{fig:qtop_hist}). The projection-free
APE/ HYP-varieties always give a number very close to zero (omitted in the
improved action panels).}
\label{fig:qtop_time}
\end{figure}

\begin{figure}
\begin{center}
\includegraphics[width=84mm,angle=0]{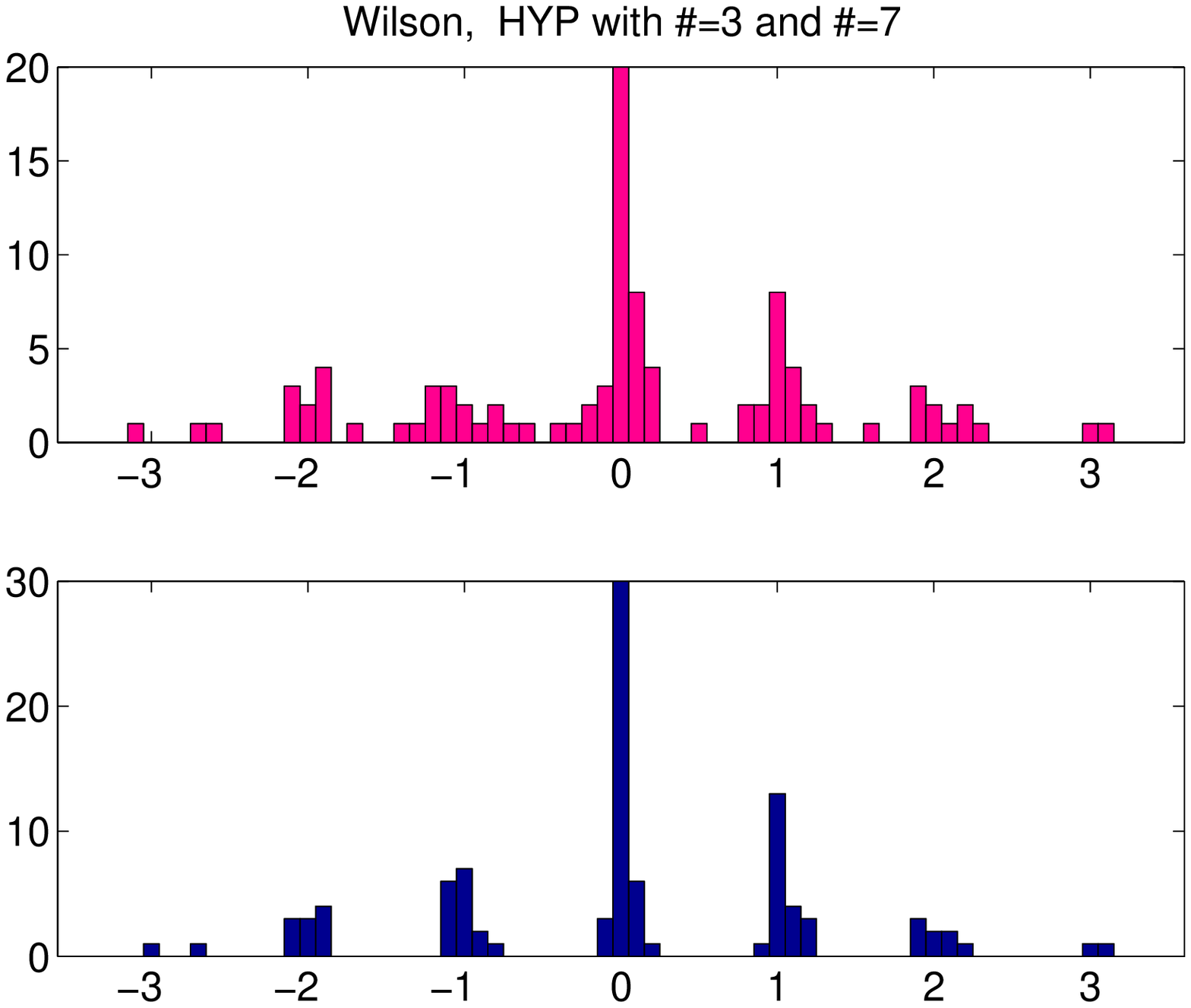}
\includegraphics[width=84mm,angle=0]{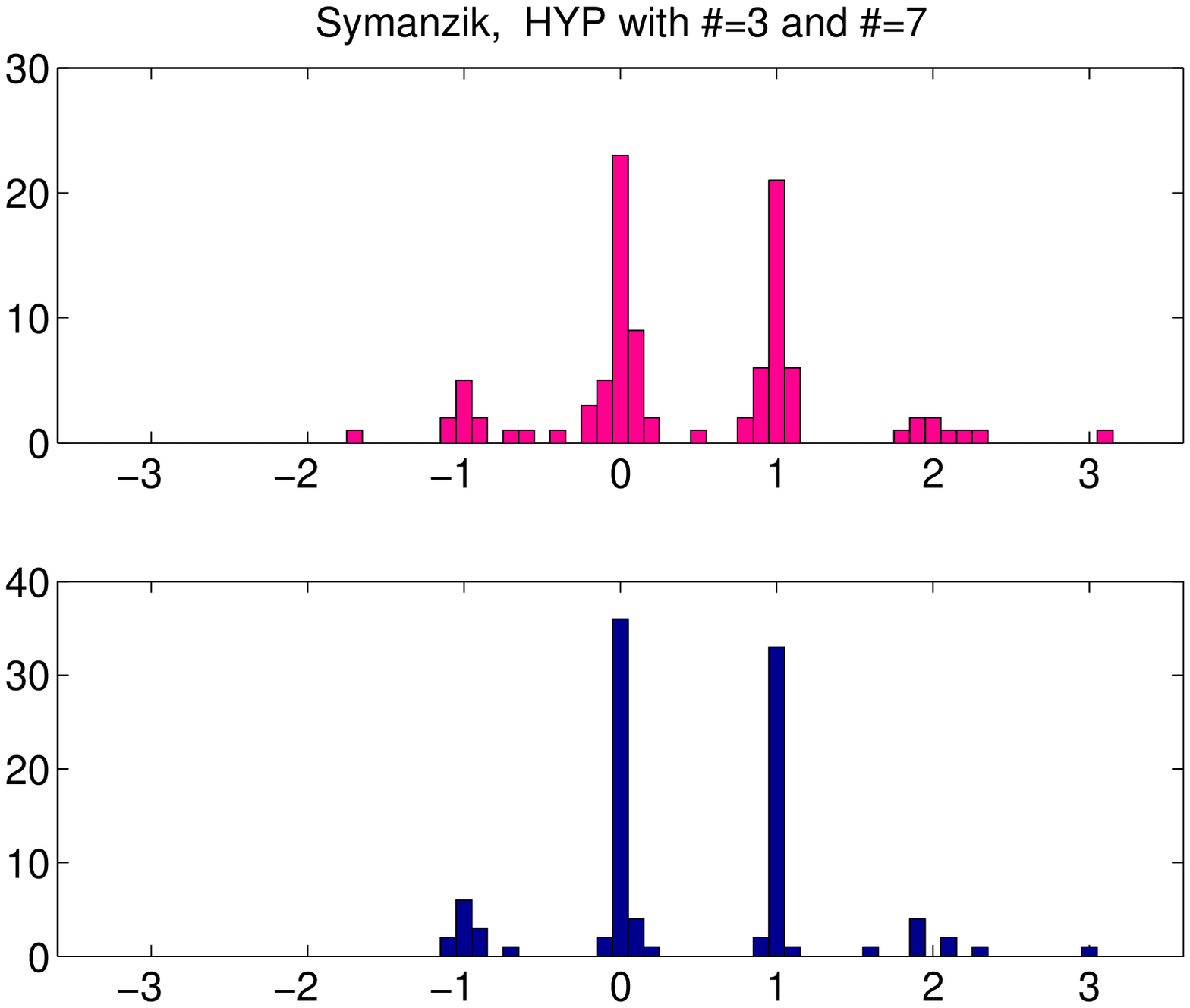}\\[4mm]
\includegraphics[width=84mm,angle=0]{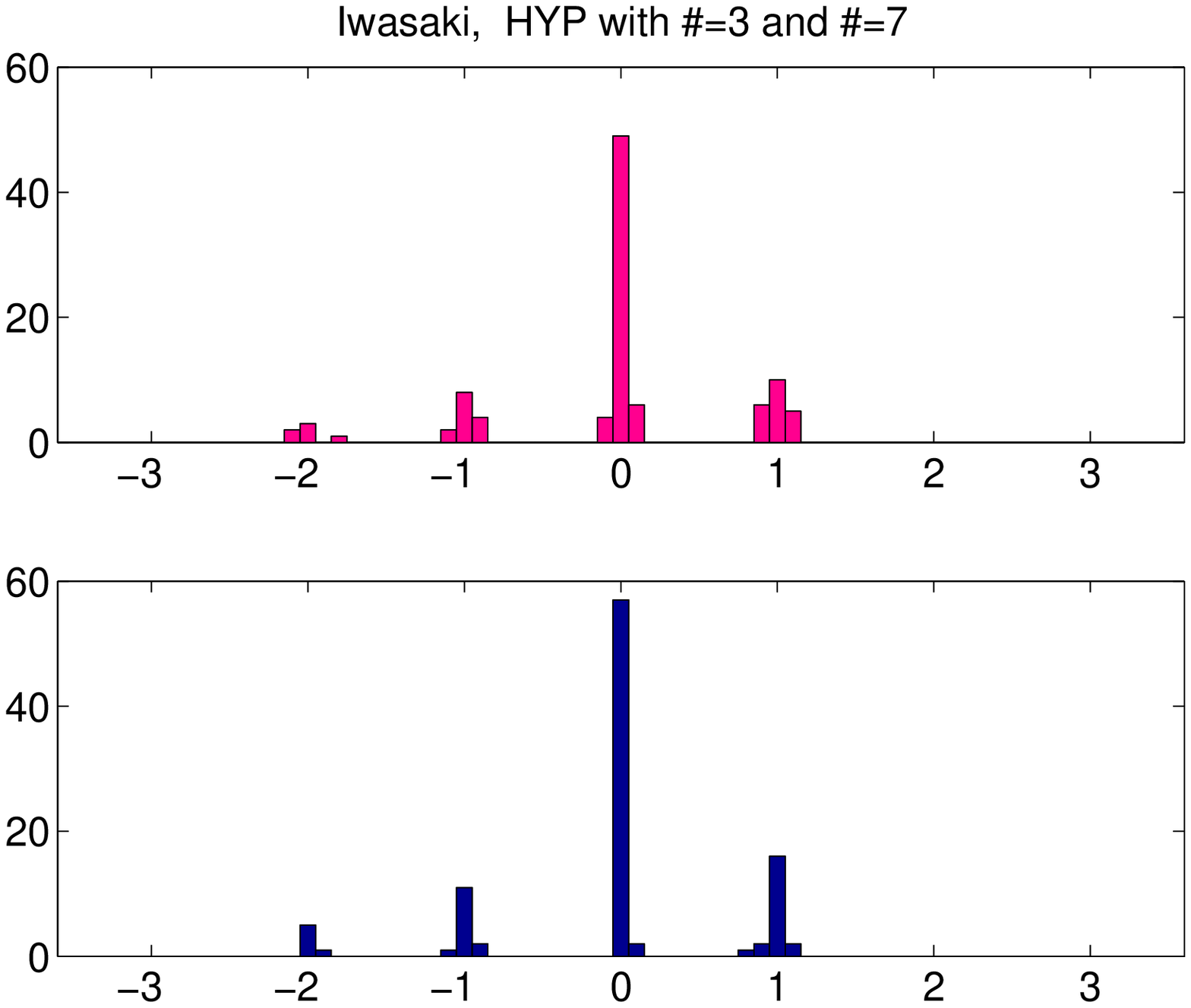}
\includegraphics[width=84mm,angle=0]{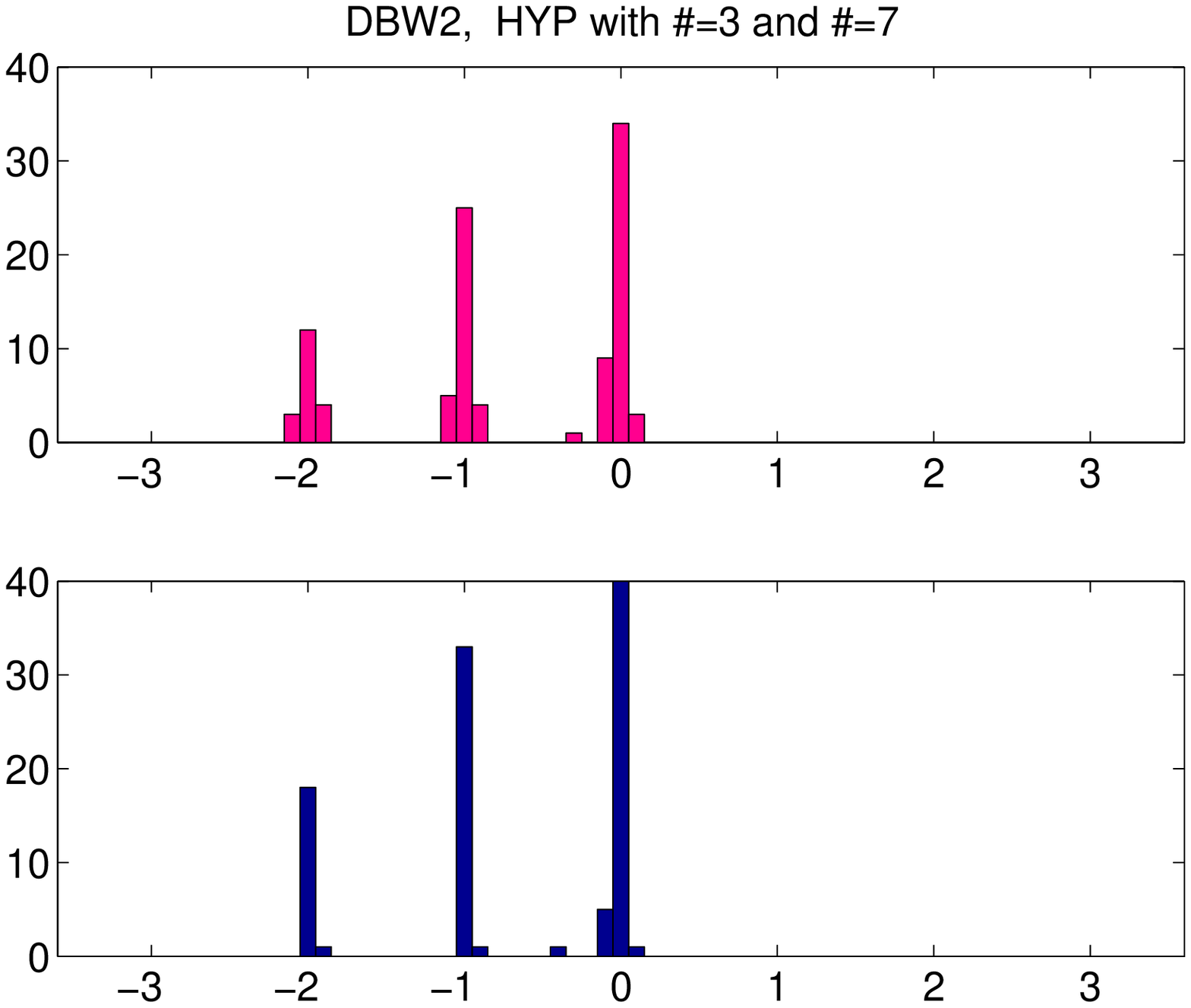}\\[4mm]
\includegraphics[width=84mm,angle=0]{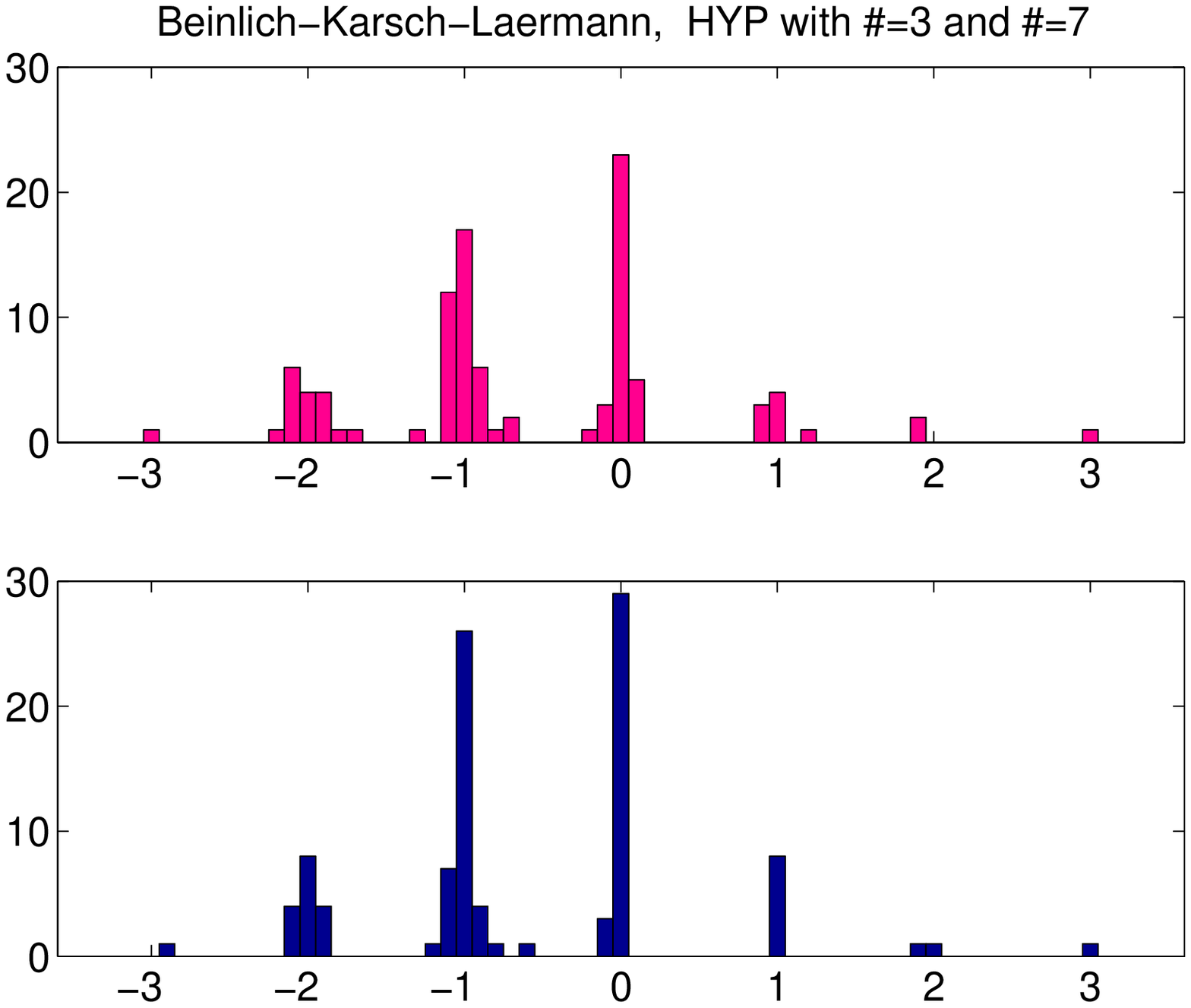}
\includegraphics[width=84mm,angle=0]{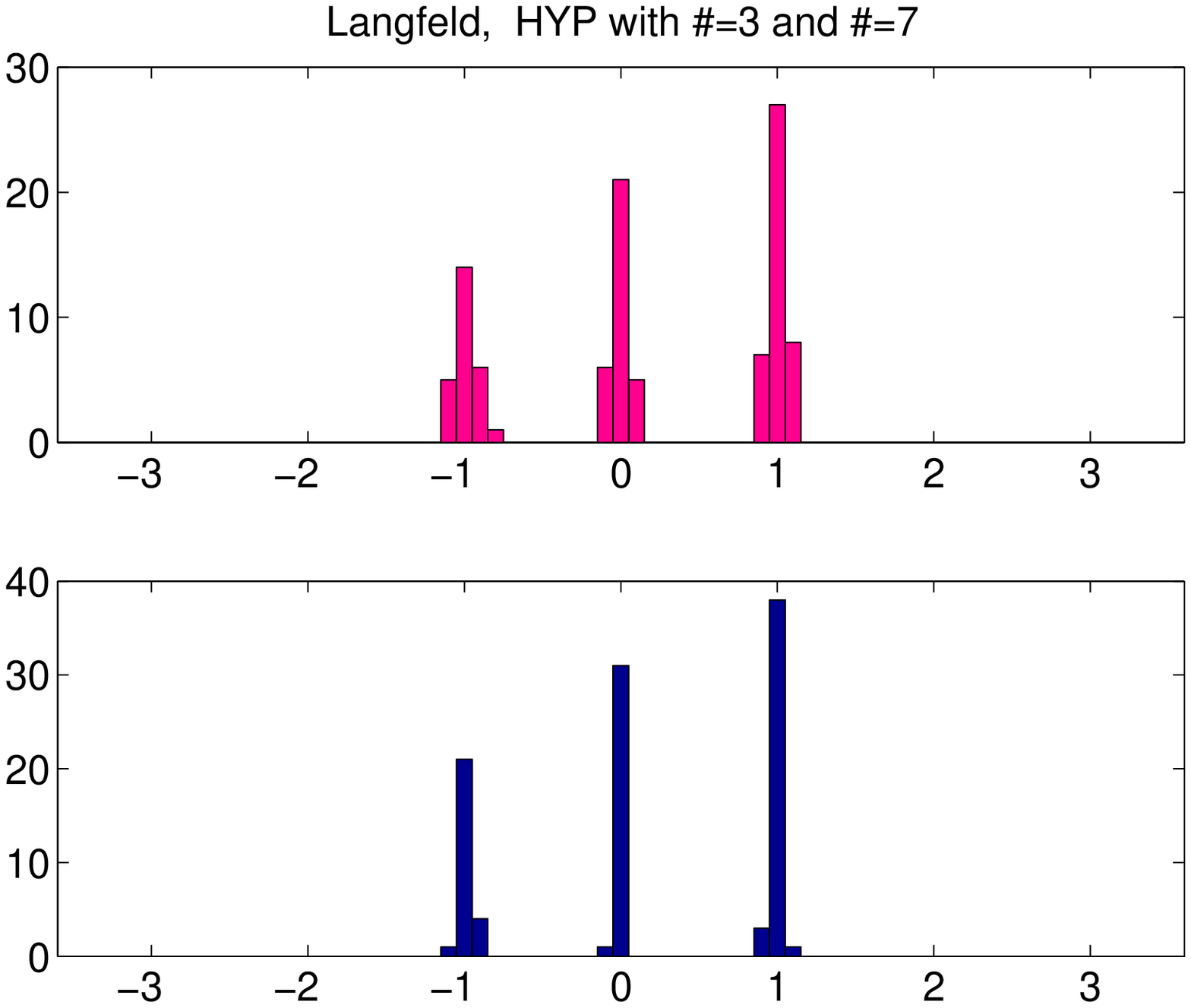}
\end{center}
\vspace{-6mm}
\caption{\sl Histograms of the semi-renormalized topological charges after 3
(light) and 7 (dark) steps of HYP-smearing. The appropriate $Z$-factors have
been applied, but not the ``cast-to-integer'' operation. The smearing
parameter is $\al_\mr{HYP}\!=\!(0.3, 0.6, 0.75)$.}
\label{fig:qtop_hist}
\end{figure}

\begin{figure}
\begin{center}
\includegraphics[width=84mm,angle=0]{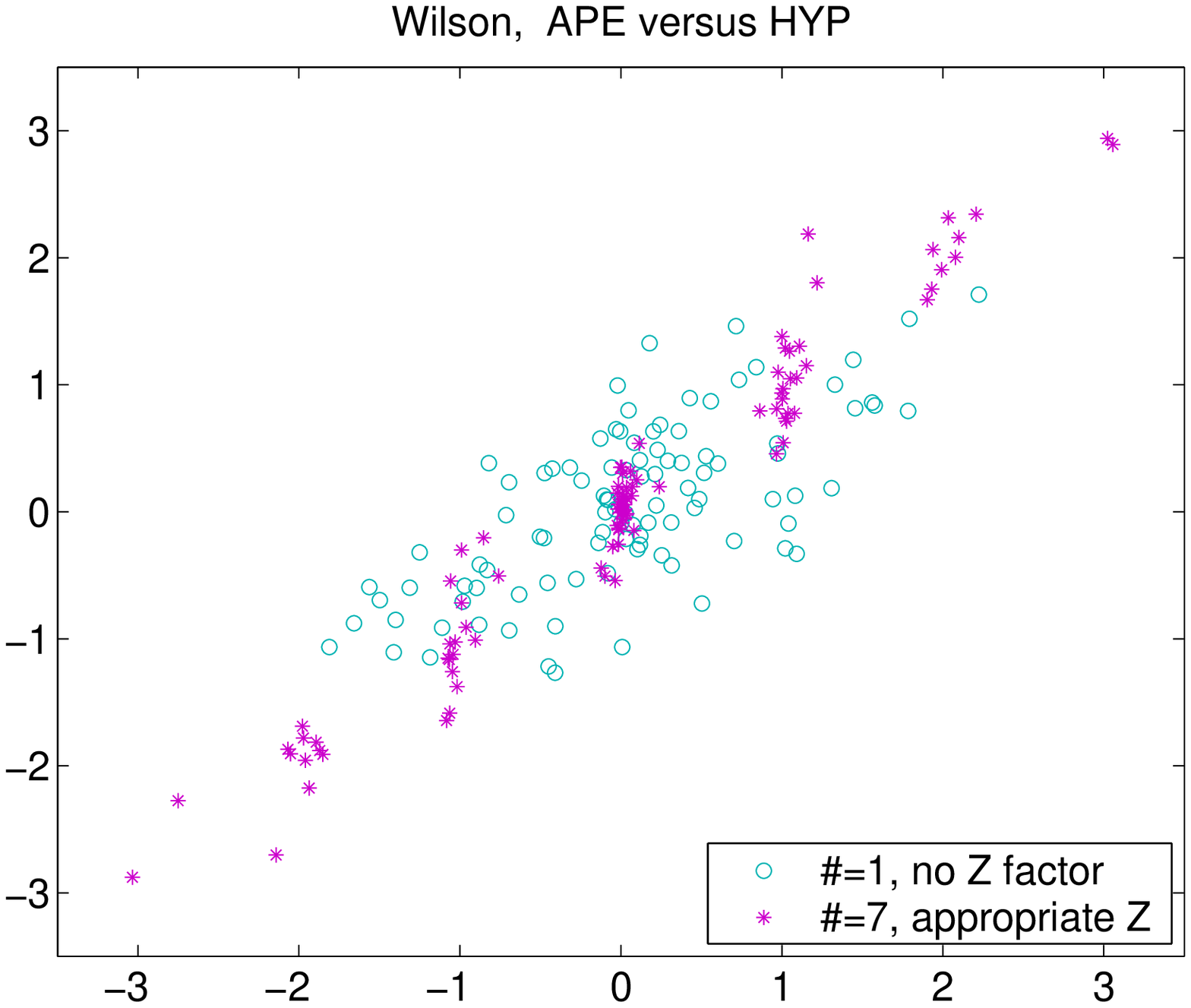}
\includegraphics[width=84mm,angle=0]{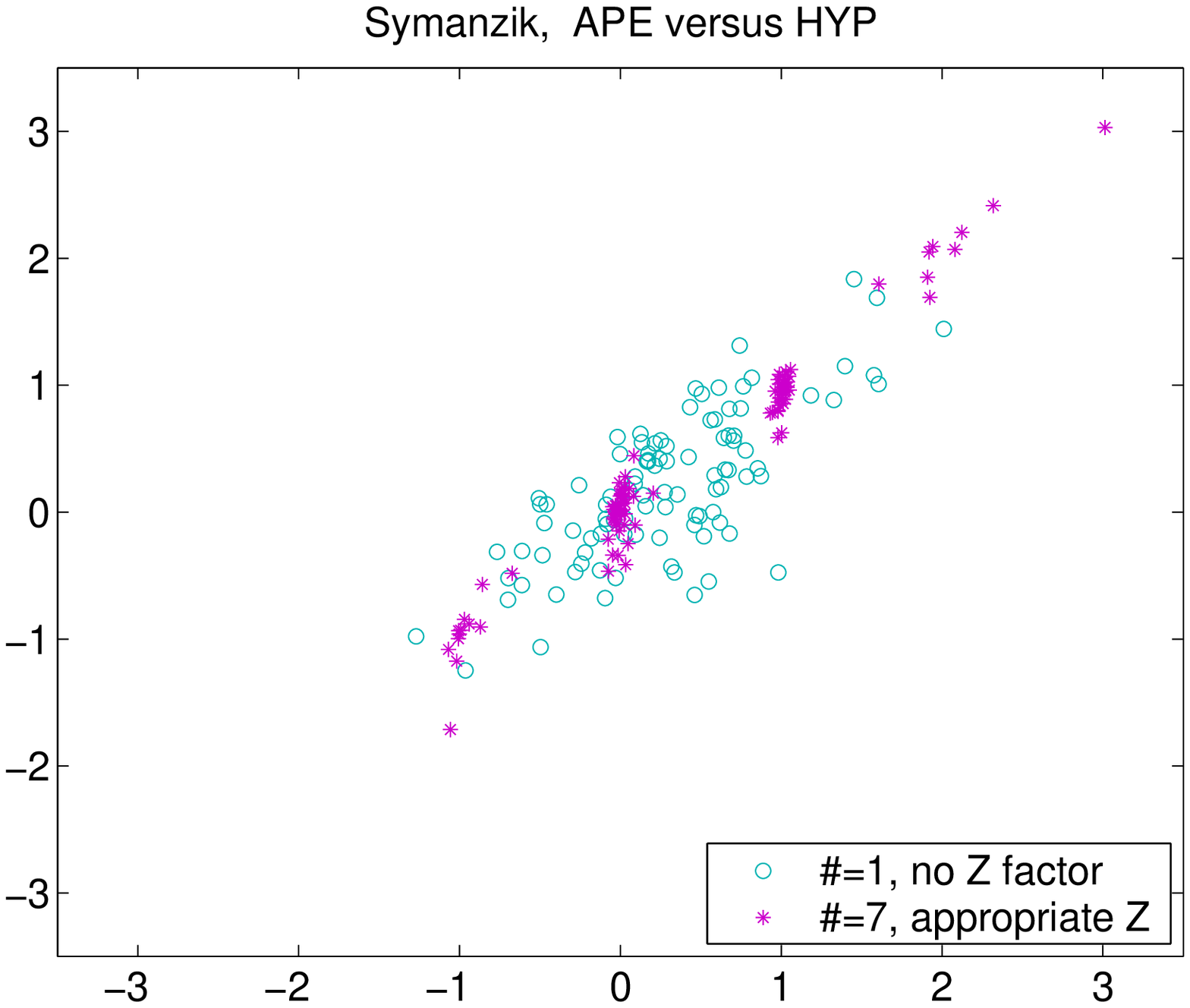}\\[4mm]
\includegraphics[width=84mm,angle=0]{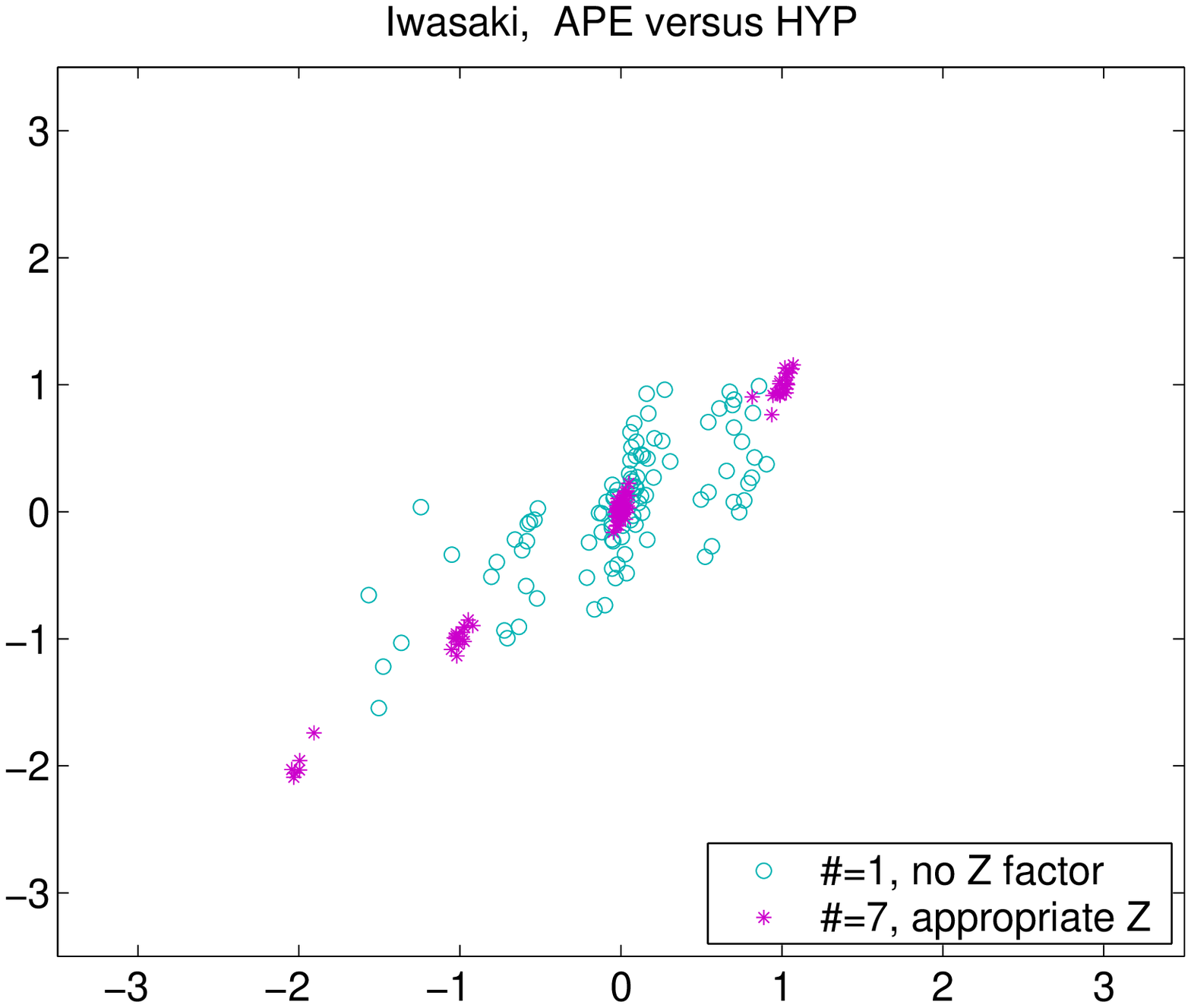}
\includegraphics[width=84mm,angle=0]{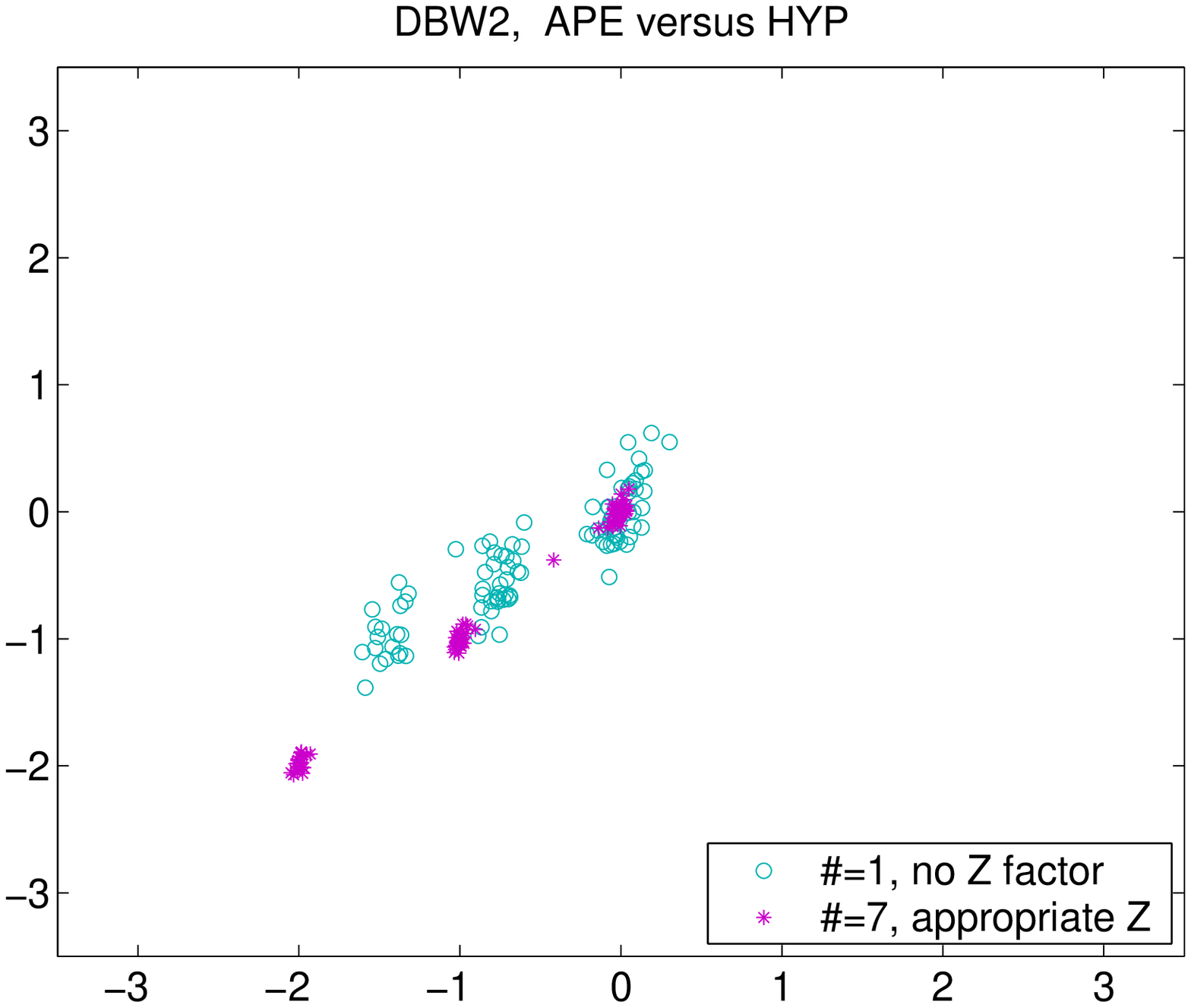}\\[4mm]
\includegraphics[width=84mm,angle=0]{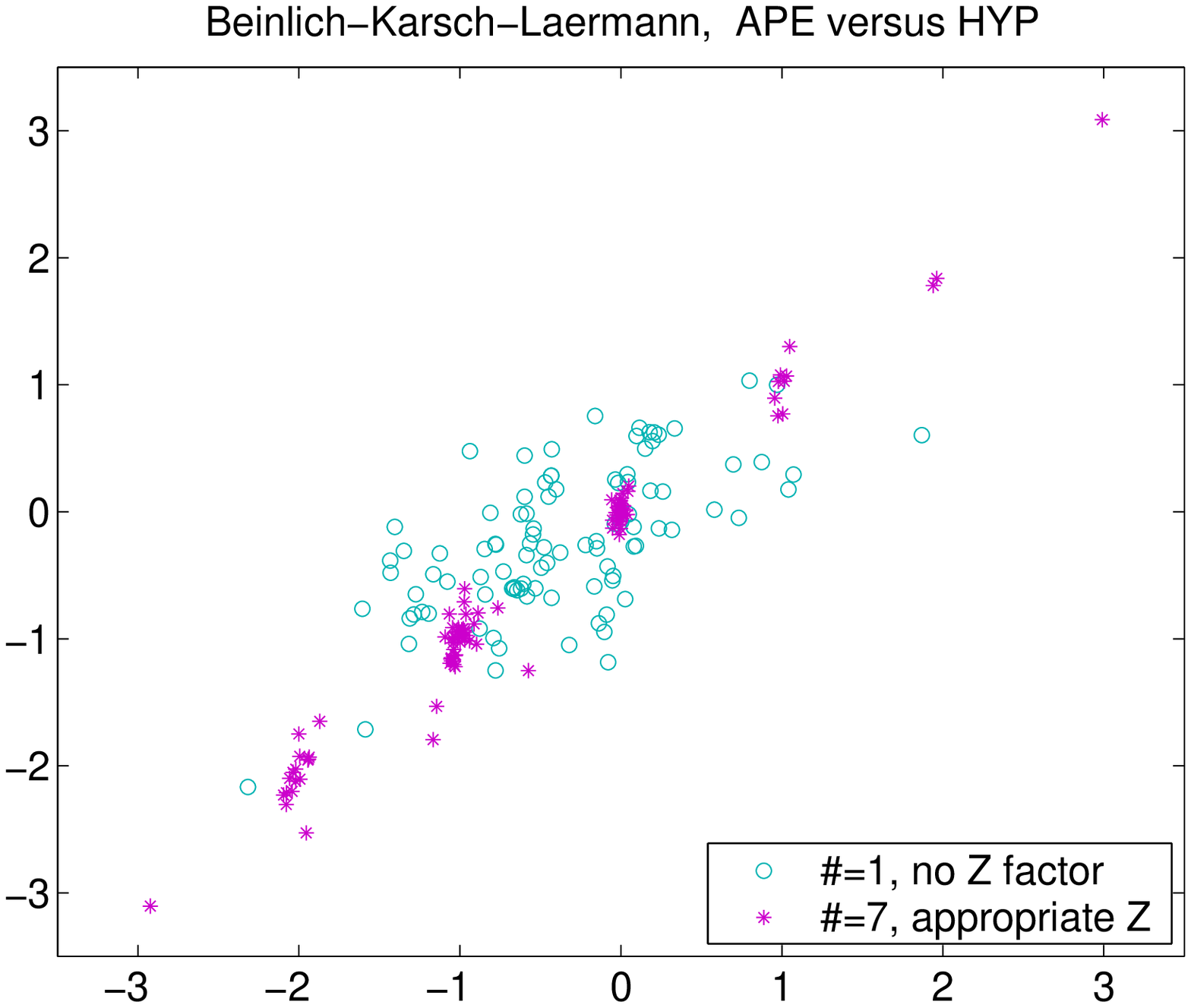}
\includegraphics[width=84mm,angle=0]{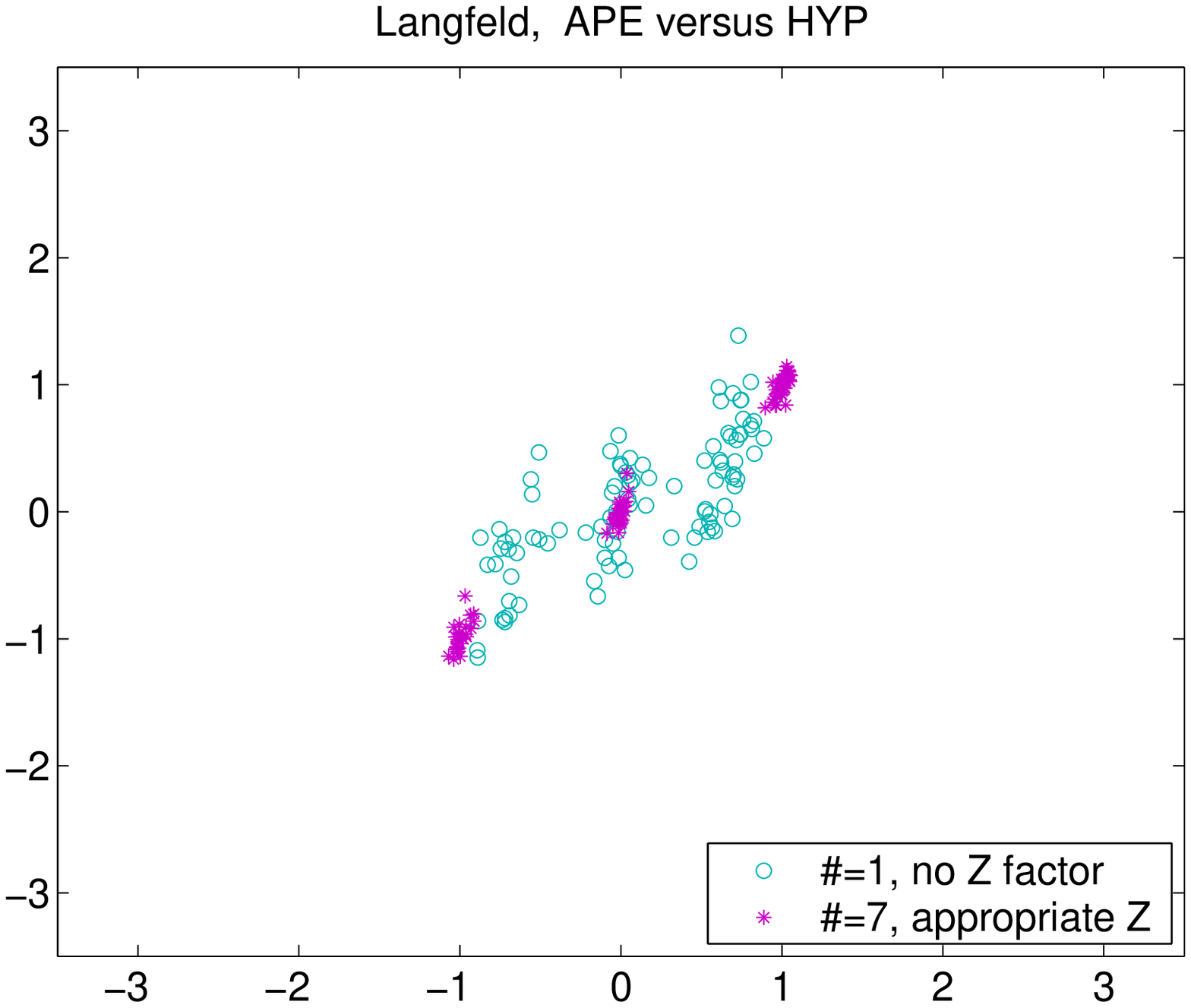}
\end{center}
\vspace{-6mm}
\caption{\sl Correlation of the unrenormalized topological charges after 1 step
of APE- or HYP-smearing $($light$)$ and of the semi-renormalized charges after
7 steps of APE- or HYP-smearing $($dark$)$. In the latter case, and with the
final ``cast-to-integer'' operation performed, the number of configurations
on which the APE- and HYP-charges disagree is $13, 2, 0, 0, 3, 0$,
respectively.}
\label{fig:qtop_corr}
\end{figure}

Fig.~\ref{fig:qtop_time} shows the time evolution of the bare topological
charge (\ref{defqnai}) after 7 smearing steps.
The APE/HYP varieties without projection tend to drive $q_\mr{nai}[U]$ to zero
for any gauge background; these symbols are shown only for the Wilson action.
The results of all other recipes tend to cluster near non-integer values
$n/Z$ with $n\!\in\!\mb{Z}$ and $Z$ a function of the smearing recipe (the
order is $1\!<\!Z_\mr{HYP}\!<\!Z_\mr{APE}\!\ll\!Z_\mr{EXP}$ for our parameters,
i.e.\ HYP and APE with projection require only moderate rescaling).
This suggests a renormalization factor $Z$ such that $Zq[U]$ tends to cluster
near integer values.
Define --~for a given smearing recipe and a given definition of $F_{\mu\nu}$ in
(\ref{defqnai})~-- the charge renormalization factor $Z$ as the solution of
\begin{equation}
\min_{Z>1}(\ch^2)
\quad\mr{where}\qquad
\ch^2=\sum_{U}\Big(Z\,q_\mr{nai}[U]-\mr{nint}(Z\,q_\mr{nai}[U])\Big)^2
\;.
\label{defzfactor}
\end{equation}
Having $Z$ at hand, one may define the renormalized ``naive'' field-theoretic
topological charge
\begin{equation}
q_\mr{ren}[U]=\mr{nint}(Z\,q_\mr{nai}[U])
\label{defqren}
\end{equation}
which, by construction, is an integer.
Employing such a definition one sees well-defined charge histories in each
graph of Fig.~\ref{fig:qtop_time}.
Obviously, we recover the standard wisdom that --~at a fixed physical lattice
spacing~-- the Wilson gauge action accomplishes topology changes more easily
than the Symanzik, Iwasaki and DBW2 action (in this order).
The same phenomenon happens for the $W^{2\times2}$-actions; the larger $c_2$
the more difficult is it to tunnel into another sector.

\bigskip

Coming back to the definition (\ref{defzfactor}, \ref{defqren}), the prime
question is whether the assigned charge $q_\mr{ren}[U]$ is really a property
of the gauge background $U$.
Fig.~\ref{fig:qtop_hist} contains the histograms of $Zq$ after 3 and 7
HYP-steps ($Z$ depends both on the smearing level and the gauge action).
Here, the valleys between the peaks are sufficiently depleted, and the
cast-to-integer operation in (\ref{defqren}) makes sense.
Unfortunately, there is no real valley structure at the levels 0 and 1.
Hence the question is to which extent such an integer-valued charge
$q_\mr{ren}[U]$ is a genuine function of the background $U$ and to which extent
it is an artifact of the smearing procedure.

The key to this issue is to compare different smearing recipes.
Fig.~\ref{fig:qtop_corr} relates the unrenormalized APE- and HYP-charges after
1 and 7 smearing steps.
In the former case there is some correlation between the unrenormalized
$q_\mr{nai}[U]$, but there is no clustering whatsoever.
In the latter case, i.e.\ after a few more smearing steps, the two rescaled
$Zq_\mr{nai}$ cluster predominantly near \emph{the same} integers.
With 100 configurations in each ensemble, on 87/98/100/100/97/100
the 7-APE-step and 7-HYP-step charges agree after renormalization.
Thus, for $a^{-1}\!\simeq\!2\GeV$ and an improved gauge action the two charges
agree almost on a configuration-by-configuration basis and this suggests that
such a charge is indeed a function of the background $U$.

Still, this does not mean that the renormalized ``naive'' field-theoretic
charge (\ref{defzfactor}, \ref{defqren}) is as much a clean concept as the
fermionic charge defined through the index of the Dirac operator
$q_\mr{ind}\!=\!{1\ovr2}\trc(\gaf\,D)$~\cite{Hasenfratz:1998ri,
Niedermayer:1998bi} with $D$ any Dirac operator that satisfies the
Ginsparg-Wilson relation~\cite{Ginsparg:1981bj}.
However, (\ref{defzfactor}, \ref{defqren}) has the virtue of being cheap in
terms of CPU-time; the $O(3...7)$ smearing steps cost less than the hundreds of
cooling steps that have been used in the past, and results are competitive with
those of the cooling era (see e.g.\ Ref.\,\cite{Grandy:1996mm} for a nice
example).
An issue that deserves further study is whether too much smearing eventually
destroys the charge.


\section{Summary}


The goal of this work has been a comparison of the newly proposed EXP smearing
\cite{exp} against the well established APE and HYP procedures \cite{ape,hyp}
on various gauge backgrounds.

I started by scanning through (part of) the space of parameters and 
iteration levels to see which combinations drive the
plaquette and larger Wilson loops to 1 rather than 0 and to see whether the
choice of the gauge action would have a big influence.
The result is that standard parameters around $\al_\mr{APE}\!\simeq\!0.65,
\al_\mr{HYP}\!\simeq\!(0.3,0.6,0.75), \al_\mr{EXP}\!\simeq\!0.15$ prove nearly
optimal, at least if $O(1..3)$ iterations are performed.
Obviously, the plaquette criterion is rather arbitrary, but for
$a^{-1}\!\simeq\!2\GeV$ the result is universal, i.e.\ independent of the
gauge action.

The only physics point of this paper has been an explicit test of the Lepage
argument in the heavy-heavy case.
This argument establishes a connection between the signal-to-noise ratio of an
observable involving a static quark and the HQ self energy, and our
Fig.\,\ref{fig:signal_noise} shows that it passed the test with flying colors.
Hence a future optimization strategy tailored to HQ physics might be to reduce
the HQ self energy as much as possible.
The key insight is that in the case of large Wilson loops various smearing
recipes differ in the signal generated, not in the noise.

Next I evaluated the squared string tension defined via the Creutz ratio at
$R\!=\!T\!\simeq\!0.5\fm$.
The main finding is that any smearing recipe drastically reduces the noise and
a detailed tuning of parameters and/or iteration level is neither needed nor
useful.
Instead there is a broad two-dimensional plateau in which changing the two does
not significantly alter the central value or the error.
Only at absurdly high levels systematic effects get visible, presumably due to
the reduced locality in terms of the original links.
In summary, moderate smearing suggests itself as a simple and efficient means
to damp the UV-noise in standard gluonic observables.

Finally, I have focused on the ``naive'' field-theoretic topological charge.
Here, the power of any smearing recipe depends a lot on the gauge action used.
The good news is that the lattice spacings where the smearing details implicit
in the definition (\ref{defqnai}, \ref{defzfactor}, \ref{defqren}) prove
irrelevant are accessible; for the Iwasaki action it is around
$a\!\simeq\!0.1\fm$, for other actions not much below.


\bigskip\noindent{\bf Acknowledgments}:
It is a pleasure to acknowledge useful correspondence with Silvia Necco as well
as the benefits from a discussion with Rainer Sommer, Michele Della Morte
and Francesco Knechtli on the Lepage argument.
Computations were done on a PC at DESY Zeuthen.
This work has been supported by the DFG in the framework SFB/TR-9.


\section*{Appendix: Projection details}


Here I spell out the projection procedure introduced in \cite{Kiskis:2003rd}.
For related ideas see \cite{Christou:1995zn,Giusti:1999be}.

The idea is to define the projection step $V_\mu(x)\to W_\mu(x)$ via the
overlap-type recipe
\begin{equation}
W_\mu(x)={V_\mu(x)\,[V_\mu(x)\dag V_\mu(x)]^{-1/2}\ovr
[\det(\mr{numerator})]^{1/3}}
\end{equation}
which leads to an eigenvector decomposition problem of the hermitean operator
$X\!\equiv\!V\dag V$.

In principle, one might find the eigenvalues $x_1,...,x_3$ of $X$ (following
Cardano) by forming
\begin{eqnarray}
a&=&-\trc(X)\nonumber\\
b&=&X_{11}X_{22}+X_{22}X_{33}+X_{33}X_{11}
   -X_{12}X_{21}-X_{23}X_{32}-X_{31}X_{13}\nonumber\\
c&=&-\det(X)\nonumber\\
p&=&b-{a^2\ovr3}\nonumber\\
q&=&{2a^3\ovr27}-{ab\ovr3}+c\nonumber
\end{eqnarray}
and continuing with
\begin{equation}
\mr{if}\;p=0\;\mr{then}\;
u_1=\left\{
\begin{array}{ll}
0&\mr{if}\;q=0\\
({1\ovr2}+\ii{\sqrt{3}\ovr2})\sqrt[3]{q})&\mr{if}\;q>0\\
\sqrt[3]{-q}&\mr{if}\;q<0
\end{array}
\right\}\;\mr{and}\;
\begin{array}{l}
x_1=u_1-{a\ovr3}
\\
x_2=(-{1\ovr2}+\ii{\sqrt{3}\ovr2})u_1-{a\ovr3}
\\
x_3=(-{1\ovr2}-\ii{\sqrt{3}\ovr2})u_1-{a\ovr3}
\end{array}
\label{casepeqzero}
\end{equation}
\begin{equation}
\mr{else}\;
\begin{array}{ll}
u_1\!&\!=\;\sqrt[3]{-{q\ovr2}+\sqrt[2]{{q^2\ovr4}+{p^3\ovr27}}}\\
v_1\!&\!=\;-{p\ovr3u_1}
\end{array}\;\;\mr{and}\;\;
\begin{array}{l}
x_1=u_1+v_1-{a\ovr3}
\\
x_2=(-{1\ovr2}+\ii{\sqrt{3}\ovr2})u_1+
    (-{1\ovr2}-\ii{\sqrt{3}\ovr2})v_1-{a\ovr3}
\\
x_3=(-{1\ovr2}-\ii{\sqrt{3}\ovr2})u_1+
    (-{1\ovr2}+\ii{\sqrt{3}\ovr2})v_1-{a\ovr3}
\label{casepnezero}
\end{array}
\end{equation}
where the roots in (\ref{casepnezero}) must be taken over the body of complex
numbers, though the result $x_1,...,x_3$ is real.

A practical problem emerges whenever $X$ is close to unity (what is generic
after a few smearing iterations), as the naive implementation cannot separate
the eigenvalues any more.
In order to reduce numerical extinction, one should implement
\begin{equation}
Y={1\ovr2}(V\dag\!-\!1)(V\!-\!1)+\mr{h.c.}+(V\dag\!-\!1)+(V\!-\!1)
\qquad\Big(=X\!-\!1\Big)
\end{equation}
rather than $X$ and trade, for the same reason, the quantities $a,b,c$ for
alternative ones designed to involve only ``small'' numbers in the case
$X\!\simeq\!1$:
\begin{eqnarray}
a_3\!&\!=\!&\!-Y_{11}-Y_{22}-Y_{33}
\qquad\Big(=a\!+\!3\simeq0\Big)
\label{dettricka}
\\
b_3&=&Y_{11}Y_{22}
     +Y_{22}Y_{33}
     +Y_{33}Y_{11}
     -Y_{12}Y_{21}-Y_{23}Y_{32}-Y_{31}Y_{13}-2a_3
\qquad\Big(=b\!-\!3\simeq0\Big)
\label{dettrickb}
\\
c_1&=&-Y_{11}Y_{22}Y_{33}
      -Y_{12}Y_{23}Y_{31}-Y_{32}Y_{21}Y_{13}\nonumber\\&&
      +Y_{11}Y_{23}Y_{32}
      +Y_{22}Y_{31}Y_{13}
      +Y_{33}Y_{12}Y_{21}-b_3-a_3
\quad\Big(=c\!+\!1\simeq0\Big)
\label{dettrickc}
\\
p&=&b_3-{a_3^2\ovr3}+2a_3
\\
q&=&{2a_3^3\ovr27}-{2a_3^2\ovr3}+a_3-{a_3b_3\ovr3}+b_3+c_1
\end{eqnarray}
before one proceeds with (\ref{casepeqzero}, \ref{casepnezero}), where again
$y_1=u_1\;[+v_1]\,-{a_3\ovr3}$, etc.\ should be formed rather than
$x_1, x_2, x_3$.
Knowing the eigenvalues (minus 1) with good precision, it is easy to find
the eigenvectors of $X$ by solving the characteristic equation
\begin{equation}
Yv_i=y_iv_i\qquad(i\in\{1,..,3\})
\end{equation}
and after normalizing them one forms
\begin{equation}
W={V\,\sum\limits_{i=1}^3 v_i{1\ovr\sqrt{y_i+1}}v_i\dag
\ovr[\det(\mr{numerator})]^{1/3}}
\;.
\end{equation}

Of course this costly projection procedure is only needed if the argument is
potentially far away from $SU(3)$; whenever $V$ is known to be close to $SU(3)$
(e.g.\ in the MC a newly proposed link is projected to $SU(3)$ before it is
subject to the accept/reject-test, but here the deviation is just due to
round-off errors) one may project via an approximate procedure, e.g.\ via
\begin{equation}
V\;\longrightarrow\;V'={3\ovr2}V-{1\ovr2}VV\dag V\;,\qquad
V'\;\longrightarrow\;V''=V'\Big(1-{1\ovr3}(\det V'-1)\Big)
\;.
\end{equation}

Finally, it is worth mentioning that the same eigenmode decomposition technique
(simplified by $a\!=\!0$) has been used in the matrix exponentiation in the EXP
recipe (\ref{defexp}).

\clearpage



\end{document}